\UseRawInputEncoding
\documentclass[12pt, a4paper, twoside, titlepage, openright]{book}

\usepackage{float}
\usepackage{amsfonts}
\usepackage{dsfont}
\usepackage{amsmath}        
\usepackage{amssymb}
\usepackage{amsthm}        
\usepackage{latexsym} 
\usepackage[mathcal]{euscript}      
\usepackage{graphicx}                 
\usepackage{color}                    
\usepackage{hyperref}              
\usepackage{setspace}      
\usepackage{textcomp}			 
\usepackage{marvosym} 
\usepackage{comment}
\usepackage{dsfont}
\usepackage{dutchcal}

\newcommand{\HDRname}{\Large Habilitation \`a Diriger des Recherches}
\newcommand{\HDRfield}{\Large (Informatique)}
\newcommand{\HDRpresented}{M\'emoire pr\'esent\'e par}
\newcommand{\HDRjury}{\textbf{Jury}}
\newcommand{\HDRpresident}{Pr\'esident du jury}

\newcommand{\hdrheader}[1]{%
	\text{#1}\\[0.8em]
}

\newcommand{\jurytitle}{%
	\vspace{1.6cm}
	\hdrheader{\HDRjury}
}

\newenvironment{jurylist}{
	\begin{tabular}{@{}p{4.5cm}p{5.5cm}p{3.5cm}@{}}
	}{
	\end{tabular}
}

\newcommand{\jurymember}[3]{%
	#1 & #2 & #3\\
}

\newcommand{\toquote}[2]{\begin{flushright}{\singlespacing\small\em #1 \\ ---#2}\end{flushright}}

\newcommand{\toabstract}[1]{\vspace{1.5cm}\begin{flushright}\rule{10.04cm}{0.05cm}\vspace{-0.35cm}\\
 \parbox{10cm}{\singlespacing\small #1}\\
\vspace{0.04cm} \rule{10.04cm}{0.05cm}\end{flushright}}

\newcommand\id{\mathds{1}}
\newcommand{\Tr}{\text{Tr}}

\newcommand{\ket}[1]{| #1 \rangle}
\newcommand{\bra}[1]{\langle #1 |}
\newcommand{\braket}[2]{\langle #1 | #2 \rangle}
\newcommand{\ketbra}[2]{|#1 \rangle\langle #2|}

\newcommand{\cH}{\mathcal{H}}

\newtheorem{Th}{Theorem}[chapter]
\newtheorem*{Thnon}{Theorem}

\newtheorem{Def}{Definition}[chapter]

\makeindex
\begin{document}

\frontmatter
\singlespace\pagestyle{plain}\pagenumbering{roman}

\begin{titlepage}
	\centering
	
	\vspace*{\fill}
	
	\hdrheader{\Large Aix-Marseille Universit\'e}
	
	\hdrheader{ \HDRname}
	\hdrheader{\HDRfield}
	
	\vspace{3cm}
	
	{\bfseries\Huge Quantumness via Discrete Structures}\par
	
	\vspace{4cm}
	
	\hdrheader{\HDRpresented}
	{\Large Ravi Kunjwal}\par
	
	\jurytitle
	
	\begin{jurylist}
		\jurymember
		{Cyril Branciard}
		{Institut N\'eel, CNRS}
		{Rapporteur}
		
		\jurymember
		{Mio Murao}
		{University of Tokyo}
		{Rapporteure}
		
		\jurymember
		{Ion Nechita}
		{Universit\'e Toulouse 3}
		{Rapporteur}
		
		\jurymember
		{Pierre Clairambault} 
		{Aix-Marseille Universit\'e}
		{\HDRpresident}
		
		\jurymember
		{Damian Markham}
		{Sorbonne Universit\'e}
		{Membre}
		
		\jurymember
		{Giuseppe Di Molfetta}
		{Aix-Marseille Universit\'e}
		{Tuteur}
	\end{jurylist}
	
	\vspace*{\fill}
	
	December 8, 2025
	
\end{titlepage}

\cleardoublepage
\section*{Abstract}\normalsize
 \addcontentsline{toc}{section}{Abstract}
Quantum theory departs from classical probabilistic theories in foundational ways. These departures---termed \textit{quantumness} here---power quantum information and computation. 
This thesis charts the role of discrete structures in assessing quantumness, synthesizing elements of my postdoctoral research through this lens. 
After an introduction to the necessary background concepts, I present my work under three broad categories. 
First, I present work on contextuality that extensively relies on (undirected) graphs and hypergraphs as the discrete structures of interest; more specifically, it relies on invariants associated with them. This work includes Kochen-Specker (KS) contextuality and its operationalization to generalized contextuality, expressed via (hyper)graph-theoretic frameworks. I also present work on KS-contextuality in multiqubit systems and an application of generalized contextuality to a one-shot communication task, both of which rely on hypergraphs.
Second, I present work on causality, where the discrete structures of interest are directed graphs. 
This includes work on indefinite causal order, specifically its connections to the gap between local operations and classical communication (LOCC) and separable operations (SEP), and a device-independent notion of nonclassicality---termed \textit{antinomicity}---that generalizes Bell nonlocality without global causal assumptions. 
Finally, I present work on the incompatibility of quantum measurements, its connection to Bell nonlocality, and its role in discriminating between quantum and almost quantum correlations in the single-system setting. The discrete structures of interest here are hypergraphs that model joint measurability relations between quantum measurements. I conclude with a summary and an overview of work that is \textit{not} covered in this thesis.
\clearpage
 \addcontentsline{toc}{section}{Acknowledgements}  
 \textit{To Ija, Baju, Hina, and Wowes}
\clearpage

\section*{Collaborations and publications}\normalsize
 \addcontentsline{toc}{section}{Collaborations and publications}  

Following is a chapter-wise summary of the collaborations and publications covered in this thesis:

\begin{itemize}
	\item Chapter \ref{chap:hypergraphframeworks} covers the essential aspects of the hypergraph frameworks for generalized contextuality that were laid out in two publications, namely, 
	
	\begin{itemize}
		\item \textit{Beyond the Cabello-Severini-Winter framework: making sense of contextuality without sharpness of measurements}, R. Kunjwal, \href{https://doi.org/10.22331/q-2019-09-09-184}{Quantum 3, 184 (2019)}, and
		\item \textit{Hypergraph framework for irreducible noncontextuality inequalities from logical proofs of the Kochen-Specker theorem}, R. Kunjwal, \href{https://doi.org/10.22331/q-2020-01-10-219}{Quantum 4, 219 (2020)}.
	\end{itemize}
	
	It also covers an application of these frameworks to the problem of entanglement-assisted one-shot classical communication considered in the following publication (completed under my supervision):
	
	\begin{itemize}
		\item \textit{Contextuality in entanglement-assisted one-shot classical communication}, S.A. Yadavalli and R. Kunjwal, \href{https://doi.org/10.22331/q-2022-10-13-839}{Quantum 6, 839 (2022)}.
	\end{itemize}
	
	This chapter is based on collaborations with Robert W.~Spekkens (Perimeter Institute, Canada) and Shiv A.~Yadavalli (Duke University, USA).
	 
	\item Chapter \ref{chap:ContextualityEntanglement} covers the essential aspects of results on the interplay between KS-contextuality, Bell nonlocality, and Gleason's theorem in the case of multiqubit systems, and how these provide insights into models of multiqubit quantum computation. This chapter is based on the following joint work with Victoria Wright (now a researcher at Quantinuum):
	
	\begin{itemize}
		\item \textit{Contextuality in composite systems: the role of entanglement in the Kochen-Specker theorem}, V.J. Wright and R. Kunjwal, \href{https://doi.org/10.22331/q-2023-01-19-900}{Quantum 7, 900 (2023)}.
	\end{itemize}
	
	This work was initiated as a project under my supervision while Victoria was a PhD student from the University of York (UK) visiting the Perimeter Institute (where I was a postdoctoral researcher).
	
	\item Chapter \ref{chap:CorrelationsICO} covers the essential aspects of work on multipartite correlations in the presence of indefinite causal order presented in following publication and preprints:
	
	\begin{itemize}
		\item \textit{Trading causal order for locality, R. Kunjwal and \"A.~Baumeler}, \href{https://doi.org/10.1103/PhysRevLett.131.120201}{Phys.~Rev.~Lett.~131, 120201 (2023)}.
		
		\item \textit{Nonclassicality in correlations without causal order}, R. Kunjwal and O. Oreshkov, \href{https://doi.org/10.48550/arXiv.2307.02565}{arXiv:2307.02565v3 (2025)}.
		
		\item \textit{Generalizing Bell nonlocality without global causal assumptions}, R. Kunjwal and O. Oreshkov, \href{https://doi.org/10.48550/arXiv.2411.11397}{arXiv:2411.11397 (2024)}.
	\end{itemize}
	
	This chapter is based on collaborations with colleagues at IQOQI-Vienna, Austria (\"Amin Baumeler, now at USI Lugano, Switzerland) and Universit\'e libre de Bruxelles (ULB), Belgium (Ognyan Oreshkov).
	
	\item Chapter \ref{chap:CorrelationsJMS} covers work on measurement incompatibility and its relationship with quantum correlations, presented in the following three publications:
	
	\begin{itemize}
		\item \textit{Almost quantum correlations are inconsistent with Specker's principle}, by T. Gonda, R. Kunjwal, D. Schmid, E. Wolfe,  and A. B. Sainz, \href{https://doi.org/10.22331/q-2018-08-27-87}{Quantum 2, 87 (2018)}.
		
		\item \textit{Joint measurability structures realizable with qubit measurements: Incompatibility via marginal surgery}, by N. Andrejic and R. Kunjwal, \href{https://doi.org/10.1103/PhysRevResearch.2.043147}{Phys.~Rev.~Research 2, 043147 (2020)}.
		
		\item \textit{Qualitative equivalence between incompatibility and Bell nonlocality}, by S.A. Yadavalli, N. Andrejic, and R. Kunjwal, \href{https://doi.org/10.1103/PhysRevA.110.L060201}{Phys. Rev. A 110, L060201 (2024)}.
	\end{itemize}
	
	This chapter is based on collaborations with colleagues at the Perimeter Institute, Canada (Tom\'a\v s Gonda, Elie Wolfe, David Schmid, Ana Bel\'en Sainz), the University of Ni\v s, Serbia (Nikola Andreji\'c), and Duke University, USA (Shiv A.~Yadavalli). All these collaborations were research projects where I (co-)supervised undergraduate and graduate students.

	\item Chapter \ref{chap:conclusion} concludes with a summary and some comments on my past and future research directions.
\end{itemize}
\hspace{1cm}

\section*{Research Supervision}

\subsection*{Master theses co-supervised}

\begin{itemize}
	\item Eliot Niedercorn (2022-2023), \textit{Quantum processes with indefinite causal order on time-delocalized subsystems: Shift basis distinguishability using time-delocalized realization of the Lugano process} 
	
	(Co-supervised with Ognyan Oreshkov and Julian Wechs at ULB, Belgium.)
	
	\item Anirudh Krishna (2015), \textit{Experimentally Testable Noncontextuality Inequalities Via Fourier-Motzkin Elimination}
	
	(Informal co-supervision, the official supervisor was Robert W. Spekkens.)
	
	\textbf{Remark:} Anirudh was an undergraduate student at Chennai Mathematical Institute(CMI), Chennai, India, and later a Master's student at the Institute for Quantum Computing (IQC), Waterloo, Canada, during 2013-2015. I informally mentored Anirudh when he was an undergraduate student and later when he was working on his Master's thesis (especially Chapter 2) at the Institute for Quantum Computing, building on my work with Rob Spekkens.
\end{itemize}

\subsection*{Research projects supervised}

\begin{itemize}
	\item Victoria Wright (2019), \textit{Contextuality in composite systems: the role of entanglement in the Kochen Specker theorem}, \href{https://doi.org/10.22331/q-2023-01-19-900}{Quantum 7, 900 (2023)}.
	
	\textbf{Remark:} Victoria was a mathematics PhD student from the University of York, United Kingdom, visiting me at the Perimeter Institute (Canada) in May 2019 when she started working under my supervision. This work was presented at Quantum Physics and Logic (QPL 2022), and Theory of Quantum Computation (TQC 2023).
	
	\item Shiv Akshar Yadavalli (2018), \textit{Contextuality in entanglement-assisted one-shot classical communication}, \href{https://doi.org/10.22331/q-2022-10-13-839}{Quantum 6, 839 (2022)}.
	
	\textbf{Remark:} Shiv visited me from the University of Texas at Austin, USA, during June-August 2018 as an undergraduate research intern at Perimeter Institute (Canada). His Undergraduate Honors Thesis (`Senior Thesis'), \textit{`Quantum Contextuality drives Quantum Advantage'}, was based on work he did during, and after, his summer internship with me (resulting in the publication listed above). This work was a contributed talk at QPL 2020 and an invited talk at ENSPM2021, annual meeting ofthe Portuguese Mathematical Society (2021). 
	
	\item Nikola Andrejic (2017), \textit{Joint measurability structures realizable with qubit measurements: Incompatibility via marginal surgery}, \href{https://doi.org/10.1103/PhysRevResearch.2.043147}{Phys.~Rev.~Research 2, 043147 (2020)}.
	
	\textbf{Remark:} Nikola visited me from the University of Ni\v s, Serbia during May-August 2017 as an undergraduate research intern at the Perimeter Institute (Canada). This work was a contributed talk at QPL 2020.
	
	\item Emily Kendall, Tom\'a\v{s} Gonda, and David Schmid (2017), \textit{Almost Quantum Correlations are Inconsistent with Specker's Principle}, \href{https://doi.org/10.22331/q-2018-08-27-87}{Quantum 2, 87 (2018)}.
	
	(Co-supervised with Ana Bel\'en Sainz and Elie Wolfe)
	
	\textbf{Remark:} This work was initiated at the Perimeter Scholars International (PSI) winter school in Canada. Emily dropped out of the project later since her research interests shifted. This work was a contributed talk at Quantum Information Processing (QIP 2019).
	
	\item Shiv Akshar Yadavalli and Nikola Andrejic (2024), \textit{Qualitative equivalence between incompatibility and Bell nonlocality}, \href{https://doi.org/10.1103/PhysRevA.110.L060201}{Phys. Rev. A 110, L060201 (2024)}.
	
	\textbf{Remark:} Both Shiv and Nikola were PhD students (in the USA and Serbia, respectively) while this project was being completed, even though the initial idea for this project was proposed during Shiv's undergraduate research internship at Perimeter Institute.
\end{itemize}

I note that, although the research projects listed above were not theses by themselves, they were all supervised by me (any co-supervision is explicitly mentioned above) and they led to publications in leading quantum information science journals and contributed or invited talks at quantum conferences. Taken together, the five papers I have co-authored with students could easily make more than one PhD thesis.

\subsection*{Ongoing PhD supervision} 

Since October 1, 2024, I have been supervising a PhD student, Nasra Daher Ahmed, in the Calcul Naturel (CANA) team at the Laboratoire d'Informatique et des Syst\`emes (LIS), Aix-Marseille University (AMU), funded by
\href{https://anr.fr/ProjetIA-22-CMAS-0001}{QuanTEdu-France}. Nasra's thesis is co-supervised by Ognyan Oreshkov from Universit\'e libre de Bruxelles (ULB), Brussels, Belgium. The PhD project is titled, \textit{`Characterizing nonclassicality in quantum networks'}, and registered at the \'Ecole Doctorale en Math\'ematiques et Informatique de Marseille (\href{https://ed184.lis-lab.fr/doku.php}{ED 184}).

\subsection*{Ongoing postdoctoral supervision}
Through my A*MIDEX Chaire d'Excellence grant (AMX-22-CEI-01), I have recruited four postdoctoral researchers to work with me on the project, \textit{`Quantumness: combinatorial, computational, and distributed' (QCCD)}. These postdocs are:

\begin{itemize}
	\item Leonardo Vaglini (PhD 2023, University of Pavia, Italy), to work on \textit{Quantum causality} (January 2025 - December 2026),
	\item Shashaank Khanna (PhD 2024, University of York, United Kingdom), to work on \textit{Nonclassicality in quantum algorithms} (March 2025 - February 2027),
	\item Jacopo Surace (PhD 2020, University of Strathclyde, United Kingdom), to work on \textit{Quantum information and foundations} (March 2025 - Februrary 2027),
	\item Shintaro Minagawa (PhD 2025, Nagoya University, Japan), to work on \textit{Nonclassicality in models of quantum computation} (May 2025 - February 2027).
\end{itemize}

\section*{Serving on PhD Jury}
I served as an external examiner (reviewer) and member of the jury for the PhD viva of Raman Choudhary, on May 13, 2025, at \href{https://mapi.map.edu.pt/}{MAP-i (Doctoral Programme in Computer Science)}, University of Minho, Braga, Portugal. The PhD  thesis,\textit{``Classical, quantum, and post-quantum aspects of Kochen-Specker contextuality scenarios"}, was supervised by Rui Soares Barbosa  (INL – International Iberian Nanotechnology Laboratory), Lu\'is Paulo Santos (University of Minho), and Shane Mansfield (Quandela).
\clearpage

\addcontentsline{toc}{section}{Contents}
\tableofcontents
\mainmatter
\singlespace\pagestyle{plain}\pagenumbering{arabic}

\chapter{Introduction}\label{chap:introduction}

\toquote{I have a friend who has a little dog that likes to mark trees. It usually bypasses trees that have not been already marked by other canines, but when it encounters a highly frequented tree, it does its best to impart its own essence. In fact, the dog tries hard to mark higher up on the tree than have other dogs previously. Since my friend's dog is small, the effort to reach high often causes the animal to tumble, rather comically, over on its back. Science is a bit like that.}{Fredric M.~Menger}

\clearpage

\section{Background}
How we come to do what we end up doing in life is largely a result of choice, chance, or circumstance. The choice in my case was to pursue my interests in physics, mathematics, and computer science after finishing high school; the chance was encountering people and ideas which led me towards quantum information science (broadly construed), where these interests converged in a remarkable manner; and the circumstance was one of coming of age with the advent of open science and virtually free access to the worldwide web of information and knowledge.

I ask the indulgence of the reader as I briefly recount below my winding career trajectory through quantum lands and how it shapes the present document.

It started with undergraduate quantum mechanics, taught in some combination of the style of Feynman and Hibbs \cite{FH10} and Shankar \cite{Shankar12}. The pedagogical emphasis was on how to go from classical physics to quantum physics through a least action principle and working up to the path integral approach to quantum physics that Feynman pioneered. The lesson I took away from this first course was that ``there are no trajectories" in quantum physics, that one has to do a weird averaging using complex numbers over the various classical trajectories a particle might take in spacetime. A lesson that wasn't taught but I made up for myself was that it's not that there are no trajectories (after all, cloud chambers are supposed to document particle trajectories) but that there are no \textit{predictable} trajectories in quantum physics; I imagined that a quantum particle must really be a very indecisive character, not making up its mind until the moment of measurement (whatever that might mean \cite{Bell90}) where exactly it is going to land up.\footnote{I hasten to add here I am not neither endorsing nor rejecting here my youthful fancies about particles' trajectories or their \href{https://en.wikipedia.org/wiki/Free_will_theorem}{free will}, merely recalling them. Surely I am not the only student of quantum mechanics to have \href{https://galileo-unbound.blog/2022/09/04/is-there-a-quantum-trajectory/}{wondered} about the \href{https://physics.stackexchange.com/questions/186170/does-quantum-mechanics-imply-that-particles-have-no-trajectories}{question}.} Much later, I remembered echoes of this ``story" I made up when I first learned about the Kochen-Specker theorem \cite{KS67}. The slogan``unperformed measurements have no outcomes" \cite{Peres78,Fuchs23} immediately clicked into place and this time with mathematical rigour.

It continued with graduate-level courses in quantum theory and its mathematics, attending a school on the \href{https://www.youtube.com/watch?v=Hj9XKmPVE-E&list=PLD3E479AB374A718F}{functional analysis} of quantum information theory \cite{GMS15}, and following the \href{https://www.youtube.com/@QplusHangouts}{Q+ seminar series} on quantum foundations. Although formally in a theoretical physics PhD program, working in an \href{https://www.imsc.res.in/}{institute} that was dedicated to theoretical abstractions meant that I was ready to abstract away from quantum \textit{physics} to quantum \textit{theory}, where the latter is an abstraction that is applicable across physics, information theory, and computer science. It was around this time that my research interests converged on the question of quantumness, \textit{i.e.}, aspects of quantum theory that set it apart from classical probabilistic theories and power its applications in information theory and computation. It was a big surprise to me that the question of what powers quantum computational speedups was not already settled, that it was hotly debated, and that there were both examples and counter-examples to many claimed sources of quantum computational advantage \cite{Bub10}. This situation convinced me that a principled account of the nonclassicality of quantum theory---its \textit{quantumness}---is necessary if we are to make any progress on such fundamental questions in quantum information, foundations, and computation. 

Key to this conviction was the idea that since the predictions of quantum theory are intrinsically probabilistic, its quantumness must be understood in probabilistic---and thus, \textit{operational}---terms \cite{CSFoils}. During my PhD, I spent my energies on developing the formalism of generalized contextuality \cite{KS15,MPK16} and better understanding measurement incompatibility \cite{KG14,KHF14}. I saw the development of these frameworks as essential to the project of unravelling the nonclassicality of quantum theory in ways that have real-world applicability, \textit{e.g.}, in allowing for noise-robust witnesses of quantumness. My postdoctoral research further consolidated this work on generalized contextuality, besides also branching out in the directions of resource theories of nonclassicality, and questions of quantum causality. 

All this brings me to my current endeavour, namely, bringing insights from developments in quantum foundations to the field of quantum information and computation, and conversely. This has historical precedent. The second quantum revolution \cite{DM03} is powered by notions of nonclassicality---ways in which quantum theory radically departs from intuitions rooted in classical physics---such as entanglement and Bell's theorem. Such notions were largely restricted to quantum foundations until the fields of quantum information and computation took off in the 1990s. \textit{`Applied quantum foundations'} \cite{DF19} is perhaps an apt term for the field of quantum information science. In turn, the emergence of the information-theoretic outlook on quantum theory has led to a resurgence of work in quantum foundations in the first quarter of this century, \textit{i.e.}, much of contemporary quantum foundations uses the tools of quantum information science.

In conceptualizing the narrative arc of this thesis and situating my work within it, the aspect that stood out the most for me was just \textit{how much} of the essential structure of quantum theory---and the ways in which it departs from classical probabilistic theories of physics---is captured by discrete structures like graphs and hypergraphs. The fundamental theorems of Bell (1964) \cite{Bell64}, Kochen-Specker (1967) \cite{KS67}, and Gleason (1957) \cite{Gleason57} lie at the heart of this \textit{quantumness}. And at the heart of these theorems lie discrete structures and their combinatorics \cite{HP04}. I have therefore chosen to focus in this thesis on those aspects of my postdoctoral research that directly use such discrete structures for the purposes of assessing quantumness.\footnote{There is also another reason I have chosen this focus. It allows me to cover research projects where I supervised or co-supervised Master's and PhD students. Insofar as this dissertation is supposed to attest to my ability to supervise research independently, I consider it pertinent to include these contributions in it.}

In the next section, I provide a lightning introduction to the mathematics of quantum theory, assuming basic knowledge of finite-dimensional vector spaces on the part of the reader. 
If the reader is already familiar with the mathematical formalism of (operational) quantum theory, they may safely skip the next section and move on to the plan for the rest of the thesis.

\section{The mathematics of quantum theory}

Below I introduce a minimalist conception of quantum theory, often called \textit{operational quantum theory} \cite{BGL97,CSFoils}, that is sufficient for the purposes of this thesis. For more details, the interested reader may consult the textbook of Nielsen and Chuang \cite{NielsenAndChuang} (in particular, Chapters 2 and 8).

\subsection{What is an operational theory?}
\begin{figure}
	\centering
	\includegraphics[scale=0.5]{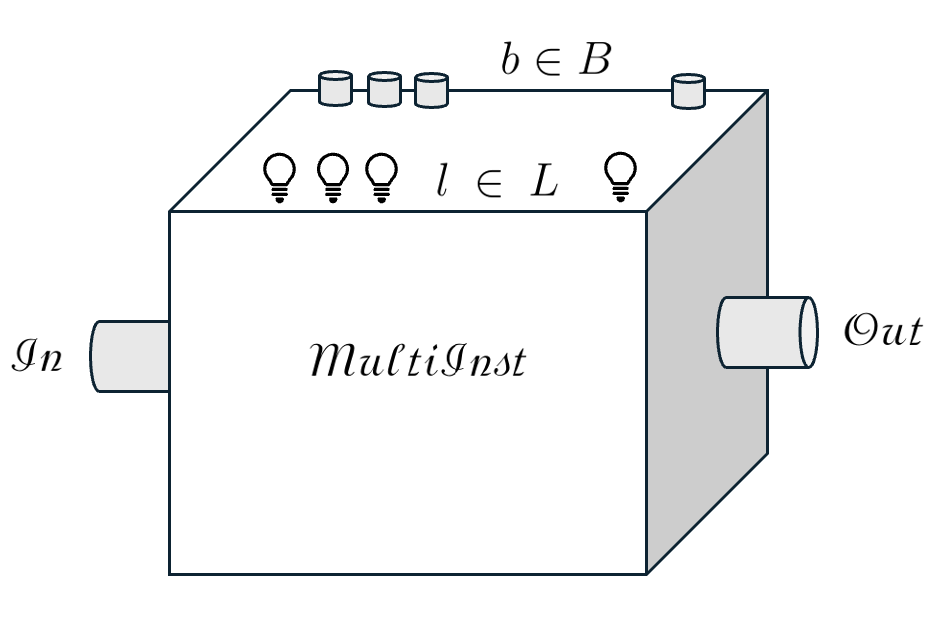}
	\caption{Schematic of a multi-instrument: the task of an operational theory is to describe its behaviour.}
	\label{fig:multiinst}
\end{figure}

Consider an experimentalist in a lab who holds a box, labelled $\mathcal{MultiInst}$ and called a \textit{multi-instrument}, that has an input port (labelled $\mathcal{In}$) and an output port (labelled $\mathcal{Out}$), together with some buttons (labelled $b\in B$) and lights on this box (labelled $l\in L$). The box can receive an input system through port $\mathcal{In}$, potentially transform it, and then send out the transformed system through port $\mathcal{Out}$. (See Figure \ref{fig:multiinst}.)

Given an incoming system through $\mathcal{In}$, the experimentalist is interested in the probability that a given light, $l\in L$, lights up when a particular button, $b\in B$, is pressed. These probabilities are captured by the following set of probability distributions 
\begin{align}
	\left\{p(l|b)\Bigg|p(l|b)\geq 0\hspace{1mm}\forall l\in L, b\in B, \sum_l p(l|b)=1 \hspace{1mm}\forall b\in B\right\}.	
\end{align}
It is the job of an operational theory to predict these probabilities based on instruction sets that describe how the box behaves when pressing a button $b\in B$. Each choice of $b$ corresponds to implementing a particular \textit{instrument}. The fact that the experimentalist chooses button $b$ and observes light $l$ means that the outgoing system through $\mathcal{Out}$ is prepared in a way that is labelled by $[l|b]$.

This multi-instrument could serve special purposes. For example,
it could be a \textit{multi-source} that receives no input system but produces an output system prepared according to $[l|b]$ with probability $p(l|b)$ when button $b$ is pressed. It could also be a \textit{multi-meter} that receives an input system and measures it in a manner specified by $b$ without producing an output system but resulting in light $l$ lighting up with probability $p(l|b)$.

\subsection{What is operational \textit{quantum} theory?}

We will now assume that the box $\mathcal{MultiInst}$ is quantum  and call it $\mathcal{M}$, a \textit{quantum} multi-instrument \cite{BLN25}. Furthermore, we will assume that it receives a quantum system through its input port $\mathcal{In}$ in the quantum state $\rho_{\rm in}$. What does this mean?

It means that we can give a story about how the box produces the observed probabilities $p(l|b)$ in terms of the mathematics of quantum theory. In short, the story says that the probability $p(l|b)$ arises from computing the trace of a matrix, \textit{i.e.},
\begin{align}
	p(l|b)=\Tr\left(\mathcal{M}_{l|b}(\rho_{\rm in})\right),
\end{align}
where $\mathcal{M}_{l|b}$ describes the procedure where button $b$ is pressed and light $l$ lights up, yielding an output state $\rho_{l|b}:=\mathcal{M}_{l|b}(\rho_{\rm in})/p(l|b)$ with probability $p(l|b)$. We introduce the relevant mathematics below.

\subsubsection{Dirac notation}

We will work with vectors and matrices using the  \textit{Dirac notation}: the symbol $\ket{\cdot}$, called a ``ket", will denote a column vector; the symbol $\bra{\cdot}$, called a ``bra", will denote a row vector;
the inner product $\braket{\cdot}{\cdot}$ is called a ``braket". The outer product, $\ket{\cdot}\bra{\cdot}$, represents a matrix obtained by taking the matrix product of a column vector $\ket{\cdot}$ and row vector $\bra{\cdot}$.

The vectors and matrices of interest to us will be defined with respect to finite-dimensional complex vector spaces (\textit{i.e.}, Hilbert spaces) equipped with an inner product defined via 
\begin{align}
	\braket{v}{w}=\sum_{i=0}^{d-1}v^*_iw_i
\end{align}
for any pair of $d$-dimensional complex vectors $\ket{v},\ket{w}\in\mathcal{H}\cong \mathbb{C}^d$. Here $\ket{v}:=(v_0,v_1,v_2,\dots,v_{d-1})^{\rm T}$, a column vector of $d$ complex numbers $v_i\in \mathbb{C}$, where $i\in\{0,1,\dots,d-1\}$ (and similarly for $\ket{w}\in\mathcal{H}$). Its conjugate transpose, $\bra{v}:=(v^*_0,v^*_1,v^*_2,\dots,v^*_{d-1})$, denotes the corresponding row vector, where $v^*_i$ is the complex conjugate of $v_i$ for all $i\in\{0,1,\dots,d-1\}$.

\subsubsection{Preparations, measurements, transformations}
\textit{States:} Physically, all the possible ways in which a quantum system of dimension $d$ can, in principle, be prepared is given by the \textit{state space} $\mathcal{S}(\mathcal{H})$, defined as
\begin{align}
	\mathcal{S}(\mathcal{H}):=\{\rho\in\mathcal{L}(\mathcal{H})|\rho\geq 0, \Tr\rho=1\},
\end{align}
where $\mathcal{L}(\mathcal{H})$ denotes (bounded) linear operators from $\mathcal{H}\rightarrow \mathcal{H}$, representable as $d\times d$ matrices.
Here, the positive semidefiniteness of $\rho$ (denoted via $\rho\geq 0$) means that the matrix $\rho$ has real and non-negative eigenvalues, or equivalently, $\langle v | \rho | v\rangle \geq 0$ for all vectors $\ket{v}\in\mathcal{H}$. Furthermore, the trace condition means that the eigenvalues are probabilities that add up to $1$.

\textit{Effects:} Similar to preparations, all the possible ways in which a quantum system of dimension $d$ can, in principle, be measured is given by the \textit{effect space} $\mathcal{E}(\mathcal{H})$, defined as
\begin{align}
	\mathcal{E}(\mathcal{H}):=\{E\in\mathcal{L}(\mathcal{H})|0\leq E \leq \id\}.
\end{align}
Here, the positive semidefiniteness of $E$ (denoted via $0\leq E$ or $E\geq 0$) means that the matrix $E$ has real and non-negative eigenvalues. Furthermore, $\id$ is the $d\times d$ identity matrix and $E\leq \id$ denotes the fact that the matrix $\id-E\geq 0$, \textit{i.e.}, it is positive semidefinite and, therefore, the eigenvalues of $E$ lie in the range $[0,1]$. 

A set of effects $\{E_k \in \mathcal{E}(\mathcal{H})\}_{k}$ denotes a quantum measurement---called a positive operator-valued measure (POVM)---if and only if it forms a resolution of the identity, \textit{i.e.},
\begin{align}
	\sum_kE_k=\id
\end{align}
Such a POVM is called a projection-valued measure (PVM) if and only if $E_k^2=E_k$ for all $k$ in the POVM.

\textit{Born rule:} Preparing a system in some quantum state $\rho$ and measuring it using some POVM $\{E_k \in \mathcal{E}(\mathcal{H})\}_{k}$ results in outcome $k$ with a probability given by the Born rule, \textit{i.e.},,
\begin{align}
	{\rm Prob}(k|\rho)=\Tr(\rho E_k).
\end{align}

\textit{Composite systems:} Two quantum systems associated with Hilbert spaces $\mathcal{H}_1$ and $\mathcal{H}_2$ are jointly described via a  composite Hilbert space $\mathcal{H}_1\otimes\mathcal{H}_2$. A quantum state $\rho_{12}\in\mathcal{S}(\mathcal{H}_1\otimes\mathcal{H}_2)$ is said to be \textit{separable} if and only if it can be expressed as a classical probabilistic mixture of quantum states on the component Hilbert spaces, \textit{i.e.}, 
\begin{align}
	\rho_{12}=\sum_k p_k\rho^{(k)}_1\otimes\rho^{(k)}_2,
\end{align}
where $\forall k:$ $p_k\geq 0$, $\rho^{(k)}_1\in\mathcal{S}(\mathcal{H}_1)$, $\rho^{(k)}_2\in\mathcal{S}(\mathcal{H}_2)$, and $\sum_kp_k=1$. A quantum state that fails to be separable is called \textit{entangled}.

\textit{Quantum instruments:} Finally, all the possible ways in which a quantum system can, in principle, be transformed is given by a \textit{quantum instrument}, which generalizes both preparations and measurements. 

A quantum instrument describes transformations of quantum states of one system (associated with Hilbert space $\mathcal{H}_1$, say) into quantum states of another system (associated with Hilbert space $\mathcal{H}_2$, say). We denote by $d_1$ the dimension of $\mathcal{H}_1$ and by $d_2$ the dimension of $\mathcal{H}_2$ (in particular, we could have $d_1=d_2$). 

Mathematically, a quantum instrument is a collection of maps $\mathcal{M}^{(b)}:=\{\mathcal{M}_{l|b}\}_{l\in L}$, where each $\mathcal{M}_{l|b}$ is a \textit{completely positive and trace non-increasing} (CPTNI) map, given by 
\begin{align}
	&\mathcal{M}_{l|b}:\mathcal{S}(\mathcal{H}_1)\rightarrow\mathcal{S}(\mathcal{H}_2),\textrm{ where}\\
	&\forall n\in \mathbb{N}: \mathcal{I}_n\otimes \mathcal{M}_{l|b}(A)\geq 0\textrm{ for all }A\geq 0, A\in \mathcal{L}(\mathbb{C}^n\otimes\mathcal{H}_1)\nonumber\\
	&\textrm{ (completely positive), }\textrm{and}\\
	&\Tr\left(\mathcal{M}_{l|b}(\rho)\right)\leq \Tr(\rho)=1\textrm{  (trace non-increasing).}
\end{align}
Here $\mathcal{I}_n$ is the identity channel that ``does nothing" to the operators in $\mathcal{L}(\mathbb{C}^n)$ and $\mathbb{N}$ is the set of natural numbers (or positive integers). The CPTNI maps in the quantum instrument $\mathcal{M}^{(b)}$ coarse grain to define a \textit{completely positive trace-preserving} (CPTP) map given by 
\begin{align}
	\mathcal{M}_b:=\sum_{l\in L}\mathcal{M}_{l|b},\textrm{ where }\Tr\left(\mathcal{M}_b(\rho)\right)=\Tr(\rho)=1.
\end{align}

\textit{Operator-sum representation:} Any CPTNI map $\mathcal{M}_{l|b}$ can be written as
\begin{align}
	&\mathcal{M}_{l|b}(\rho)=\sum_k K_k\rho K^{\dagger}_k,\textrm{ where}\\
	&\forall k, K_k\in\mathcal{L}(\mathcal{H}_1\rightarrow\mathcal{H}_2) \textrm{ and }\sum_k K^{\dagger}_k K_k \leq \id_{\mathcal{H}_1}.
\end{align}
Here $\id_{\mathcal{H}_1}$ is the $d_1\times d_1$ identity matrix on $\mathcal{H}_1$ and $\mathcal{L}(\mathcal{H}_1\rightarrow\mathcal{H}_2)$ denotes the set of (bounded) linear operators, represented by $d_2\times d_1$ matrices, that map vectors in  $\mathcal{H}_1$ to those in $\mathcal{H}_2$. The set of operators $\{K_k\}_k$, called \textit{Kraus operators}, in the operator-sum representation of a CPTNI map is not unique; in general, a CPTNI map can admit multiple such \textit{Kraus decompositions}.

A preparation corresponds to an instrument where the input system is trivial and a (destructive) measurement corresponds to an instrument where the output system is trivial. The standard case of unitary evolution for closed quantum systems, in particular, corresponds to a quantum instrument with exactly one CPTP map $\mathcal{U}:\mathcal{S}(\mathcal{H}_1)\rightarrow \mathcal{S}(\mathcal{H}_2)$, given by 
\begin{align}
	\mathcal{U}(\rho)=U\rho U^{\dagger},
\end{align}
where $U^{\dagger}U=UU^{\dagger}=\id_{\mathcal{H}_1}$.

\section{The plan}
In Chapter \ref{chap:hypergraphframeworks}, we review hypergraph frameworks for generalized contextuality \cite{Spekkens05} that systematically turn witnesses of Kochen-Specker contextuality \cite{KS67} into noise-robust witnesses of generalized contextuality \cite{Kunjwal19,Kunjwal20} and consider an application of these to the problem of entanglement-assisted one-shot classical communication \cite{YK22}. In Chapter \ref{chap:ContextualityEntanglement}, we review work on the logical relationship between multiqubit entanglement and proofs of the Kochen-Specker theorem, as well as what this tells us about their interplay in restricted models of quantum computation with state injection \cite{BDB17}. After thus covering applications of graphs and hypergraphs, we turn towards work on causality in Chapter \ref{chap:CorrelationsICO}, where directed graphs and their (a)cyclicity captures the (in)definiteness of causal order and its information-theoretic implications. In Chapter \ref{chap:CorrelationsJMS}, we review work on joint measurability structures that uses hypergraphs representing abstract simplicial complexes \cite{KHF14} and how they can help reveal the incompatibility structure of quantum measurements as well as its consequences for quantum correlations. We conclude in Chapter \ref{chap:conclusion} with a summary and some comments on future work.
\newpage
 
\chapter{Graphs and hypergraphs for contextuality}\label{chap:hypergraphframeworks}

\toabstract{I survey two related hypergraph frameworks that provide a recipe to turn witnesses of Kochen-Specker contextuality into witnesses of generalized contextuality. These hypergraph frameworks rely on the introduction of a hypergraph invariant---the \textit{weighted max-predictability}---that is essential to obtaining bounds on generalized noncontextuality inequalities that operationalize  KS-noncontextuality inequalities. I conclude with a brief account of an application of this approach to generalized noncontextuality to the problem of entanglement-assisted one-shot classical communication.}

\clearpage

\section{Introduction}
Contextuality is a notion of nonclassicality that puts constraints on conceivable classical models that can reproduce the operational predictions of quantum theory (or, more generally, of other operational theories). Before formally defining contextuality in subsequent sections, I provide some intuitions for it starting from simple quantum examples. 

Consider the Born rule probability that an outcome $a$ for measurement $x$ clicks when the input state is prepared according to some preparation procedure $P$: 
\begin{equation}
	p(a|x,P)=\Tr(E_{a|x}\rho_P),
\end{equation}
where $E_{a|x}$ is the positive operator associated with the measurement procedure and $\rho_P$ is the density matrix associated with the preparation procedure. There are two properties of the Born rule that stand out:
\begin{enumerate}
	\item \textbf{Preparation context doesn't matter \textit{operationally}}, \textit{i.e.}, the probability $\Tr(E_{a|x}\rho_P)$ depends only on the density matrix $\rho_P$ and no other information (``preparation context") about how the preparation procedure was carried out.
	
	\textit{Example:} The maximally mixed state on a qubit, $\rho=\frac{\id}{2}$, can be prepared by (at least) two distinct preparation procedures: $P_0$, which involves a uniform mixture of $Z$ basis states ($\ketbra{0}{0}$ or $\ketbra{1}{1})$),	
	\begin{equation}
		\rho_{P_0}=\frac{1}{2}\left(\ketbra{0}{0}+\ketbra{1}{1}\right)
	\end{equation}
	and $P_1$, which involves a uniform mixture of $X$ basis states ($\ketbra{+}{+}$ or $\ketbra{-}{-})$),
	\begin{equation}
		\rho_{P_1}=\frac{1}{2}\left(\ketbra{+}{+}+\ketbra{-}{-}\right).
	\end{equation}
	The Born rule, however, does not care about the difference in implementation details---that is, difference of \textit{preparation context}---between $\rho_{P_0}$ and $\rho_{P_1}$, and we have 
	\begin{equation}
		\rho_{P_0}=\rho_{P_1}=\frac{\id}{2}
	\end{equation}
	The preparation procedure $P_0$ is therefore operationally equivalent to the preparation procedure $P_1$, \textit{i.e.}, for any choice of measurement procedure, they yield identical outcome probabilities. We denote this operational equivalence by $P_0\simeq P_1$.
	
	\item \textbf{Measurement context doesn't matter \textit{operationally}}, \textit{i.e.}, the probability $\Tr(E_{a|x}\rho_P)$ depends only on the positive operator $E_{a|x}$ and no other information (``measurement context") about how the measurement procedure was carried out.
	
	\textit{Example:} Consider a qutrit measured in one of two orthonormal bases: measurement procedure $x=0$ given by $\{\ket{0},\ket{1},\ket{2}\}$, with the outcomes respectively labelled $a=0,1,2$, or measurement procedure $x=1$ given by $\{\ket{+},\ket{-},\ket{2}\}$, with the outcomes respectively labelled $a=0,1,2$. The measurement procedure resulting in the event $[a=2|x=0]$ is distinct from the one resulting in $[a=2|x=1]$.
	
	The Born rule, however, does not care about the difference in implementation details---that is, difference of \textit{measurement context}---between $E_{2|0}$ and $E_{2|1}$ and we have 
	\begin{equation}
		E_{2|0}=E_{2|1}=\ketbra{2}{2}.
	\end{equation}
	The event $[a=2|x=0]$ is thus operationally equivalent to the event $[a=2|x=1]$, \textit{i.e.}, for any choice of preparation procedure, they occur with identical probabilities. We denote this operational equivalence by $[2|0]\simeq [2|1]$.
\end{enumerate}

\textit{Noncontextuality} is the idea that implementation details that are irrelevant to the operational statistics of quantum theory should remain irrelevant in a classical model that reproduces the predictions of quantum theory. In the examples above, this means that operationally equivalent procedures ($P_0\simeq P_1$, $[2|0]\simeq [2|1]$) should be treated as equivalent in a classical model. Classical models that fail to satisfy this condition of noncontextuality are said to be \textit{contextual}. Quantum theory exhibits \textit{contextuality} in the sense that it is impossible to build a noncontextual classical model of it.

In the rest of this chapter, I will describe the frameworks set up in Refs.~\cite{Kunjwal19, Kunjwal20} with a focus on their key conceptual and technical aspects, referring the reader to \cite{Kunjwal19,Kunjwal20} for more details (\textit{e.g.}, proof details) on these aspects.

 \section{Generalized contextuality}
Although contextuality has its roots in the Kochen-Specker (KS) theorem \cite{KS67}, I have chosen to lead with generalized contextuality \cite{Spekkens05} in this chapter. The reasons for this ahistorical approach are pedagogical: the operational approach of generalized contextuality will help to clarify the limitations of KS-contextuality and how these limitations can be overcome when KS-contextuality is appropriately reformulated in operational terms.

\begin{figure}
	\centering
	\includegraphics[scale=0.3]{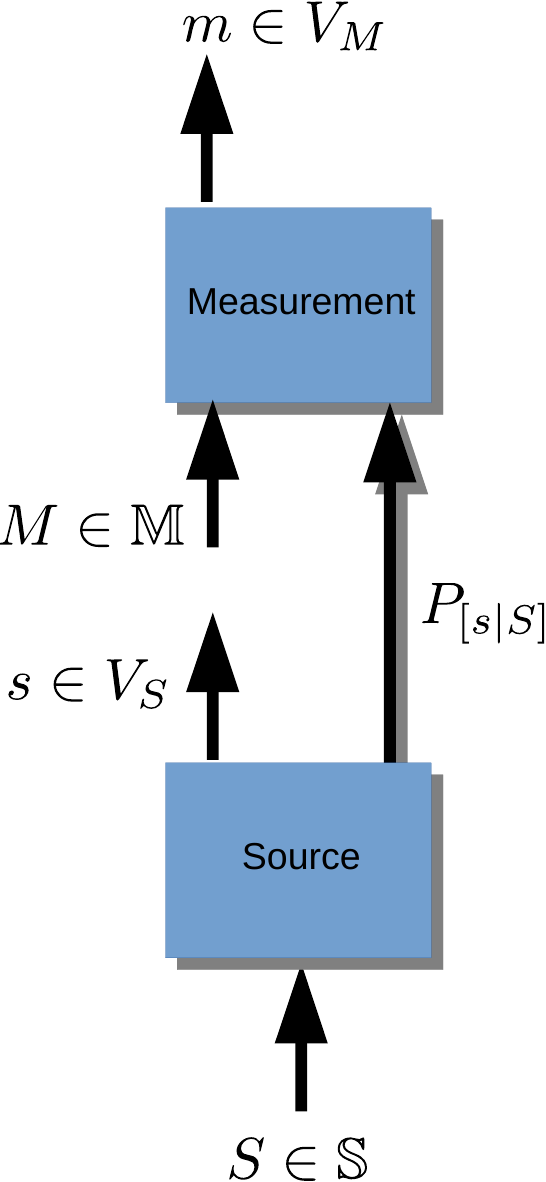}
	\caption{A prepare-and-measure experiment.}
	\label{fig:pmsetup}
\end{figure}

Generalized noncontextuality has three aspects, each covering a particular type of operational procedure, namely, \textit{preparation noncontextuality}, \textit{transformation noncontextuality}, and \textit{measurement noncontextuality}. In this chapter, we will focus entirely on preparation and measurement noncontextuality since they are the ones of direct relevance for our purposes.\footnote{For examples of transformation noncontextuality, see Refs.~\cite{KLP19,SBS24}.} We therefore consider the following schematic of a general experimental set-up that prepares a system and subsequently measures it (Fig.~\ref{fig:pmsetup}): 
\begin{itemize}
	\item The experimenter uses a source device to implement various preparation procedures. Each preparation procedure is specified by a source setting $S\in\mathbb{S}$ and each setting $S$ has possible outcomes $s\in V_{S}$. We denote by $[s|S]$ the source event: it describes the preparation procedure according to which the system released by the source device is prepared.
		
	\item The experimenter then subjects the system---prepared according to $[s|S]$---to a measurement procedure, \textit{i.e.}, the system enters a measurement device which has a setting $M\in\mathbb{M}$ and each setting $M$ has possible outcomes $m\in V_M$. We denote by $[m|M]$ the measurement event: it describes the measurement procedure to which the system released by the source device is subjected.
\end{itemize}

The composition of a preparation procedure with a measurement procedure results in the outcomes $m$ and $s$ with some probability given the settings $M$ and $S$. The empirical content of an operational theory is summarized by the following probabilities:
\begin{equation}
	p(m,s|M,S), \forall M\in\mathbb{M}, S\in\mathbb{S}, m\in V_M, s\in V_S.
\end{equation}
Operational equivalences are given by 
\begin{align}
	[s|S]\simeq [s'|S'] &\Leftrightarrow p(m,s|M,S)=p(m,s'|M,S')\quad\forall M\in\mathbb{M}, m\in V_M.\\
	[m|M]\simeq [m'|M'] &\Leftrightarrow p(m,s|M,S)=p(m',s|M',S)\quad\forall S\in\mathbb{S}, s\in V_S.
\end{align}

In quantum foundations parlance, a classical model is termed an \textit{ontological model} and we describe such models within the \textit{ontological models framework} \cite{HS10}. In an ontological model, the probability that outcomes $m,s$ occur for settings $M,S$ is given by 
\begin{equation}
	p(m,s|M,S)=\sum_{\lambda}p(m|M,\lambda)p(\lambda,s|S)=\sum_{\lambda}p(m|M,\lambda)p(s|S,\lambda)p(\lambda|S),
\end{equation}
where $p(s|S,\lambda), p(m|M,\lambda)\geq 0$ for all $\lambda$, $\sum_{\lambda}p(\lambda|S)=1$, and $\sum_{s}p(s|S,\lambda)=\sum_{m}p(m|M,\lambda)=1$ for all $\lambda$.

It is always possible to write down such a model for any operational statistics, including quantum statistics \cite{Kunjwal16}. No-go theorems like Bell's theorem and the Kochen-Specker theorem rule out such models only under additional assumptions on them. The assumption of interest to us here is generalized noncontextuality, \textit{i.e.}, the ontological model of an operational theory must treat operationally equivalent procedures equivalently. It can be stated for the case of preparation and measurement procedures as follows:
\begin{enumerate}
	\item Preparation noncontextuality: 
	\begin{equation}
     [s|S]\simeq [s'|S'] \Rightarrow p(s,\lambda|S)=p(s',\lambda|S)\quad\forall\lambda.
	\end{equation}
	
	\item Measurement noncontextuality: 
	\begin{equation}
	 [m|M]\simeq [m'|M'] \Rightarrow p(m|M,\lambda)=p(m'|M',\lambda)\quad\forall\lambda.
	\end{equation}
\end{enumerate} 

In this chapter, we will present hypergraph frameworks for witnessing the  failure of the joint assumption of preparation and measurement noncontextuality. These frameworks do not assume the validity of operational quantum theory and are therefore applicable to arbitrary operational theories. 
 
 \section{Kochen-Specker contextuality}
 KS-noncontextuality can be seen as a conjunction of two assumptions---\textit{measurement noncontextuality} and \textit{outcome determinism}---applied to a particular type of measurement context, namely, when a measurement outcome is shared between two distinct measurement procedures, \textit{e.g.}, the case of qutrit measurements we previously considered, where projective measurements in the bases $\{\ket{0},\ket{1},\ket{2}\}$ and $\{\ket{+},\ket{-},\ket{2}\}$ share the outcome $\ket{2}$. 
 
 The assumption of \textit{outcome determinism} in an ontological model simply means that given the ontic state $\lambda$ of a system, the outcome of any measurement is deterministic, \textit{i.e.}, $p(m|M,\lambda)\in\{0,1\}$ for all $\lambda$ and for all measurement events $[m|M]$. 
 
The failure of KS-noncontextuality for ontological models of quantum theory using only a finite set of projective measurements was first witnessed by the KS theorem \cite{KS67}. We recall below a formulation of the KS theorem as presented in Ref.~\cite{WK21}:
\begin{Th}[Kochen--Specker \cite{KS67}]\label{thm:KS}
	Given a separable Hilbert space $\mathcal{H}$ of dimension at least three, there does not exist a map
	\begin{equation}
		c:\mathcal{R}\rightarrow\{0,1\}
	\end{equation}
	such that for any complete set of rays $\{\ket{\psi_1},\ket{\psi_2},\ldots\}$, we have $c(\ket{\psi_j})=1$ for exactly one value of $j\in\{1,2,\ldots\}$.\footnote{Such a map $c$ is also called a \textit{KS-colouring}, with the ``colours" referring to the values $0$ and $1$.}
\end{Th}
In the language of ontological models, the existence of a map $c:\mathcal{R}\rightarrow\{0,1\}$ is necessary to be able to define measurement noncontextual and outcome-deterministic response functions for all possible projectors in quantum theory. This is because such a model would require that for any complete set of rays (\textit{i.e.}, an orthonormal basis defining a projective measurement) $\{\ket{\psi_1},\ket{\psi_2},\ldots\}$, we have for each $\lambda$, $p(\ket{\psi_k}|\lambda)=\delta_{c_{\lambda}(\ket{\psi_k}),1}$ for some map $c_{\lambda}:\mathcal{R}\rightarrow \{0,1\}$ (satisfying the same properties as $c$ in Theorem \ref{thm:KS} above). The KS theorem therefore implies that a measurement noncontextual and outcome-deterministic ontological model of quantum theory is impossible, \textit{i.e.}, quantum theory exhibits KS-contextuality.
 
Interestingly, for ontological models of quantum theory, KS-noncontextuality can be shown to be strictly implied by preparation noncontextuality (see Refs.~\cite{LM13, Kunjwal16} for a proof of this), \textit{i.e.}, we have 
 \begin{equation}
 	\textrm{Preparation noncontextuality}\Rightarrow \textrm{KS-noncontextuality}.
 \end{equation}
Hence, for quantum theory, every proof of KS-contextuality can also be seen as a proof of preparation contextuality (but not conversely).
 
\section{The CSW and AFLS frameworks}\label{sec:probmodels}

We will now consider prepare-and-measure experiments with a fixed preparation procedure but multiple measurement procedures that can be implemented following the preparation. This is the type of prepare-and-measure experiment envisaged by Bell and Kochen-Specker theorems. Such a scenario---termed a \textit{contextuality scenario}\cite{AFL15}---is represented by a hypergraph $H$ with vertices $V(H)$ and hyperedges $E(H)$. 
Each vertex $v\in V(H)$ corresponds to a measurement outcome and each hyperedge $e\in E(H)$ corresponds to a measurement setting (or procedure). A \textit{probabilistic model} on a contextuality scenario is a map $p:V(H)\rightarrow[0,1]$, where $\sum_{v\in e}p(v)=1$ for all $e\in E(H)$. The fixed preparation procedure defines a probabilistic model for all the measurement procedures considered in a contextuality scenario. A vertex $v\in V(H)$ that appears in multiple hyperedges (say, $e,e'\in E(H)$) represents an operational equivalence between two measurement events, namely, $[v|e]\simeq [v|e']$: this is reflected in the fact that the probabilistic model depends only on $v$ but not on the hyperedge it might be considered a part of.

We define the following families of probabilistic models that will appear repeatedly in this chapter:

\begin{itemize}
	\item \textit{KS-colouring} or \textit{deterministic classical model}: Here we have $p: V(H)\rightarrow \{0,1\}$, \textit{i.e.}, the vertices are assigned probabilities $0$ or $1$ but nothing in between. In other words, all measurement settings have deterministic outcomes in such a model. We denote the set of such models on $H$ by $\mathcal{KS}(H)$.
	
	\item  \textit{KS-noncontextual} or \textit{classical model}: Here we have that $p: V(H)\rightarrow [0,1]$ can be decomposed into a probabilistic mixture of KS-colourings, \textit{i.e.}, even though the measurement settings may have probabilistic outcomes, they can be understood as arising from coarse-graining over fundamentally deterministic assignments specified by underlying KS-colourings. We denote the set of such models on $H$ by $\mathcal{C}(H)$.
	
	For any contextuality scenario $H$, $\mathcal{C}(H)$ forms a convex polytope whose vertices are given by $\mathcal{KS}(H)$.
	
	\item \textit{Quantum model}: Here we have that $p: V(H)\rightarrow [0,1]$  can be realized via some separable Hilbert space $\mathcal{H}$ such that (i) for each $v\in V(H)$, $\exists$ a projector $\Pi_v$, so that $\sum_{v\in e}\Pi_v=I_\mathcal{H}$ for all $e\in E(H)$, and (ii) $\exists$ a quantum state $\rho$ on $\mathcal{H}$ such that $p(v)=\Tr(\Pi_v\rho)$ for all $v\in V(H)$. We denote the set of such models on $H$ by $\mathcal{Q}(H)$.
\end{itemize}
We denote the full set of general probabilistic models on $H$ by $\mathcal{G}(H)$, so that, for any $H$, $\mathcal{KS}(H)\subseteq \mathcal{C}(H)\subseteq \mathcal{Q}(H)\subseteq \mathcal{G}(H)$.

We will distinguish between two qualitatively distinct proofs of the KS theorem in the rest of this chapter:

\begin{itemize}
	\item \textit{Logical proofs of the KS theorem}: These are witnessed by any  contextuality scenario $H$ where $\mathcal{KS}(H)=\mathcal{C}(H)=\varnothing \subsetneq \mathcal{Q}(H)$. That is, the measurement events in the scenario are such that they \textit{cannot} be assigned deterministic outcomes that define a KS-colouring, but they can be assigned projectors that define a quantum model.
	
	\item \textit{Statistical proofs of the KS theorem}: These are witnessed by any contextuality scenario $H$ where $\varnothing\neq \mathcal{C}(H)\subsetneq\mathcal{Q}(H)$. That is, the measurement events in the scenario are such that they \textit{can} be assigned deterministic outcomes that define KS-colourings (and thus classical models), but such classical models are not enough to account for the quantum models that arise from assigning projectors to the measurement events.
	
\end{itemize}

\subsection{A logical proof of the KS theorem}\label{sec:cega}
\begin{figure}
	\centering
	\includegraphics[scale=0.5]{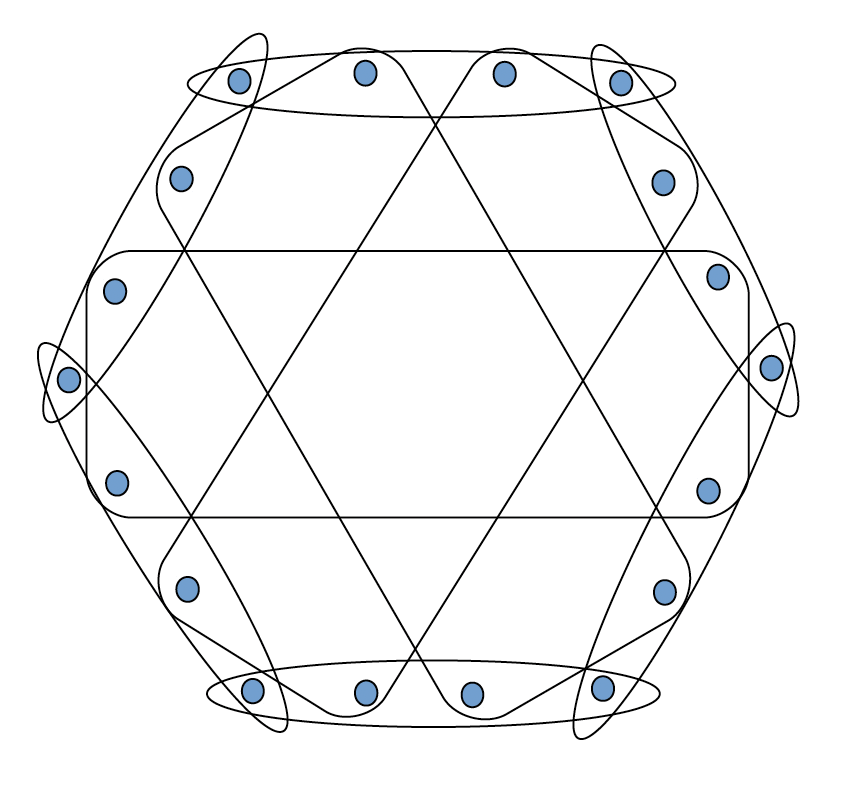}
	\caption{The $18$-ray contextuality scenario, $\Gamma_{18}$, used in Ref.~\cite{CEGA96}.}
	\label{fig:18ray}
\end{figure}

We will use the 18-ray proof of the KS theorem due to Cabello \textit{et al.}~\cite{CEGA96} as our working example of a logical proof of the KS theorem, illustrated in Fig.~\ref{fig:18ray}. It will serve as a paradigmatic example to illustrate features of our hypergraph framework for instances of generalized contextuality motivated by such proofs \cite{KS15,Kunjwal20}.

We denote this contextuality scenario by the hypergraph $\Gamma_{18}$ (Fig.~\ref{fig:18ray}), its vertices by the set $V(\Gamma_{18}):=\{v_i\}_{i=1}^{18}$, and its hyperedges by the set $E(\Gamma_{18}):=\{e_j\}_{j=1}^9$, where 
\begin{align}
	&e_1:=\{v_1,v_2,v_3,v_4\}, e_2:=\{v_4,v_5,v_6,v_7\}, e_3:=\{v_7,v_8,v_9,v_{10}\},\nonumber\\
	&e_4:=\{v_{10},v_{11},v_{12},v_{13}\}, e_5:=\{v_{13},v_{14},v_{15},v_{16}\}, e_6:=\{v_{16},v_{17},v_{18},v_1\},\nonumber\\
	&e_7:=\{v_{18},v_2,v_9,v_{11}\}, e_8:=\{v_3,v_5,v_{12},v_{14}\}, e_9:=\{v_6,v_8,v_{15},v_{17}\}.\label{eq:hyperedges}
\end{align}
Hence, any probabilistic model $p\in\mathcal{G}(\Gamma_{18})$ on must satisfy the following set of linear equations:
\begin{align}
	&p(v_1)+p(v_2)+p(v_3)+p(v_4)=1\nonumber\\
	&p(v_4)+p(v_5)+p(v_6)+p(v_7)=1\nonumber\\
	&p(v_7)+p(v_8)+p(v_9)+p(v_{10})=1\nonumber\\
	&p(v_{10})+p(v_{11})+p(v_{12})+p(v_{13})=1\nonumber\\
	&p(v_{13})+p(v_{14})+p(v_{15})+p(v_{16})=1\nonumber\\
	&p(v_{16})+p(v_{17})+p(v_{18})+p(v_1)=1\nonumber\\
	&p(v_{18})+p(v_2)+p(v_9)+p(v_{11})=1\nonumber\\
	&p(v_3)+p(v_5)+p(v_{12})+p(v_{14})=1\nonumber\\
	&p(v_6)+p(v_8)+p(v_{15})+p(v_{17})=1.\label{eq:18rayeqs}
\end{align}
It is straightforward to see that no $p\in\mathcal{G}(\Gamma_{18})$ is a KS-colouring, \textit{i.e.}, $\mathcal{KS}(\Gamma_{18})=\mathcal{C}(\Gamma_{18})$: a KS-colouring would require that $p(v_i)\in\{0,1\}$ for all $i\in\{1,2,\dots,18\}$; summing all the linear equations in Eq.~\eqref{eq:18rayeqs} we have that $2\sum_ip(v_i)=9$, which cannot hold if $p$ is a KS-colouring (since an even number cannot equal an odd number); hence, no $p\in\mathcal{G}(\Gamma_{18})$ is a KS-colouring, and we have $\mathcal{KS}(\Gamma_{18})=\mathcal{C}(\Gamma_{18})=\varnothing$. 

To complete this logical proof of the KS theorem, all we need to argue now is that $\mathcal{Q}(\Gamma_{18})\neq \varnothing$. The argument for this follows from using $18$ of the $24$ vectors first identified by Peres \cite{Peres91}, namely, those used in the construction of Cabello \textit{et al.}~\cite{CEGA96}.
The rays associated with the vertices of $\Gamma_{18}$ in this construction are given by:

\begin{align}
	&v_1: (0,0,0,1), v_2: (0,0,1,0), v_3: (1,1,0,0), v_4: (1,-1,0,0),\nonumber\\
	&v_5: (0,0,1,1), v_6: (1,1,-1,1), v_7: (1,1,1,-1), v_8: (-1,1,1,1),\nonumber\\
	&v_9: (1,0,0,1), v_{10}: (0,1,-1,0), v_{11}: (1,0,0,-1), v_{12}: (1,-1,-1,1),\nonumber\\
	&v_{13}: (1,1,1,1), v_{14}: (1,-1,1,-1), v_{15}: (0,1,0,-1), v_{16}: (1,0,-1,0),\nonumber\\
	&v_{17}: (1,0,1,0), v_{18}: (0,1,0,0).\label{eq:quantumcega}
\end{align}
It is then easy to verify that each hyperedge in Eq.~\eqref{eq:hyperedges} defines a projective measurement when the vertices are associated with projections onto the vectors in Eq.~\eqref{eq:quantumcega}. Hence, we have that $\mathcal{Q}(\Gamma_{18})\neq \varnothing$ since, given these projective measurements, any choice of density matrix on a Hilbert space of dimension $4$ will lead to a valid quantum probabilistic model on $\Gamma_{18}$.

\subsection{A statistical proof of the KS theorem}\label{sec:kcbs}
\begin{figure}
	\centering
	\includegraphics[scale=0.4]{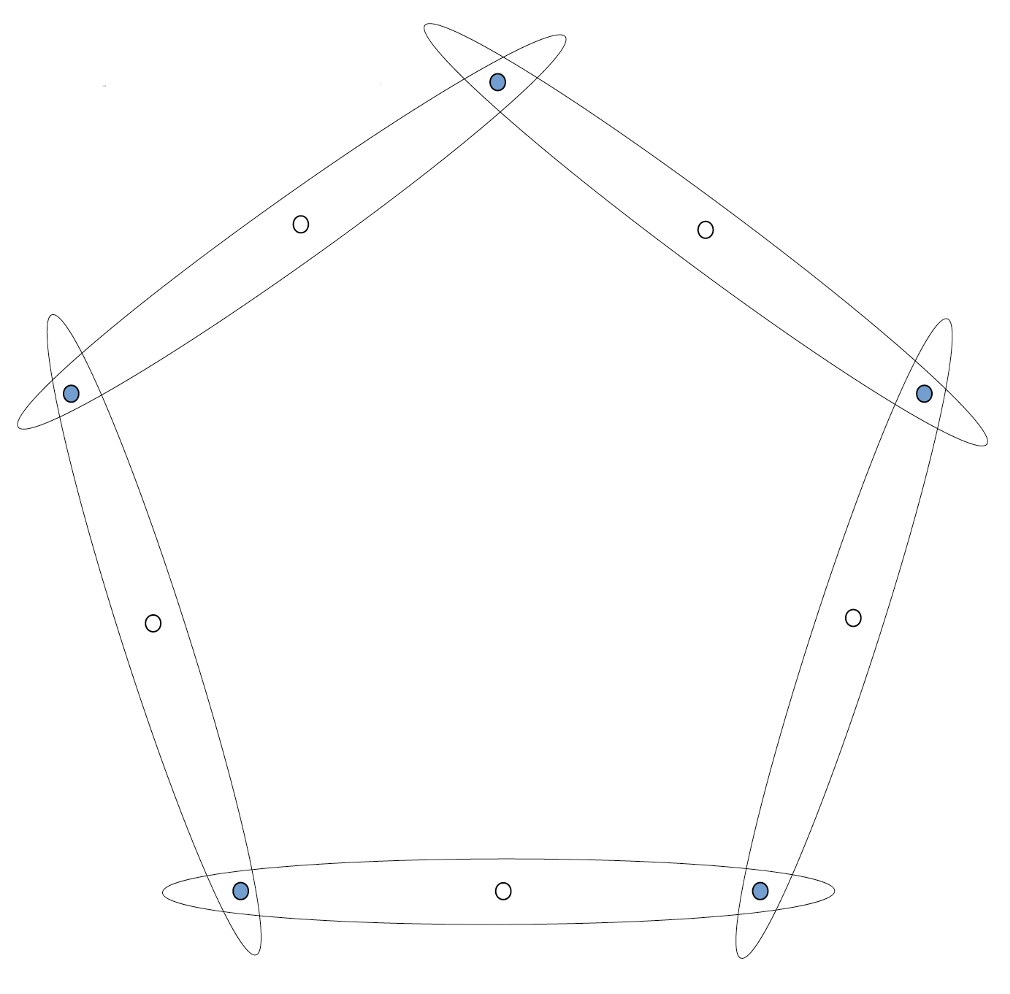}
	\caption{The KCBS contextuality scenario, $\Gamma_5$ \cite{KCBS}.}
	\label{fig:gamma5}
\end{figure}

We will use the proof due to Klyachko \textit{et al.} (or KCBS) \cite{KCBS} as our working example of a statistical proof of the KS theorem, illustrated in Fig.~\ref{fig:gamma5}. It will serve as a paradigmatic example to illustrate features of our hypergraph framework for instances of generalized contextuality motivated by such proofs \cite{KS18,Kunjwal19}.

\begin{figure}
	\centering
	\includegraphics[scale=0.3]{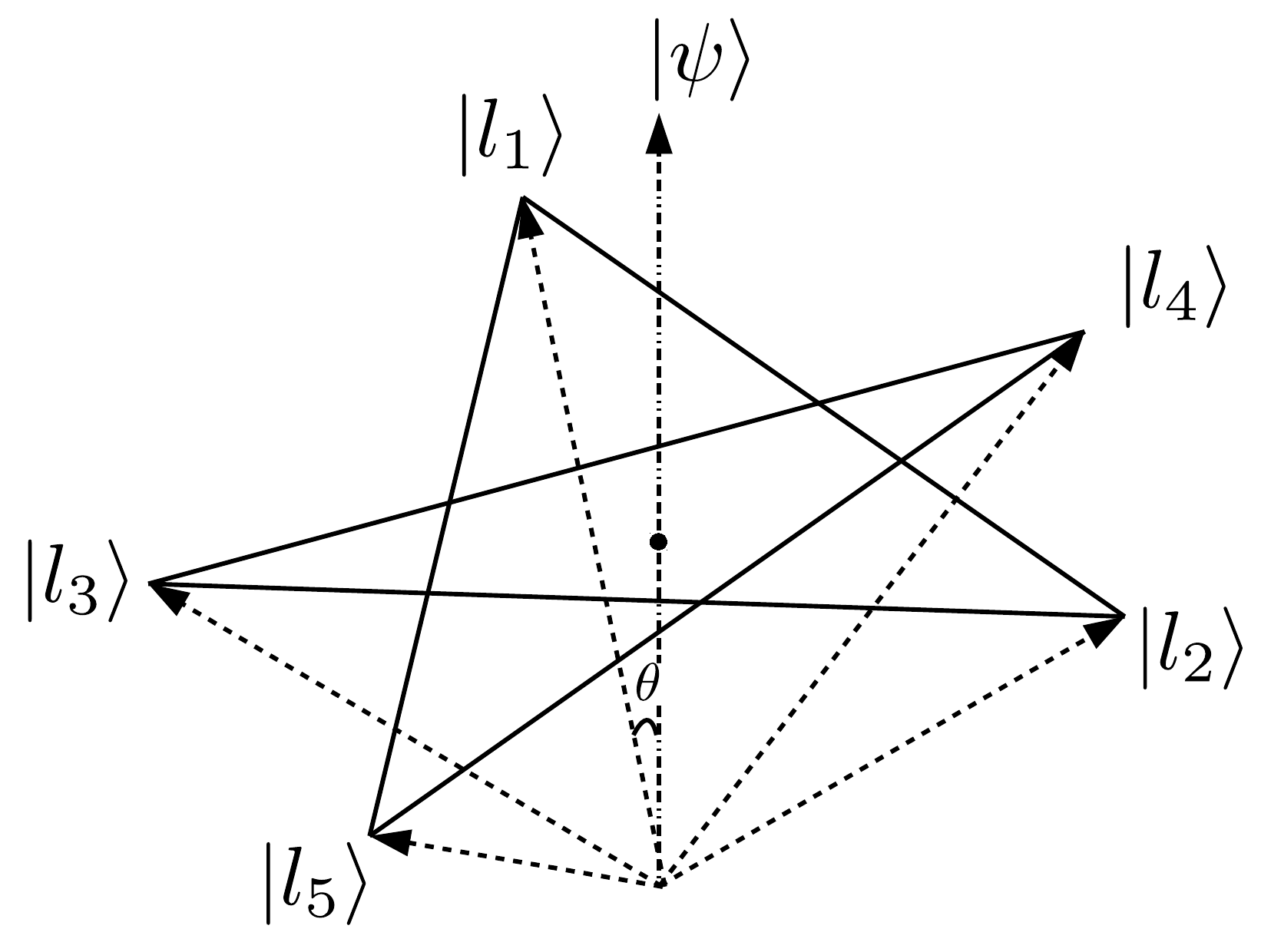}
	\caption{The KCBS construction \cite{KCBS}.}
	\label{fig:kcbsfig}
\end{figure}

We denote this contextuality scenario by the hypergraph $\Gamma_5$ (Fig.~\ref{fig:gamma5}), its vertices by the set $V(\Gamma_5):=\{v_i,u_i\}_{i=1}^5$, and its hyperedges by the set $E(\Gamma_5):=\{e_j\}_{j=1}^5$, where 
\begin{align}
	&e_1:=\{v_1,u_1,v_2\}, e_2:=\{v_2, u_2, v_3\}, e_3:=\{v_3,u_3,v_4\},\nonumber\\
	&e_4:=\{v_4,u_4,v_5\}, e_5:=\{v_5,u_5,v_1\}.\label{eq:kcbsedges}
\end{align}
Hence, any probabilistic model $p\in\mathcal{G}(\Gamma_5)$ must satisfy the following set of linear equations:
\begin{align}
	&p(v_1)+p(u_1)+p(v_2)=1\nonumber\\
	&p(v_2)+p(u_2)+p(v_3)=1\nonumber\\
	&p(v_3)+p(u_3)+p(v_4)=1\nonumber\\
	&p(v_4)+p(u_4)+p(v_5)=1\nonumber\\
	&p(v_5)+p(u_5)+p(v_1)=1\label{eq:kcbseqs}
\end{align}
It is not difficult to see that there exist $p\in\mathcal{G}(\Gamma_5)$ that are KS-colourings, \textit{e.g.}, $p(v_1)=p(v_3)=1$, $p(u_1)=p(v_2)=p(u_2)=p(u_3)=p(v_4)=p(v_5)=p(u_5)=0$, and $p(u_4)=1$. Hence, we have that $\mathcal{KS}(\Gamma_5)\neq \varnothing$. Indeed, the constraints of Eq.~\eqref{eq:kcbseqs} define the classical (Bell-KS) polytope for this scenario, \textit{i.e.}, the set $\mathcal{C}(\Gamma_5)$. The following inequality defines a facet of this polytope, \textit{i.e.}, it is satisfied by all probabilistic models in $\mathcal{C}(\Gamma_5)$ (and saturated by some of them): 
\begin{align}
	\sum_{i=1}^5p(v_i)\leq 2.
\end{align}
The statistical proof of the KS theorem due to KCBS \cite{KCBS} then follows from identifying a quantum probabilistic model that violates this inequality and is, therefore, outside the set $\mathcal{C}(\Gamma_5)$. Indeed, KCBS identify a quantum probabilistic model that achieves the maximum possible quantum violation of this inequalitiy and consists of the following assignments of vectors $\ket{l_i}$ to vertices $v_i$ (for all $i\in\{1,2,\cdots,5\}$) in Fig.~\ref{fig:gamma5} and a particular quantum state $\ket{\psi}$ that achieves the maximum quantum value:
\begin{align}
	&\forall i\in\{1,2,\cdots,5\}:\nonumber\\
	&\ket{l_i}:=(\sin\theta \cos\phi_i,\sin\theta \sin\phi_i,\cos\theta), \phi_i=\frac{4\pi i}{5}, \cos\theta=\frac{1}{\sqrt[4]{5}},\textrm{ and}\\
	&\ket{\psi}=(0,0,1).\label{eq:quantumkcbs}
\end{align}
Fig.~\ref{fig:kcbsfig} depicts the geometry of these vectors. It is then easy to verify that we have 
\begin{align}
	\sum_{i=1}^5p(v_i)=\sum_{i=1}^5|\braket{l_i}{\psi}|^2=\sqrt{5}.
\end{align}
\subsection{Three graph invariants of CSW}

The object of interest in the CSW framework \cite{CSW14} is not the contextuality scenario directly but rather the \textit{orthogonality graph} constructed from it. The orthogonality graph, $G$, is constructed out of the contextuality scenario, $H$, with the mapping denoted by $G:=O(H)$, via
\begin{align}
	&V(G):=V(H),\\
	&E(G):=\{\{v,v'\}|\{v,v'\}\subseteq e, \textrm{ for some }e\in E(H)\}.
\end{align}
Consider the linear combination of probabilities of measurement outcomes relative to a given source event, $[s|S]$, given by
\begin{align}\label{eq:linearfnl}
	R([s|S]):=\sum_{v\in V(G)}w_vp(v|S,s),
\end{align}
where $w_v\geq 0$ for all $v\in V(G)$, \textit{e.g.}, in the KCBS scenario (Fig.~\ref{fig:gamma5}) we have $w_{v_i}=1$ for all $v_i$, $i\in \{1,2,3,4,5\}$. The CSW framework identifies upper bounds on such functions (\textit{cf.}~Eq.~\eqref{eq:linearfnl}) for classical, quantum, and general probabilistic models with graph invariants of $G=O(H)$ for any contextuality scenario $H$. Specifically, we have 

\begin{align}\label{eq:cswbounds}
\forall [s|S]: R([s|S])\overset{\rm \mathcal{C}(H)}{\leq}\alpha(G,w)\overset{\mathcal{Q}(H)}{\leq}\theta(G,w)\overset{\mathcal{CE}^1(H)}{\leq}\alpha^*(G,w).
\end{align}
Here, $\alpha(G,w)$ is the \textit{independence number} of the weighted graph $(G,w)$, $\theta(G,w)$ is the Lovasz theta number of $(G,w)$, and $\alpha^*(G,w)$ is the \textit{fractional packing number} of $(G,w)$. Furthermore, $\mathcal{CE}^1(H)\subseteq\mathcal{G}(H)$ is the set of probabilistic models satisfying the principle of \textit{consistent exclusivity}, defined as follows:

\begin{Def}[Consistent Exclusivity.]
	A probabilistic model $p\in\mathcal{G}(H)$ satisfies the principle of consistent exclusivity, \textit{i.e.}, $p\in\mathcal{CE}^1(H)$, if and only if $\sum_{v\in c}p(v)\leq 1$ for all cliques\footnote{A {\em clique} in a graph $G$ is a subset of vertices such that every pair in the subset is an edge of $G$.} $c$ in the orthogonality graph $G=O(H)$.\footnote{We are using the terminology of Ref.~\cite{AFL15} here. $\mathcal{CE}^1(H)$ is the same as the set of $E1$ probabilistic models in Ref.~\cite{CSW14}.}
\end{Def}

Although for an arbitrary contextuality scenario $H$ the sets $\mathcal{CE}^1(H)$ and $\mathcal{G}(H)$ can be distinct,\footnote{A simple example is a contextuality scenario $H$ with $V(H)=\{v_1,v_2,v_3\}$ and $E(H)=\{\{v_1,v_2\}, \{v_2,v_3\}, \{v_3,v_1\}\}$. Here $\mathcal{CE}^1(H)=\varnothing$ but $\mathcal{G}(H)=\{p: p(v_i)=\frac{1}{2}\textrm{ for all }i\in\{1,2,3\}\}$.} for the contextuality scenarios that can arise in quantum theory with projective measurements (that is, $H$ such that $\mathcal{Q}(H)\neq\varnothing$) we have $\mathcal{CE}^1(H)=\mathcal{G}(H)$. This is because for any set of projections in quantum theory, pairwise orthogonality implies that they can appear as outcomes of a single projective measurement. Following terminology introduced in Ref.~\cite{Kunjwal19}, theories where pairwise compatibility of measurement events\footnote{That is, where \textit{each pair} can occur as outcomes of a single measurement.} implies their global compatibility\footnote{That is, where they can \textit{all} occur as outcomes of a single measurement.} will be said to satisfy \textit{Structural Specker's Principle} (SSP). Quantum theory, restricted to the case where measurement events are projections, is an example of such a theory. In such theories, consistent exclusivity follows as a corollary of SSP: given a set of pairwise compatible measurement events $\{v\}_{v\in c}$, we have that they are pairwise exclusive, \textit{i.e.}, for any $p\in\mathcal{G}(H)$,  $p(v)+p(v')\leq 1$ for all $v,v'\in c$; now, from SSP, we have that the set of pairwise compatible measurement events are globally compatible, which implies that they satisfy global exclusivity, \textit{i.e.}, for any $p\in\mathcal{G}(H)$, $\sum_{v\in c}p(v)\leq 1$. Hence, given that a theory satisfies SSP, for any contextuality scenario $H$ arising in such a theory we have that 
$\mathcal{CE}^1(H)=\mathcal{G}(H)$, \textit{i.e.}, all probabilistic models in the theory satisfy the principle of consistent exclusivity.

The graph invariants appearing in Eq.~\eqref{eq:cswbounds} are defined as follows:

\begin{Def}[Independence number.]
	The independence number of $(G,w)$ is defined as 
	\begin{align}\label{eq:defnin}
		\alpha(G,w)=\max_{\mathcal{I}}\sum_{v\in\mathcal{I}}w_v,
	\end{align}
	where $\mathcal{I}\subseteq V(G)$ is an independent set of vertices of $G$, \textit{i.e.}, these vertices do not share any edges among them.
\end{Def}

\begin{Def}[Lovasz theta number.]
	The Lovasz theta number of $(G,w)$ is defined as 
	\begin{align}\label{eq:defnlt}
		\theta(G,w):=\max_{\{\ket{\phi_v}\}_{v\in V(G)}, \ket{\psi}}\sum_{v\in V(G)}w_v|\braket{\psi}{\phi_v}|^2,
	\end{align}
	where $\{\ket{\phi_v}\}_{v\in V(G)}=\{\ket{\phi_v}\}_{v\in V(\bar{G})}$ (each $\ket{\phi_v}$ a unit vector in $\mathbb{R}^d$ for some integer $d\geq 1$) is an orthonormal representation of $\bar{G}$, the complement of $G$, and the unit vector $\ket{\psi}$ is called a handle. 
	
	Here $V(\bar{G}):=V(G)$ and $E(\bar{G}):=\{(v_1,v_2):v_1,v_2\in V(G), (v_1,v_2)\notin E(G)\}$, and in an orthonormal representation of $\bar{G}$, we have $\braket{\phi_{v}}{\phi_{v'}}=0$ for all $(v,v')\notin E(\bar{G})$, or equivalently, for all $(v,v')\in E(G)$.
\end{Def}

\begin{Def}[Fractional packing number.]
	The fractional packing number of $(G,w)$ is defined as 
	\begin{align}\label{eq:defnfp}
		\alpha^*(G,w):=\max_{\{p(v)\}_{v\in V(G)}}\sum_{v\in V(G)}w_vp(v),
	\end{align}
where $\{p(v)\}_{v\in V(G)}$ is such that $p(v)\geq 0$ for all $v\in V(G)$ and $\sum_{v\in c}p(v)\leq 1$ for all cliques $c$ in $G$.
\end{Def}
When $G=O(H)$ (as it is in our case, $H$ being the underlying contextuality scenario), we have that 
\begin{align}
		\alpha^*(G,w)=\max_{p\in\mathcal{CE}^1(H)}\sum_{v\in V(G)}w_vp(v).
\end{align}

\section{Hypergraph frameworks for generalized contextuality}
We are now in a position to outline how proofs of the KS theorem can be upgraded to witnesses of noise-robust or generalized contextuality within a hypergraph-theoretic approach. The key additional ingredients (beyond those in the CSW \cite{CSW14} and AFLS \cite{AFL15} frameworks) that we need to be able to do so are
\begin{itemize}
	\item a hypergraph invariant that we term the \textit{weighted max-predictability}, and
	\item an operational quantity, ${\rm Corr}$, that captures source-measurement correlations as a proxy for the noise that might affect measurements in a real experiment.
\end{itemize} 

\subsection{Weighted max-predictability}
Given a contextuality scenario $H$ and probabilistic weights $\{q_e\}_{e\in E(H)}$ assigned to its hyperedges such that $q_e\geq 0$ for all $e\in E(H)$ and $\sum_{e\in E(H)}q_e=1$, we define the weighted max-predictability of $(H,q)$ as 
\begin{align}\label{eq:defbeta}
	\beta(H,q):=\max_{p\in\mathcal{G}(H)|_{\rm ind}}\sum_{e\in E(H)}q_e\left(\max_{v\in e}p(v)\right),
\end{align}
where $\mathcal{G}(H)|_{\rm ind}\subseteq \mathcal{G}(H)$ is the set of probabilistic models that are in the convex hull of indeterministic vertices---namely, vertices of the polytope that assign probabilities in $(0,1)$ to at least one vertex in the hypergraph---of the full polytope of general probabilistic models $\mathcal{G}(H)$.\footnote{Note that we use the word ``vertex" in two senses: vertex of a hypergraph and vertex of a polytope that describes probabilistic models on the hypergraph. We will make the distinction explicit whenever it is not obvious from the context.}

\subsection{Source-measurement correlations, {\rm Corr}}
The source-measurement correlations are defined as 
\begin{align}\label{eq:defcorr}
	{\rm Corr}:=\sum_{e\in E(H)}q_e\sum_{m_e,s_e}\delta_{m_e,s_e}p(m_e,s_e|M_e,S_e),
\end{align}
where $\{q_e\}_{e\in E(H)}$ are probabilistic weights satisfying $q_e\geq 0$ for all $e\in E(H)$ and $\sum_{e\in E(H)}q_e=1$, chosen such that $\beta(H,q)<1$. Furthermore, we assume the following labelling conventions in Eq.~\eqref{eq:defcorr}:
\begin{itemize}
	\item The measurement event $[m_e|M_e]$ corresponds to the vertex $v\in e$, where the hyperedge $e\in E(H)$. To be concrete, we choose labels $m_e\in\{1,2,\dots,|e|\}$ and $M_e\in\{1,2,\dots,|E(H)|\}$, where operationally equivalent measurement events are represented by the same vertex $v\in V(H)$.
	
	The general form of the operational equivalence between measurement events $[m_e|M_e]$ and $[m_{e'}|M_{e'}]$ (representing the same vertex $v\in e,e'$, where $e,e'\in E(H)$) is given by 
	\begin{align}\label{eq:opeqsmmts}
		\forall [s|S]: p(m_e,s|M_e,S)&=p(m_{e'},s|M_{e'},S),\nonumber\\
		\textrm{ or equivalently, }
		[m_e|M_e]&\simeq [m_{e'}|M_{e'}]
	\end{align}
	
	\item $S_e\in\mathbb{S}:=\{1,2,\dots,|E(H)|\}$ represent source settings (one for each $e\in E(H)$) and $s_e\in V_{S_e}:=\{1,2,\dots,|e|\}$ denote source outcomes (one for each $v\in e$), \textit{i.e.}, we mirror the measurement events $[m_e|M_e]$ by source events $[s_e|S_e]$ and we measure the correlation $p(m_e,s_e|M_e,S_e)$ between them. Crucially, though, we do not need to mirror the operational equivalences between measurement events at the level of the source events. 
	
	The operational equivalence  of interest for source events is the one resulting from coarse graining over the source outcomes, \textit{i.e.},
	
	\begin{align}\label{eq:opeqspreps}
		\forall [m|M]: \sum_{s_e}p(m,s_e|M,S_e)&=\sum_{s_{e'}}p(m,s_{e'}|M,S_{e'}),\textrm{ for all }S_{e},S_{e'}\in\mathbb{S},\nonumber\\
		\textrm{ or equivalently, }
		[\top|S_e]&\simeq [\top|S_{e'}]\textrm{ for all }S_e,S_{e'}\in\mathbb{S}.
	\end{align}

\end{itemize}

\subsection{Generalized contextuality from logical proofs of the KS theorem}
Logical proofs of the KS theorem rely on contextuality scenarios $H$ for which $\mathcal{Q}(H)\neq\varnothing$ but $\mathcal{C}(H)=\varnothing$. For these scenarios the full set of probabilistic models $\mathcal{G}(H)=\mathcal{G}(H)|_{\rm ind}$ since, in the absence of any deterministic vertices (as $\mathcal{C}(H)=\mathcal{KS}(H)=\varnothing$), all the vertices of the polytope of probabilistic models must be indeterministic.

Applying the assumptions of noncontextuality---for measurement events and for source events---with respect to the operational equivalences of Eqs.~\eqref{eq:opeqsmmts} and \eqref{eq:opeqspreps}, we arrive at the following general form for a witness of generalized contextuality based on a logical proof of the KS theorem arising from contextuality scenario $H$:

\begin{align}\label{eq:logicalineqs}
	{\rm Corr}\leq \beta(H,q)<1,
\end{align}
where ${\rm Corr}$ is defined by Eq.~\eqref{eq:defcorr}, $\beta(H,q)$ is defined by Eq.~\eqref{eq:defbeta}, and the probability distribution $q$ over hyperedges is chosen such that $\beta(H,q)<1$ (the simplest choice being a uniform probability distribution).

We refer the reader to Ref.~\cite{Kunjwal20} for more details of this framework and proceed to describe concrete applications next.

\subsubsection{Example: the 18-ray proof \cite{CEGA96}}
Here we illustrate how the general form of Eq.~\eqref{eq:logicalineqs} applies to the special case of the the 18-ray proof of Cabello \textit{et al.}~\cite{CEGA96} that we described in Section \ref{sec:cega}. This was originally done in Ref.~\cite{KS15}, but was later shown to be a special case of the general form of Eq.~\eqref{eq:logicalineqs} in Ref.~\cite{Kunjwal20}. The contextuality scenario $\Gamma_{18}$ associated with this proof is given in Fig.~\ref{fig:18ray}. 

Assuming a uniform probability distribution $q_e=\frac{1}{9}$ for all $e\in E(\Gamma_{18})$, we have that 
\begin{align}
	\beta(\Gamma_{18},q)=\frac{5}{6}.
\end{align}
The corresponding noise-robust witness of contextuality is then given by 
\begin{align}
	{\rm Corr}=\frac{1}{9}\sum_{e\in E(\Gamma_{18})}\sum_{m_e,s_e}\delta_{m_e,s_e}p(m_e,s_e|M_e,S_e)\leq \frac{5}{6}.
\end{align}

In the ideal quantum case where all the measurement events are projections (i.e., noiseless sharp measurements) and the source event associated to each measurement event is an eigenstate of the projection, we have that 
\begin{align}
	{\rm Corr}=1,
\end{align}
maximally violating our inequality. However, our witness of contextuality is noise-robust in the sense that in the non-ideal case of noisy measurements, it is still possible to violate our inequality (and witness generalized contextuality) as long as the noise is low enough and the source-measurement correlations high enough that ${\rm Corr}>\frac{5}{6}$. An experiment in the noisy regime where $\frac{5}{6}<{\rm Corr}<1$ is therefore able to rule out the existence of a generalized-noncontextual ontological model of the experiment under conditions far removed from those of the KS theorem (which requires projective measurements and an assumption of outcome determinism in any ontological model for them).

\subsection{Generalized contextuality from statistical proofs of the KS theorem}

Statistical proofs of the KS theorem rely on contextuality scenarios $H$ for which $\varnothing \subsetneq \mathcal{C}(H) \subsetneq \mathcal{Q}(H)\subseteq \mathcal{G}(H)$.

We apply the assumptions of noncontextuality---for measurement events and for source events---with respect to the operational equivalences of Eqs.~\eqref{eq:opeqsmmts} and \eqref{eq:opeqspreps} together with an additional operational equivalence corresponding to a special source setting $S_*\notin\mathbb{S}$ with source outcomes $s_*\in\{0,1\}$, given by
\begin{align}
	\forall [m|M]:\sum_{s_*=0}^1p(m,s_*|M,S_*)&=\sum_{s_e}p(m,s_e|M,S_e),\textrm{ for all }S_e\in\mathbb{S},\nonumber\\
	\textrm{or equivalently, }
	[\top|S_*]&\simeq [\top|S_e],\textrm{ for all }S_e\in\mathbb{S}.
\end{align}
This leads us to the following general form for a witness of generalized contextuality based on a statistical proof of the KS theorem associated with contextuality scenario $H$ with orthogonality graph $G=O(H)$:

\begin{align}\label{eq:statksgen}
	R([s_*=0|S_*])\leq \alpha(G,w)+\frac{\alpha^*(G,w)-\alpha(G,w)}{p_0}\frac{1-{\rm Corr}}{1-\beta(H,q)},
\end{align}
where $R([s_{*}=0|S_*])$ is a (non-negative) linear combination of probabilities (as defined in Eq.~\eqref{eq:linearfnl}) given by  
\begin{align}
	R([s_*=0|S_*]):=\sum_{v\in V(G)}w_vp(v|S_*,s_*=0)
\end{align}
with
\begin{align}
	p(v|S_*,s_*=0)=\frac{p(m_e,s_*=0|M_e,S_*)}{p(s_*=0|S_*)}=\frac{p(m_e,s_*=0|M_e,S_*)}{p_0},
\end{align}
where measurement event $[m_e|M_e]$ corresponds  to the vertex $v\in e$, $e\in E(H)$, and $p_0:=p(s_*=0|S_*)$. The graph invariants $\alpha(G,w)$ and $\alpha^*(G,w)$ are those defined in Eqs.~\eqref{eq:defnin} and \eqref{eq:defnfp}, respectively. The hypergraph invariant $\beta(H,q)$ is defined in Eq.~\eqref{eq:defbeta} and ${\rm Corr}$ is defined in Eq.~\eqref{eq:defcorr}.

\subsubsection{Example: the KCBS proof \cite{KCBS}}
Given the pentagon graph $G$ associated with the KCBS scenario (Fig.~\ref{fig:gamma5}), and given $w_v=1$ for all $v\in V(G)$, we have that $\alpha(G,w)=2$ and $\alpha^*(G,w)=\frac{5}{2}$. Assuming a uniform probability distribution $q_e=\frac{1}{5}$ for all $e\in E(\Gamma_5)$, we have $\beta(\Gamma_5,q)=\frac{1}{2}$. On substituting these values in Eq.~\eqref{eq:statksgen}, and denoting $R:=R([s_*=0|S_*])$, we obtain 
\begin{align}
	R\leq 2+\frac{1-{\rm Corr}}{p_0}
\end{align}
as the noise-robust witness of generalized contextuality that generalizes the statistical witness of KS-contextuality given by $R\leq 2$.

\section{Application: entanglement-assisted one-shot classical communication}

We now outline an application of generalized contextuality to the problem of entanglement-assisted one-shot classical communication, summarizing key results of Ref.~\cite{YK22}. In providing a concrete application of generalized contextuality, Ref.~\cite{YK22} also generalized the results of Ref.~\cite{CLM10}---based on KS-contextuality---to the noisy regime using the hypergraph framework of Ref.~\cite{Kunjwal20}. 

\begin{figure}[htb!]
	\centering
	\includegraphics[scale=0.15]{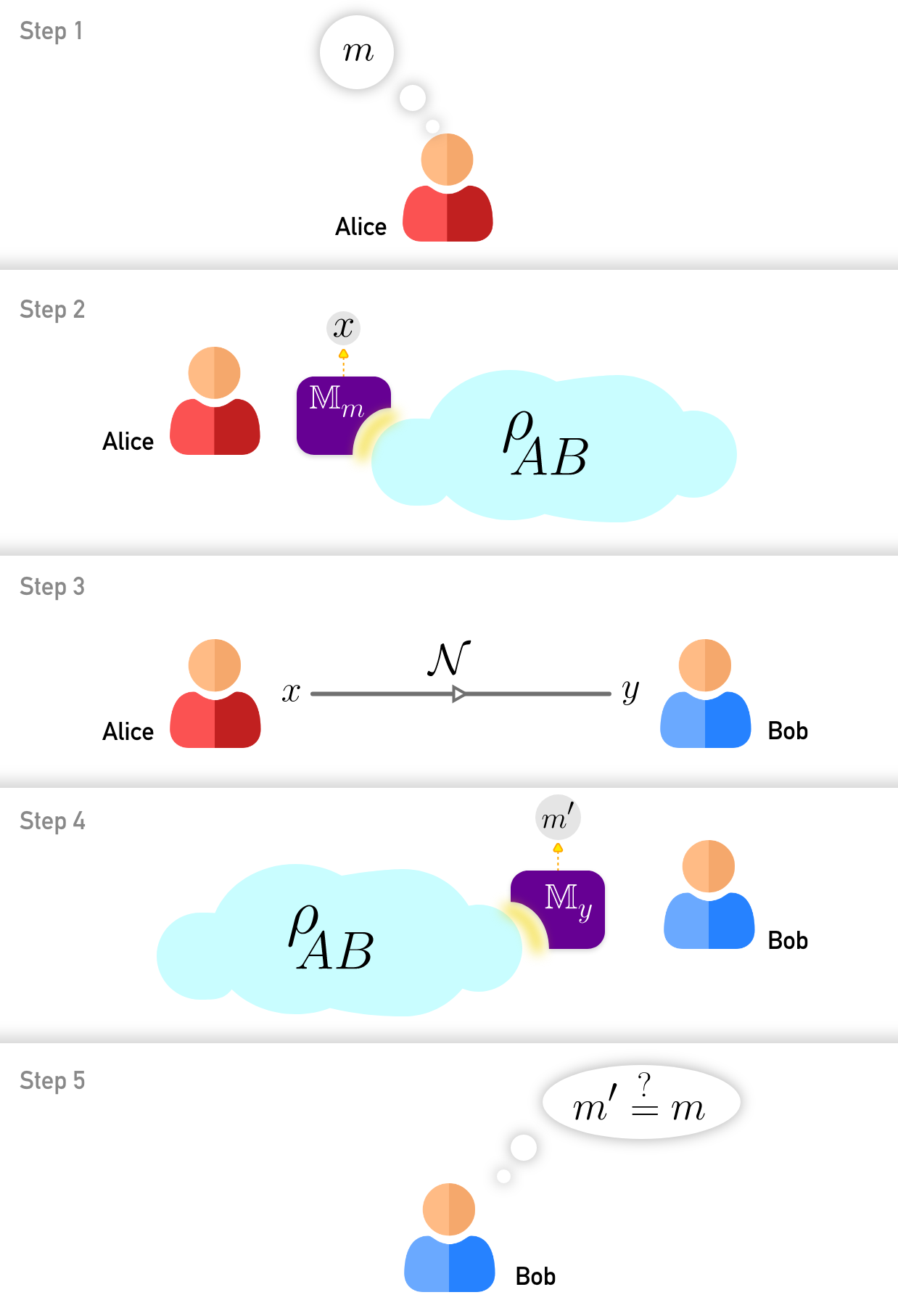}
	\caption{A schematic of the general protocol. Alice and Bob are connected via a classical channel $\mathcal{N}$ and they share an entangled state $\rho_{AB}$. Once Alice decides to send a message $m$ in Step $1$, she encodes this message in her measurement choice in Step $2$ and obtains outcome $x$. In Step $3$, this outcome serves as the channel input  and Bob obtains channel output $y$ with probability $\mathcal{N}(y|x)$. Based on $y$, in Step $4$, Bob measures his quantum system and, after some classical post-processing, obtains his guess $m'$ for Alice's message $m$. They succeed if $m'=m$ (Step $5$).}
	\label{fig:schematic}
\end{figure}

The problem concerns the general protocol illustrated in Figure \ref{fig:schematic}, where Alice and Bob share an entangled state $\rho_{\rm AB}$ and Alice can send messages to Bob via a noisy classical channel $\mathcal{N}$.\footnote{Formally, $\mathcal{N}$ is a discrete, memoryless classical channel that receives inputs $x\in X$ and outputs $y\in Y$, where the probability of output $y$ given input $x$ is $\mathcal{N}(y|x)\geq 0$ such that $\sum_{y\in Y}\mathcal{N}(y|x)=1$ for all $x\in X$.} Their goal is to improve the success probability of classical communication through the channel using shared entanglement. The general protocol consists of the following steps:
\begin{enumerate}
	\item Alice picks a message $m\in {\rm \bf Msg}$ that she wants to send to Bob from a set of messages ${\rm \bf Msg}$ with probability $p(m)$.
	\item Alice's encoding strategy for $m$ is as follows: she implements a POVM $\mathbb{M}_m$ on her part of the entangled state to obtain an outcome $x$ (corresponding to the input alphabet of $\mathcal{N}$). She feeds $x$ into the channel.
	\item Bob receives the symbol $y$ in the output alphabet of $\mathcal{N}$ with probability $\mathcal{N}(y|x)$.
	\item Bob's decoding strategy for $m$ is as follows: he implements a POVM $\mathbb{M}_y$ on his part of the entangled state and obtains and outcome $m'$ that takes values in the set of messages that Alice wanted to send.
	\item Alice and Bob succeed in the communication task if $m'=m$. The one-shot success probability is then given by 
	\begin{align}\label{eq:oneshotsucc}
		S:=\sum_m p(m)\sum_xp(x|m)\sum_y\mathcal{N}(y|x)\sum_{m'}\delta_{m',m}p(m'|m,y,x).
	\end{align}
\end{enumerate}

We obtain the following results:
\begin{enumerate}
	\item Assuming a preparation noncontextual model for Bob's system bounds the success probability exactly by the classical success probability. Hence, preparation contextuality relative to Bob's system is necessary for a quantum advantage in the task.
	
	\item We show mapping from the one-shot communication task to a nonlocal game such that preparation contextuality powers a quantum advantage in the communication task if and only if Bell nonlocality powers an advantage in the nonlocal game. This generalizes the connection between the communication task and nonlocal (pseudotelepathy) games noted in Ref.~\cite{CLM10} for the ideal zero-error case.
	
	\item We motivate a constraint, termed \textit{context-independent guessing} (CIG), on the communication task: this constraint makes operational sense when Bob trusts the possibilistic structure of the channel $\mathcal{N}$ (\textit{i.e.}, which outputs $y$ are possible for which inputs $x$) but does not trust the exact channel probabilities (\textit{i.e.}, the probabilities $\mathcal{N}(y|x)$). We then show that for some classical channels (\textit{e.g.}, the one studied in Ref.~\cite{CLM10}), contextuality witnessed by our hypergraph invariant (Eqs.~\eqref{eq:defbeta}, \eqref{eq:logicalineqs}) implies an enhancement in the one-shot success probability (Eq.~\eqref{eq:oneshotsucc}) beyond the classical bound.
\end{enumerate}

We refer the interested reader to Ref.~\cite{YK22} for a detailed account of our results, in particular the role of weighted max-predictability in detecting quantum advantage in this task.

\section{Conclusion}
In this chapter, we first reviewed the notions of Kochen-Specker and generalized contextuality \cite{KS67,Spekkens05}. We then presented an account of the complementary hypergraph frameworks of Refs.~\cite{Kunjwal19, Kunjwal20} that systematically turn logical/statistical proofs of KS-contextuality into proofs of generalized contextuality. We also discussed an application of witnesses of generalized contextuality to the problem of noisy one-shot classical communication assisted by entanglement \cite{YK22}. We expect these frameworks to be useful for turning applications of KS-contextuality into noise-robust applications of generalized contextuality beyond those envisaged in Ref.~\cite{YK22}.
\newpage

\chapter{Contextuality vs.~Entanglement}\label{chap:ContextualityEntanglement}

\toabstract{
We study the logical relationship between Gleason's theorem, KS-contextuality, and entanglement in the case of multiqubit systems. Specifically, we show the necessity of entanglement for demonstrating KS-contextuality in systems of many qubits and the implications of this for models of multiqubit quantum computation.}

\clearpage

\section{Introduction}
Entanglement and KS-contextuality are both notions of nonclassicality central to quantum theory. While entanglement is an intrinsically composite notion of nonclassicality, KS-contextuality is applicable to indivisible systems. Witnessing the former notion requires at least a pair of qubits while the latter requires at least a qutrit. To understand the interplay between entanglement and KS-contextuality requires us, therefore, to consider quantum systems of dimension at least $d=4$. In this chapter we will provide an overview of results from Ref.~\cite{WK21} on the interplay between entanglement and KS-contextuality for multiqubit systems. 

The starting point of our investigation is Gleason's theorem, which lies at the heart of quantum theory,  showing how the combinatorics of quantum measurements constrains quantum probabilities \cite{HP04}. Beginning with the observation that Gleason's theorem implies the KS theorem, we show the following key results:

\begin{enumerate}
	\item For multiqubit systems, any logical proof of the KS theorem requires entangled measurements (Theorem \ref{thm:logicalent}).
	
	This implies, in particular, that Gleason's theorem for multiqubit systems cannot be proved with unentangled measurements (a result previously proved by Wallach \cite{Wallach02}).
	
	\item For multiqubit systems, any statistical proof of the KS theorem requires an entangled state (Theorem \ref{thm:BellKS}).
	
	\item Not every entangled state yields a statistical proof of the KS theorem with unentangled projective measurements.
	
	This follows as a consequence of the fact (which we prove) that an entangled state yields a statistical proof of the KS theorem with unentangled measurements if and only if it yields a Bell inequality violation with product projective measurements. 
	
	\item We outline the relationship between Gleason's theorem and KS theorem for unentangled measurements in Figure \ref{fig:unentksgleason}.
\end{enumerate}

We also apply these results to better understand resource aspects of restricted schemes for multiqubit quantum computation with state injection (QCSI) proposed by Bermejo-Vega \textit{et al.}~\cite{BDB17}. 

\section{Gleason's theorem vs.~the Kochen-Specker theorem}

\subsection{What Gleason's theorem says}
The structure of quantum measurements determines much of the probabilistic character of quantum theory. This is evident from Gleason's theorem \cite{Gleason57} where, given minimal assumptions on a map from projections to probabilities, one can obtain the Born rule and, in the process, the structure of quantum states. In this way Gleason's theorem rules out the existence of states in quantum theory that can assign deterministic outcomes to all quantum measurements, \textit{i.e.}, quantum measurements are intrinsically probabilistic. A key feature that this theorem hinges on is the possibility of the same projection appearing as an outcome of different projective measurements: without this feature, it is impossible to prove Gleason's theorem, \textit{e.g.}, projective measurements on a qubit do not suffice to prove Gleason's theorem since every qubit projection can appear in exactly one projective measurement.

The formal statement of Gleason's theorem is the following (in the version presented in Ref.~\cite{WK21}):

\begin{Th}[Gleason \cite{Gleason57}]\label{thm:Gleason}
	Let $\cH$ be a separable Hilbert space of dimension at least three. Any map $f:\mathcal{P}(\cH)\rightarrow[0,1]$ satisfying
	\begin{equation}
		f(\Pi_1)+f(\Pi_2)+\cdots=f(\Pi_1+\Pi_2+\cdots)\,,
	\end{equation}
	for any set of mutually orthogonal projections $\left\{\Pi_1,\Pi_2,\ldots\right\}$, and $f(\id_{\mathcal{H}})=1$ where $\id_\cH$ is the identity operator on $\cH$, admits an expression
	\begin{equation}\label{bornrule}
		f(\Pi)=\Tr(\Pi\rho)\,,
	\end{equation} 
	for some density operator $\rho$ on $\cH$.
\end{Th}

\subsection{What the KS theorem says}

The Kochen-Specker (KS) theorem \cite{KS67} provides a key insight into the logical structure of quantum theory, namely, that it is impossible to deterministically assign outcomes to projective measurements in a manner that is independent of which other measurements are implemented simultaneously with a given measurement, \textit{i.e.}, their measurement context. What this means at the level of projections is that there exists \textit{no map} from projections to the set $\{0,1\}$ such that satisfies similar constraints as in Gleason's theorem (\textit{cf.}~Theorem \ref{thm:Gleason}). We recall here the formal statement of the KS theorem (already stated in Chapter \ref{chap:hypergraphframeworks}, Theorem \ref{thm:KS}, but rephrased here for easy comparison with Gleason's theorem):
\begin{Thnon}[Kochen--Specker \cite{KS67}]
	Let $\cH$ be a separable Hilbert space of dimension at least three. There does not exist any map $c:\mathcal{P}(\cH)\rightarrow\{0,1\}$ satisfying
	\begin{equation}
		c(\Pi_1)+c(\Pi_2)+\cdots=c(\Pi_1+\Pi_2+\cdots)\,,
	\end{equation}
	for any set of mutually orthogonal projections $\left\{\Pi_1,\Pi_2,\ldots\right\}$, and $c(\id_{\mathcal{H}})=1$ where $\id_\cH$ is the identity operator on $\cH$.
\end{Thnon}

\subsection{How Gleason's theorem implies the KS theorem}
It is now easy to see that Gleason's theorem implies the KS theorem: the only difference between the function $f$ in Gleason's theorem and the function $c$ in the KS theorem is in their co-domains, with the co-domain of $c$ being a strict subset of the co-domain of $f$. The fact that Gleason's theorem---under the same assumptions on $f$ as on $c$ (except for their co-domains)---shows that $f(\Pi)=\Tr(\Pi\rho)$ for some density operator $\rho$ on $\cH$ is sufficient to conclude that no map $c$ exists. 

The argument runs as follows: Assume such a map $c$ exists; it would then assign deterministic outcomes to all projective measurements (\textit{i.e.}, probabilities in $\{0,1\}$ to all projections); since $c$ is a particular instance of $f$, by Gleason's theorem we must be able to write $c(\Pi)=\Tr(\Pi\rho)$ for some density operator $\rho$; however, and this is the crucial point, every density operator must necessarily fail to be deterministic on some projection, and so must the density operator that defines $c$, \textit{i.e.}, the co-domain of $c$ can't be $\{0,1\}$; we thus have a contradiction between the existence of $c$ and Gleason's theorem; hence, the map $c$ cannot exist.

\subsection{Why care about the KS theorem: finite KS sets}
Since Gleason's theorem implies the KS theorem, one might wonder why we care about the KS theorem at all. Let us take a moment to address this point. There is a qualitative distinction between the two theorems: while the KS theorem is a \textit{no-go theorem}, excluding probability rules of type $c$ for projections on a Hilbert space (of dimension $\geq 3$), Gleason's theorem is a \textit{go-theorem}, exactly specifying the most general form of probability rules of type $f$ for projections on a Hilbert space (of dimension $\geq 3$). While proving Gleason's theorem therefore requires one to consider an uncountably infinite number of projections on a Hilbert space, it is in principle possible to prove the KS theorem with a finite set of projections on a Hilbert space, \textit{e.g.}, by showing that this finite set is already incompatible with the existence of a map $c$, without recourse to an uncountably infinite set of projections required for Gleason's theorem. That the latter possibility is indeed realized was, in fact, the key contribution of Kochen and Specker \cite{KS67}, showing that $117$ projections on a qutrit Hilbert space suffice to rule out the existence of a map $c$. Modern proofs of the KS theorem require as few as $18$ projections on a four-dimensional Hilbert space \cite{CEGA96}. Such sets of projections that rule out the existence of map $c$ (thus proving Theorem \ref{thm:KS}) are called \textit{Kochen-Specker (or KS) sets}.

In the rest of this chapter we will focus on the KS theorem and its interplay with entanglement. We will also mention consequences of this interplay for variations on Gleason's theorem when we  restrict the domain of $f$ to unentangled projections.
\section{KS theorem vs.~Entanglement}
\begin{figure}
	\centering
	\begin{tabular}{|c|c|c|}
		\hline
		$X\otimes I$ & $I\otimes X$ & $X\otimes X$\\
		\hline
		$I\otimes Y$ & $Y\otimes I$ & $Y\otimes Y$\\
		\hline
		$X\otimes Y$ & $Y\otimes X$ & $Z\otimes Z$\\
		\hline
	\end{tabular}\vspace{0.3cm}\\
	\includegraphics[scale=0.45]{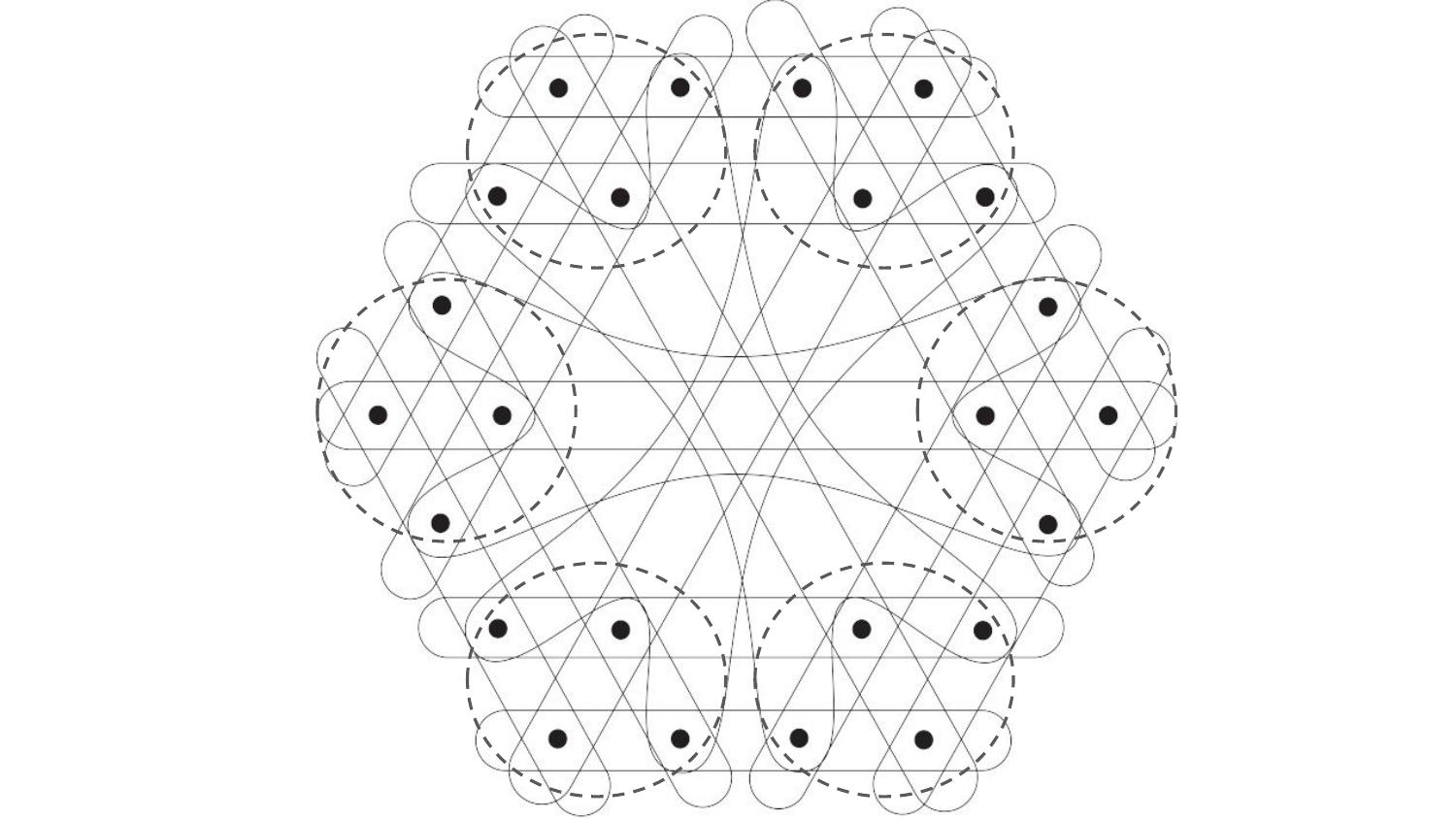}
	\caption{The Peres-Mermin square \cite{Mermin93} consisting of two-qubit Pauli matrices together with the contextuality scenario defined by the orthogonality relations between Peres's 24 rays \cite{Peres91}.}
	\label{fig:peresmermin}
\end{figure}

Consider the well-known proof of the KS theorem that follows from the Peres-Mermin square construction: we have nine two-qubit Pauli measurements arranged in a $3\times 3$ square such that every row/column forms a maximal set of commuting measurements (\textit{cf.}~Figure \ref{fig:peresmermin}).  The proof proceeds by assuming the existence of a valuation map $v$ that assigns to each Hermitian operator a real value lying in its spectrum such that the value assignments respect the algebraic relations between commuting operators. In the case of the Peres-Mermin square, we have the following algebraic relations of interest (obtained by multiplying the commuting operators in each row/column):

\begin{align}
	(X\otimes I)(I\otimes X)(X\otimes X)&=I\otimes I\nonumber\\
	(I\otimes Y)(Y\otimes I)(Y\otimes Y)&=I\otimes I\nonumber\\
	(X\otimes Y)(Y\otimes X)(Z\otimes Z)&=I\otimes I\nonumber\\
	(X\otimes I)(I\otimes Y)(X\otimes Y)&=I\otimes I\nonumber\\
	(I\otimes X)(Y\otimes I)(Y\otimes X)&=I\otimes I\nonumber\\
	(X\otimes X)(Y\otimes Y)(Z\otimes Z)&=-I\otimes I.
\end{align}

The values assigned by the map $v$ must therefore satisfy the following constraints:
\begin{align}
	v(X\otimes I)v(I\otimes X)v(X\otimes X)&=1\nonumber\\
	v(I\otimes Y)v(Y\otimes I)v(Y\otimes Y)&=1\nonumber\\
	v(X\otimes Y)v(Y\otimes X)v(Z\otimes Z)&=1\nonumber\\
	v(X\otimes I)v(I\otimes Y)v(X\otimes Y)&=1\nonumber\\
	v(I\otimes X)v(Y\otimes I)v(Y\otimes X)&=1\nonumber\\
	v(X\otimes X)v(Y\otimes Y)v(Z\otimes Z)&=-1,
\end{align}
where we have that $v(I\otimes I)=1,v(-I\otimes I)=-1$ and $v(P_1\otimes P_2)\in\{+1,-1\}$ for any qubit Pauli operators $P_1\neq I$ and $P_2 \neq I$. These constraints follow from the fact that $v$ assigns values in the spectrum of a Hermitian operator. On multiplying the left-hand-side of these equations, we note that each term appears in two equations and, as such, squares to $1$, yielding the value $1$ for the product. On the other hand the right-hand-side yields the value $-1$, showing that this system of equations is not satisfiable, \textit{i.e.}, no valuation map $v$ exists. Although this looks superficially different from the statement of the KS theorem we stated, note that the map $c$ in Theorem \ref{thm:KS} is an instance of a map of type $v$ since it assigns values in the spectrum of each projection (namely, in $\{0,1\}$) and respects the algebraic relation of additivity between mutually orthogonal (and thus commuting) projections. Ruling out the existence of $c$, therefore, also rules out the existence of $v$. Conversely, if a map $c$ did exist, then a map $v$ could be constructed by using $c$ and linearly extending it to Hermitian operators (using there spectral decompositions). We refer the interested reader to Appendix A of Ref.~\cite{WK21} for more details on the relationship between these two styles of proving the KS theorem.

Of particular interest to us here is the following fact: Each row or column of the Peres-Mermin square is associated to an underlying two-qubit orthonormal basis in which all the observables in that row or column are diagonal, \textit{i.e.}, the observables in each row/column can be obtained by coarse graining this underlying four-outcome projective measurement into three mutually commuting two-outcome projective measurements.  These underlying orthonormal bases (there are six of them, one for each row/column) are non-overlapping, resulting in a contextuality scenario with the $24$ vectors of Peres \cite{Peres91}. These vectors, in turn, have additional orthogonality relations between them, which result in $18$ more orthonormal bases that can be carved out of them. In the resulting contextuality scenario (Figure \ref{fig:peresmermin}), the six dashed hyperedges denote the six diagonalizing orthonormal bases corresponding to the rows and columns of the Peres-Mermin square. The remaining $18$  hyperedges correspond to the the additional orthonormal bases that can be carved out of the $24$ vectors.

\textit{A priori}, all the observables in the Peres-Mermin square are product measurements, \textit{i.e.}, individually, none of them require entanglement to be implemented. However, this does not mean that the Peres-Mermin square itself carries no entanglement. Consider, for example, the set of measurements  $\{XX,YY,ZZ\}$ in the third column. In order to implement these measurements in the way envisaged in the Peres-Mermin square---that is, simultaneously---one needs to do a Bell measurement since the orthonormal basis that simultaneously diagonalizes them is the Bell basis. Any other implementation of one of these measurements (e.g., implementing $XX$ via single qubit measurements on the two qubits separately) will fail to also implement the remaining measurements (e.g., $YY$ or $ZZ$).

It turns out that this feature of the Peres-Mermin square---namely, the necessity of entangled multiqubit measurements---is not accidental. In Ref.~\cite{WK21}, we showed that any multiqubit proof of the KS theorem requires entanglement, either in the measurements (for logical proofs of the KS theorem) or in the state (for statistical proofs of the kS theorem). We describe these results below.

\subsection{Logical proofs}

Our first result is the following theorem.

\begin{Th}\label{thm:logicalent}
	Any multiqubit Kochen-Specker (KS) set necessarily contains entangled projections.
\end{Th}

Recall that a KS set is a finite set of projections for which it is impossible to assign values in $\{0,1\}$ via a map $c$ satisfying the constraints in Theorem \ref{thm:KS}. In this way, a KS set rules out the existence of map $c$. Theorem \ref{thm:logicalent} therefore proves that to demonstrate the non-existence of the map $c$ in Theorem \ref{thm:KS} via a logical proof of the KS theorem, we necessarily need to consider its action on entangled multiqubit projections. Equivalently, it proves that the map $c$ always exists if we restrict the domain of $c$ to unentangled multiqubit projections, \textit{i.e.}, no multiqubit KS set can be constructed out of only unentangled projections. Theorem \ref{thm:logicalent} is the rigorous generalization of the observation that the Peres-Mermin square requires entangled measurements to arbitrary logical proofs of the KS theorem on multiqubit systems (such as generalizations of the Peres-Mermin square \cite{WA13}). We refer the reader to Ref.~\cite{WK21} for a detailed proof of Theorem \ref{thm:logicalent}.

\subsection{Statistical proofs}
In Ref.~\cite{WK21}, we also construct a KS-noncontextual ontological model for multiqubit systems restricted to separable states and unentangled projective measurements, extending the single-qubit Kochen-Specker model \cite{KS67} to account for unentangled measurements that cannot be implemented with local operations and classical communication (LOCC) \cite{BDF99}. This allows to make the following statement: 

\textit{Any multiqubit statistical proof of the KS theorem with unentangled projective measurements necessarily requires an entangled state.}

Hence, entanglement is necessary, whether in the measurements (logical proofs) or in the state (statistical proofs) involved in a proof of the KS theorem.

One might then ask which entangled states are useful for such a statistical proof of the KS theorem. We address this in the following theorem proved in Ref.~\cite{WK21} and restated here:\footnote{Note that we have strengthened the statement of the theorem relative to Ref.~\cite{WK21} to account for the fact that every proof of Bell's theorem is also a statistical proof of the KS theorem.}

\begin{Th}\label{thm:BellKS}
A multiqubit quantum state can yield a statistical proof of the KS theorem with a finite set of unentangled projective measurements if and only if it can violate a Bell inequality with local projective measurements.	
\end{Th} 

The fact that there are entangled states (\textit{e.g.}, Werner states \cite{Werner89}) that cannot violate Bell inequalities with respect to projective measurements means that such states are also useless for statistical proofs of the KS theorem. Hence, an entangled quantum state is a necessary but not sufficient condition for statistical proofs of the KS theorem.

\subsection{Entanglement in KS theorem with multiqudit systems}
Between multiqubit systems (\textit{i.e.}, all systems are qubits) and multiqudit systems where all systems have a Hilbert space dimension at least $3$, we have the possibility of multiqudit systems including both qubits and higher-dimensional qudits. In Ref.~\cite{Wallach02}, Wallach proved that restricting the domain of the function $f$ in Theorem \ref{thm:Gleason} to unentangled multiqudit projections on such systems would not be enough to constrain $f$ to take the form of the Born rule, \textit{i.e.}, having a single qubit in a multiqudit system excludes the possibility of using unentangled projections on this system to obtain Gleason's theorem. This implies, in particular, that unentangled multiqubit projections are not sufficient to prove Gleason's theorem. Our Theorem \ref{thm:logicalent} implies Wallach's for the case of multiqubit systems (recall that Gleason implies KS). However, for multiqudit systems with at least one qudit of dimension at least $3$, it remains possible to prove the KS theorem with unentangled projections even as Gleason's theorem can't be obtained using them \cite{Wallach02}, \textit{i.e.}, we have the following theorem (proved in Ref.~\cite{WK21}):

\begin{Th}\label{thm:unentd3}
	There exists a KS set consisting entirely of product projection on any separable Hilbert space $\cH_1\otimes\cdots\otimes\cH_n$ where $\dim(\cH_j)\geq3$ for some $1\leq j \leq n$.
\end{Th}

Besides the requirement of entangled projections, another curious feature of the Peres-Mermin square is the fact that the entangled measurement (Bell measurement) that appears in it is a fully-entangled measurement, \textit{i.e.}, none of its projections are product projections. Does this property generalize to other multiqubit logical proofs of the KS theorem? In Ref.~\cite{WK21}, we show that this is not the case: using the $33$-ray construction of Peres \cite{Peres91}, we construct a two-qubit KS set such that every orthonormal basis contains at least one product projection.

\begin{figure}
	\centering
	\includegraphics[scale=0.3]{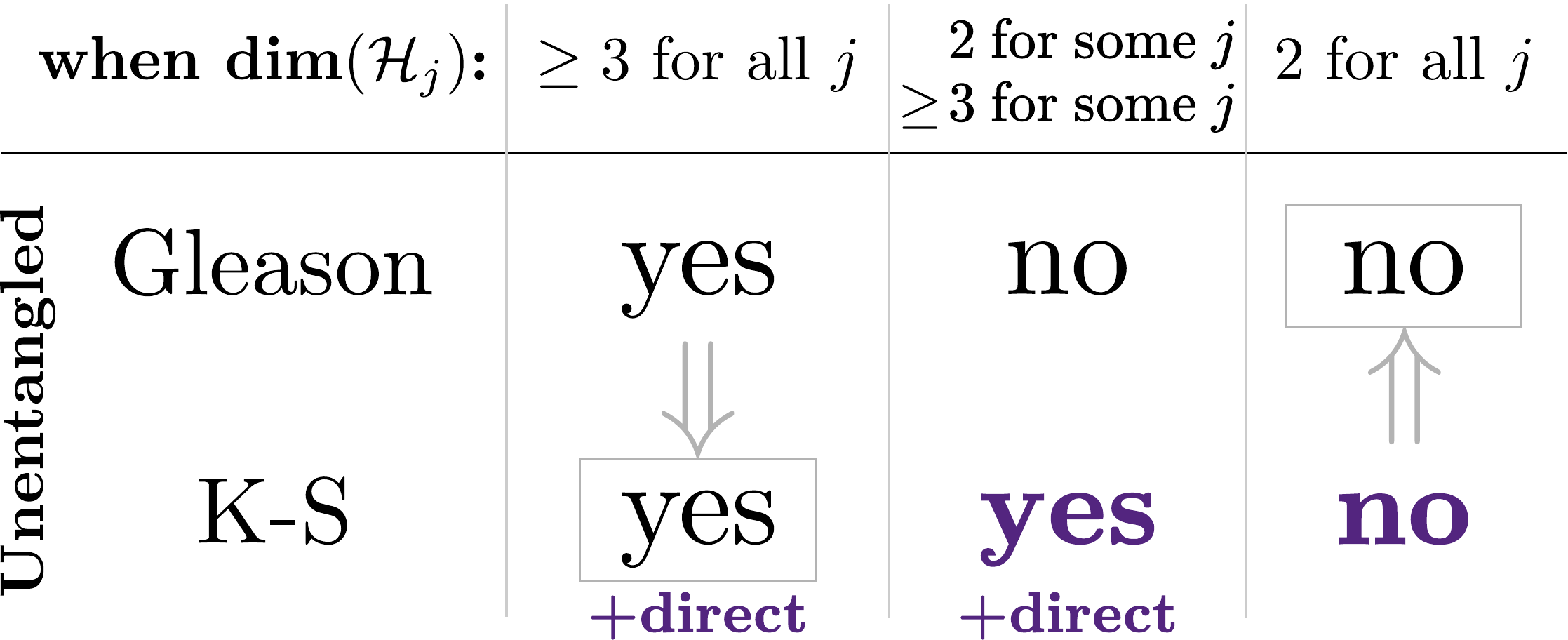}
	\caption{The existence of logical proofs of the KS theorem and proofs of Gleason's theorem without entanglement on a Hilbert space $\cH_1\otimes\cdots\otimes\cH_n$, where $1\leq j\leq n$. The implication arrows show which results follow from each other. The addition of +direct signifies the result also holds for the subset of unentangled measurements given by direct product bases, \textit{i.e.}, using only bases that can be understood as a product of local operations, without requiring the use of bases that require LOCC or separable (SEP) operations beyond LOCC \cite{BDF99,WK21}. The results in bold and purple are our results.}\label{fig:unentksgleason}
\end{figure}

Our investigation of the relationship between Gleason's theorem and the KS theorem for unentangled projections on multiqudit sytems leads us to the overall picture outlined in Figure \ref{fig:unentksgleason}.

\subsection{Models of multiqubit quantum computation with state injection}

The model of quantum computation with state injection (QCSI) \cite{HWV14, BDB17} works by lifting stabilizer quantum circuits, which can be efficiently classically simulated \cite{GK98,AG04}, to universal quantum computation via the injection of so-called {\em magic states} that cannot be prepared via stabilizer circuits. 

For circuits of odd-prime dimensional qudits (or, {\em quopits} \cite{BDB17}) it has been shown that contextuality is a necessary resource for universal quantum computation, \textit{i.e.}, the required magic states must yield a statistical proof of the KS theorem with respect to stabilizer measurements \cite{HWV14}. It has been conjectured (but not proven) that converse also holds, \textit{i.e.}, such KS-contextuality is sufficient for universal quantum computation \cite{HWV14}.

However, the case of multiqubit systems poses a fundamental challenge to making the claim that contextuality of magic states drives universal quantum computation in this model \cite{HWV14}. Multiqubit stablizer circuits are classically efficiently simulable \cite{GK98, AG04} despite the possibility of implementing KS sets with these circuits (\textit{e.g.}~the Peres-Mermin square, Fig.~\ref{fig:peresmermin}). Now, any state, stabilizer or not, exhibits contextuality with respect to a KS set. So while magic states are still a necessary resource for universal quantum computation, their contextuality does not single them out (unlike the quopit case) since {\em all} states can exhibit contextuality. To address this challenge, Ref.~\cite{BDB17} introduced {\em restricted} QCSI schemes that restore contextuality of magic states as a resource for universal quantum computation. Such a QCSI scheme $\mathcal{M}_{\mathcal{O}}$ is subject to the following restrictions:

\begin{enumerate}
	\item[(C1)] \emph{Resource character.} There exists a quantum state that
	does not exhibit contextuality with respect to measurements available in $\mathcal{M}_{\mathcal{O}}$.
	\item[(C2)] \emph{Tomographic completeness.} For any
	state $\rho$, the expectation value of any Pauli observable can be inferred via the allowed operations of the scheme.
\end{enumerate}

(C1) excludes the possibility the scheme can implement KS sets, \textit{i.e.}, state-independent contextuality (of the Peres-Mermin type) is impossible. A sufficient condition for satisfying requirement (C1), following our Theorem \ref{thm:logicalent}, is that every measurement in the scheme is unentangled. In this case, the QCSI scheme cannot implement KS sets. It turns out that the explicit schemes proposed in Ref.~\cite{BDB17} satisfy exactly this condition; they contain no entangled measurements. All contextuality in these schemes therefore stemps from entanglement of the injected magic states. Furthermore, it follows from Theorem~\ref{thm:BellKS} that the injected state can promote such a scheme to universality only if it can violate a Bell inequality with some local projective measurements. 

\section{Conclusion}

The results reported in this chapter clarify the role of entanglement in multiqubit proofs of the KS theorem. They also clarify the relationship between KS and Gleason's theorems viewed through the lens of entanglement in the measurements. We expect this deeper understanding of how these two notions of nonclassicality---entanglement and contextuality---play with each other to inform future research on their role in universal quantum computation. Hints of this are already present in the existing literature of restricted QCSI schemes \cite{BDB17} that we briefly discussed in the context of our results.
\chapter{Correlations without causal order}\label{chap:CorrelationsICO}

\toabstract{
We survey two lines of inquiry: 1) How classical instances of indefinite causal order (\textit{i.e.}, process functions) combined with local operations allow us to implement tasks that are impossible with LOCC (local operations and classical communication), 2) How Bell nonlocality as a notion of nonclassicality admits a natural generalization to another notion, antinomicity, even in the absence of global causal assumptions between parties. We present key results in both directions and discuss future prospects for research.}

\clearpage

\section{Introduction}

Causal reasoning is fundamental to the practice of science. To explain an observation causally is, roughly speaking, to provide an account of how it came about. Usually such an account assumes that causal relations are acyclic, \textit{i.e.}, it is possible to model them with directed acyclic graphs (DAGs) where the direction of an edge represents causal influence from one node of the DAG to another. 
However, what if causal relations were fundamentally quantum, \textit{i.e.}, subject to quantum indefiniteness in a similar sense as other physical properties like position and momentum?\footnote{Aside from its intrinsically foundational motivations, this question is also motivated by considerations of what a theory of quantum gravity that combines the probabilistic aspects of quantum theory with the dynamical aspects of general relativity might look like \cite{Hardy05, Hardy07, Hardy16}.} 

The process-matrix framework \cite{OCB12} offers a possible answer \cite{BLO19, BLO21} to this question by extending the formalism of standard quantum theory to a scenario without global causal assumptions. 
In doing so, it allows for the violation of statistical constraints called causal inequalities that hold under the assumption of a  definite causal order, \textit{i.e.}, it can achieve correlations between parties that are stronger than those that can be explained via an account of causal relations based on DAGs. It turns out \cite{BLO21} that it is possible to explain some such correlations within a framework for quantum causal models that allows for cyclic causality. 
Allowing for such cyclic causation---that is, directed cyclic graphs which admit a path from some node back to itself---can in general lead to time-travel antinomies. However, the process matrix framework evades this possibility by requiring a minimal constraint of logical consistency \cite{OCB12,BW16}. Surprisingly, there also exist process functions---classical deterministic counterparts of process matrices---whose causal structure is modelled exactly by directed cyclic graphs and which violate causal inequalities.

In this chapter, we review two lines of inquiry that explore the information-theoretic implications of indefinite causal order witnessed by causal inequality violations. The first line of inquiry provides an operational meaning to process functions by showing they are intimately related to the phenomenon of quantum nonlocality without entanglement (QNLWE) \cite{BDF99}. This allows us to transfer techniques and results between these two seemingly unrelated fields of inquiry while providing a more fine-grained understanding of the gap between local operations and classical communication (LOCC) and separable operations (SEP) \cite{KB22}. The second line of inquiry asks: What feature of multipartite correlations can witness their nonclassicality in the presence of indefinite causal order? Causal inequality violations do not suffice for this purpose since they admit classical violations \cite{BW16}. Instead, we introduce and investigate a notion of nonclassicality for multipartite correlations---termed  \textit{antinomicity}---that doesn't require any global causal assumptions.

\section{Indefinite Causal Order}
To rigorously formulate the notion of indefinite causal order as used in this chapter, we will follow in the footsteps of Oreshkov, Costa, and Brukner \cite{OCB12}.

\subsection{The operational paradigm}\label{sec:oplpar} We work within the following operational paradigm (introduced in Ref.~\cite{OCB12}): consider $N$ isolated labs embedded in some environment; each lab can receive an input system from the environment and subsequently (according to a local notion of definite causal order) send an output system to the environment; party $S_k$ (in the $k$th lab) receives a classical setting (or question), $a_k\in A_k$, and reports a classical outcome (or answer), $x_k\in X_k$. To determine $x_k$, $S_k$ can implement some local intervention based on $a_k$ on the input system it receives from the environment and, depending on the result of this local intervention, $S_k$ reports the answer $x_k$ and sends an output system to the environment. Crucially, each party can apply arbitrary local interventions on the input system it receives. The communication between the labs is mediated entirely by the environment, with the parties limited to local interventions in their respective labs. The central object of investigation is the observed multipartite correlation $p(\vec{x}|
\vec{a}):=p(x_1,\dots,x_N|a_1,\dots,a_N)$. Note that the parties can execute at most a single-round communication protocol since each party receives or sends out a system at most once, where receiving a system causally precedes sending out a system (see Ref.~\cite{HO21} for a generalization to multiple rounds).\footnote{Our operational paradigm specializes to a Bell scenario if the parties do not communicate and merely receive input systems once from the environment.} This operational paradigm is  \textit{a priori} agnostic about i) the nature of the input/output systems received/sent by the local labs, ii) the nature of the local interventions on them, and iii) the nature of communication between the labs that is mediated by the environment. To specify the nature of these three aspects amounts to specifying a particular operational theory that fixes them, \textit{e.g.}, the (quantum) process-matrix framework, where the input/output systems are quantum systems, the local interventions are quantum instruments, and the nature of (quantum) communication between the labs is dictated by the process describing the environment. In general, the  specification of a particular operational theory would restrict the set of multipartite correlations $p(\vec{x}|\vec{a})$ to a subset of the set of all correlations.

\subsection{The process-matrix framework} Within the operational paradigm outlined above, we now assume that the parties perform local \textit{quantum} operations, \textit{i.e.}, party $S_k$ has an incoming quantum system $I_k$ with Hilbert space $\mathcal{H}^{I_k}$, an outgoing quantum system $O_k$ with Hilbert space $\mathcal{H}^{O_k}$, and can perform arbitrary quantum operations from $I_k$ to $O_k$. A local quantum operation is described by a quantum instrument, \textit{i.e.}, a set of completely positive (CP) maps $\{\mathcal{M}^{S_k}_{x_k|a_k}:\mathcal{L}(\mathcal{H}^{I_k})\rightarrow\mathcal{L}(\mathcal{H}^{O_k})\}_{x_k\in X_k}$, where the setting $a_k\in A_k$ labels the instrument and $x_k\in X_k$ labels the classical outcome associated with each CP map.\footnote{Without loss of generality, we assume that the outcome set $X_k$ is identical for all settings $a_k\in A_k$: this can be ensured by including, if needed, outcomes that never occur, \textit{i.e.}, those represented by the null CP map, for settings that have fewer non-null outcomes than some other setting.} Summing over the classical outcomes yields a completely positive and trace-preserving (CPTP) map $\mathcal{M}^{S_k}_{a_k}:=\sum_{x_k}\mathcal{M}^{S_k}_{x_k|a_k}$. The correlations between the classical outcomes of the different labs given their classical settings are given by
\begin{align}\label{eq:corrpm}
	&p(x_1,x_2,\dots,x_N|a_1,a_2,\dots,a_N)\nonumber\\
	=&\Tr(W M^{I_1O_1}_{x_1|a_1}\otimes M^{I_2O_2}_{x_2|a_2}\otimes\dots\otimes M^{I_NO_N}_{x_N|a_N}),
\end{align}
where 
$M_{x_k|a_k}^{I_kO_k}:=[\mathcal{I}^{I_k}\otimes\mathcal{M}^{S_k}_{x_k|a_k}(d_{I_k}\ket{\Phi^+}\bra{\Phi^+})]^{\rm T} \in\mathcal{L}(\mathcal{H}^{I_k}\otimes\mathcal{H}^{O_k})$
is the Choi-Jamiołkowski (CJ) matrix associated to the CP map $\mathcal{M}^{S_k}_{x_k|a_k}$, $\ket{\Phi^+}\in\mathcal{H}^{I_k}\otimes\mathcal{H}^{I_k}$ is the maximally entangled state, \textit{i.e.}, $\ket{\Phi^+}=\frac{1}{\sqrt{d_{I_k}}}\sum_{j=1}^{d_{I_k}}\ket{j}\ket{j}$, and $\mathcal{I}^{I_k}:\mathcal{L}(\mathcal{H}^{I_k})\rightarrow\mathcal{L}(\mathcal{H}^{I_k})$ is the identity channel. We have that
$M_{a_k}^{I_kO_k}\geq 0$, $\Tr_{O_k}M_{a_k}^{I_kO_k}=\id_{I_k}$.

The operator $W\in\mathcal{L}(\mathcal{H}^{I_1}\otimes\mathcal{H}^{O_1}\otimes\mathcal{H}^{I_2}\otimes\mathcal{H}^{O_2}\otimes\dots\otimes\mathcal{H}^{I_N}\otimes\mathcal{H}^{O_N})$ is called a process matrix and it establishes correlations between the local interventions of the labs. $W$ satisfies the following constraints of \textit{logical consistency}:\footnote{Intuitively, logical consistency of a process is the requirement that ``probabilities add up"---in the sense of non-negativity and normalization of the probabilities $p(\vec{x}|\vec{a})$ for all settings---for arbitrary local interventions that might be carried out by the parties.} $W\geq 0$, and $\forall M^{S_k}\geq 0$ where $M^{S_k}\in\mathcal{L}(\mathcal{H}^{I_k}\otimes\mathcal{H}^{O_k})$ and $\Tr_{O_k}M^{S_k}=\id_{I_k}: \Tr(W \bigotimes_{k=1}^N M^{S_k})=1$.

\subsection{Classical processes and process functions} We now consider two other instantiations of the operational paradigm, the \textit{classical process framework} and the \textit{process-function framework}, both of which can be shown to arise in the diagonal limit of the process-matrix framework (\textit{i.e.}, where the process matrix is assumed to be diagonal in a product basis resulting from some fixed choice of local bases) \cite{BW16}.
In both cases, the input and output systems are classical random variables, the local operation of each party is a classical stochastic map, and the environment is described by a conditional probability distribution dictating how the inputs received by the parties are affected by the outputs sent out by the parties.

More concretely, each party $S_k$ has an incoming classical system represented by a random variable $I_k$ that takes values $i_k\in\{0,1,\dots,d_{I_k}-1\}$ and an outgoing classical system represented by a random variable $O_k$ that takes values $o_k\in\{0,1,\dots,d_{O_k}-1\}$.\footnote{With a slight but standard abuse of notation, we will often also use $I_k$ and $O_k$ to represent the respective sets in which these random variables take values.} The local operations of party $S_k$ are specified by the conditional probability distribution $p(x_k,o_k|a_k,i_k)\in[0,1]$, where $a_k$ and $x_k$ denote, respectively, the setting and outcome for party $S_k$. Using the notation $\vec{x}:=(x_1,x_2,\dots,x_N)$, and  $\vec{a}:=(a_1,a_2,\dots,a_N)$, the multipartite correlations, $p(\vec{x}|\vec{a})$, are then given by 
\begin{align}\label{eq:loccons1}
	p(\vec{x}|\vec{a})=\sum_{\vec{i},\vec{o}}\prod_{k=1}^Np(x_k,o_k|a_k,i_k)p(\vec{i}|\vec{o}),
\end{align}
where $p(\vec{i}|\vec{o})$ is the conditional probability distribution describing the environment and, as such, is not arbitrary but constrained to satisfy the requirement of \textit{logical consistency}, \textit{i.e.}, $p(\vec{i}|\vec{o})$ should be such that for any arbitrary choices of local interventions by the parties, $\{p(x_k,o_k|a_k,i_k)\}_{k=1}^N$, the correlation defined by Eq.~\eqref{eq:loccons1} satisfies non-negativity ($p(\vec{x}|\vec{a})\geq 0$ for all $\vec{x},\vec{a}$) and  normalization ($\sum_{\vec{x}}p(\vec{x}|\vec{a})=1$ for all $\vec{a}$). This condition of logical consistency is necessary and sufficient to exclude the possibility of time-travel antinomies \cite{BW16,BT21}. A logically consistent $p(\vec{i}|\vec{o})$ is a \textit{classical process} in the sense of Ref.~\cite{BW16} and correlations that are achievable by such a process via Eq.~\eqref{eq:loccons1} are said to belong to the \textit{classical process framework}.

On the other hand, correlations achievable via classical processes of the following form are said to belong to the \textit{process-function framework} \cite{BW16,BT21}: $p(\vec{i}|\vec{o})=\sum_{\lambda}p(\lambda)\delta_{\vec{i},\omega^{\lambda}(\vec{o})}$, where $\lambda$ labels the process function $\omega^{\lambda}:\vec{O}\rightarrow\vec{I}$ \cite{BW16, BW16fp, BT21, TB22} defined via $\omega^{\lambda}:=(\omega^{\lambda}_k:\vec{O}\rightarrow I_k)_{k=1}^N$. A process function is a map from the outputs of the parties to their inputs that satisfies logical consistency when written as a conditional probability distribution $p_{\lambda}(\vec{i}|\vec{o}):=\delta_{\vec{i},\omega^{\lambda}(\vec{o})}$. The set of classical processes defining the process-function framework corresponds to the \textit{deterministic-extrema polytope} of Ref.~\cite{BW16}.

\subsection{Classical quasi-processes}We will refer to an arbitrary conditional probability distribution $p(\vec{i}|\vec{o})$ as a \textit{classical quasi-process} and when this distribution is deterministic, we will represent it via a \textit{quasi-process function} $\omega:\vec{O}\rightarrow\vec{I}$, where  $\omega:=(\omega_1,\omega_2,\dots,\omega_N)$, $\omega_k:\vec{O}\rightarrow I_k$ for all $k\in\{1,2,\dots,N\}$, and $p(\vec{i}|\vec{o})=\delta_{\vec{i},\omega(\vec{o})}=\prod_{k=1}^N\delta_{i_k,\omega_k(\vec{o})}$.\footnote{This is non-standard terminology, but we will later find it useful in describing the most general set of correlations in multipartite scenarios.} A classical quasi-process that satisfies logical consistency is a \textit{classical process} \cite{BW16}. If a classical process is deterministic, the quasi-process function associated with it is a \textit{process function} \cite{BT21}. The correlations achievable by a classical quasi-process $p(\vec{i}|\vec{o})$ are given by Eq.~\eqref{eq:loccons1}, with the caveat that local interventions cannot be arbitrary when the classical quasi-process fails to be a classical process, \textit{i.e.}, some local interventions must be disallowed in that case to ensure that the left-hand-side of Eq.~\eqref{eq:loccons1} is a conditional probability distribution that is normalized for all settings. This restriction on local interventions means that classical quasi-processes do not, in general, fall within the operational paradigm we envisage.\footnote{A concrete example outside our operational paradigm is the single-party quasi-process function $p(i_1|o_1)=\delta_{i_1,o_1}$ (where $i_1,o_1\in\{0,1\}$) which results in the grandfather antinomy \cite{BT21}---corresponding to $p(x_1|a_1)=0$ (for all $x_1,a_1$)---for the intervention $p(x_1,o_1|a_1,i_1)=\delta_{o_1,i_1\oplus 1}\delta_{x_1,a_1}$ (where $x_1,a_1\in\{0,1\}$).}

\section{Correlations under a definite causal order}
Within the operational paradigm outlined above in Section \ref{sec:oplpar}, one can ask: 

\textit{Does adding the assumption of a definite causal order between the $N$ parties impose any constraints on the correlations they can achieve via local interventions?}

Oreshkov \textit{et al.}~\cite{OCB12} showed that this is indeed the case and termed these constraints \textit{causal inequalities}. Here we briefly recall the notions we will need to state and explain the key results of Refs.~\cite{KB22, KO23}.

\subsection{Correlational scenario}
A \textit{correlational scenario} consists of $N$ parties, where party $S_k$ has settings $A_k$ and each setting $a_k\in A_k$ has a set of possible outcomes $X_{a_k}$. Here $k\in[N]:=\{1,2,\dots,N\}$. Without loss of generality, we consider the situation where $|A_k|=M$ and $|X_{a_k}|=D$ for all $k, a_k$.\footnote{There is no loss of generality in the following sense: In any correlational scenario, one can define 
	$D:=\max_{a_k\in A_k, k\in[N]} |X_{a_k}|$ and $M:=\max_{k\in [N]}|A_k|$. One can then always add trivial outcomes---those that never occur---for settings that have fewer outcomes than $D$ in order to make sure that all settings have $D$ outcomes; similarly, one can always add trivial settings---those with a fixed outcome that always occurs, supplemented with $D-1$ trivial outcomes that never occur---for any party that has fewer settings than $M$ settings.} Hence, we will denote a given correlational scenario via the triple $(N,M,D)$, in a similar manner as in the case of Bell scenarios \cite{BCP14}. Here $N$ is the number of parties, $M$ is the number of measurement settings per party, and $D$ is the number of outcomes for each measurement setting.
For party $S_k$, the setting $A_k$ takes values $a_k\in\{0,1,\dots,M-1\}$, the outcome $X_k$ takes values $x_k\in\{0,1,\dots,D-1\}$. We summarize the $N$-party settings and outcomes below:
\begin{align}
	\textrm{Settings: }\vec{A}&:=(A_1,A_2,\dots,A_N),\nonumber\\
	\vec{a}&:=(a_1,a_2,\dots,a_N)\nonumber\\
	\textrm{Outcomes: }\vec{X}&:=(X_1,X_2,\dots,X_N)\nonumber\\
	\vec{x}&:=(x_1,x_2,\dots,x_N).
\end{align}
The observed probabilistic behaviour, or correlation, between the parties is given by $P_{\vec{X}|\vec{A}}: \vec{X}\times \vec{A} \rightarrow [0,1]$, satisfying non-negativity and normalization, \textit{i.e.},
\begin{align}
	P_{\vec{X}|\vec{A}}(\vec{x}|\vec{a})&\geq 0 \quad\forall \vec{x},\vec{a},\nonumber\\
	\sum_{\vec{x}}P_{\vec{X}|\vec{A}}(\vec{x}|\vec{a})&=1\quad\forall\vec{a}.
\end{align}
It will be convenient to think of this correlation as a $D^N\times M^N$ column-stochastic matrix of probabilities. Since there are no constraints beyond non-negativity and normalization on the correlation, the different probability distributions (columns of the matrix) comprising the correlation are independent. The column-stochasticity means that we can take each correlation $P_{\vec{X}|\vec{A}}$ as defining a point in the $M^N(D^N-1)$-dimensional real vector space of multipartite correlations. The set of all such correlations defines a \textit{correlation polytope} with deterministic vertices that are given by Boolean column-stochastic matrices, \textit{i.e.}, those given by $P_{\vec{X}|\vec{A}}:\vec{X}\times \vec{A} \rightarrow \{0,1\}$ satisfying normalization for each choice of setting $\vec{A}$. As noted earlier, we will often use the shorthand $p(\vec{x}|\vec{a}):=P_{\vec{X}|\vec{A}}(\vec{x}|\vec{a})$ to denote the entries of the correlation matrix $P_{\vec{X}|\vec{A}}$. 

\subsection{Causal correlations}

The correlation $P_{\vec{X}|\vec{A}}$ is said to be causal if and only if it can be expressed as
\begin{align}\label{eq:causalcorr}
	p(\vec{x}|\vec{a})=\sum_{k=1}^Nq_k p(x_k|a_k)p_{x_k}^{a_k}(\vec{x}_{\backslash k}|\vec{a}_{\backslash k}),
\end{align}
where $q_k\geq 0$, $\sum_{k=1}^Nq_k=1$, and $p_{x_k}^{a_k}(\vec{x}_{\backslash k}|\vec{a}_{\backslash k})$ is a causal correlation between $(N-1)$ parties. Here $\vec{x}_{\backslash k}$ denotes the tuple $(x_1,x_2,\dots,x_{k-1},x_{k+1},\dots,x_N)$ (\textit{i.e.}, $\vec{x}$ without $x_k$) and similarly for $\vec{a}_{\backslash k}$. A $1$-party correlation $p(x_k|a_k)$ is, by definition, causal. This recursive form of causal correlations was originally derived from a principle of causality which essentially says that a freely chosen setting cannot be correlated with properties of its causal past or causal elsewhere \cite{OG16}. An intuitive way to interpret Eq.~\eqref{eq:causalcorr} is the following: In a scenario with definite (but possibly unknown) causal order, (at least) one party must be in the causal past or causal elsewhere of every other party. The probability $q_k$ for this to be the case for a specific party $S_k$ cannot depend on anyone's settings, while the probability $p(x_k|a_k)$ for the outcome of that party to take a particular value cannot depend on the settings of the others \cite{OG16}.

For the case of deterministic correlations---namely, those where $p(\vec{x}|\vec{a})\in\{0,1\}$ for all $\vec{x},\vec{a}$---the above definition reduces to 
\begin{align}\label{eq:causalcorrdet}
	&\exists k\in\{1,2,\dots, N\}:\nonumber\\
	&p(\vec{x}|\vec{a})=\delta_{x_k, f_k(a_k)}\delta_{\vec{x}_{\backslash k},f_{\backslash k}^{a_k}(\vec{a}_{\backslash k})},
\end{align}
where $p(\vec{x}_{\backslash k}|\vec{a}):=\delta_{\vec{x}_{\backslash k},f_{\backslash k}^{a_k}(\vec{a}_{\backslash k})}$ is a deterministic causal correlation between $N-1$ parties, specified by the set of functions 
\begin{align}
	f_{\backslash k}^{a_k}:=(f_1^{a_k},f_2^{a_k},\dots,f_{k-1}^{a_k},f_{k+1}^{a_k},f_{k+2}^{a_k},\dots, f_N^{a_k}),
\end{align}
where $f_j^{a_k}:\vec{A}_{\backslash k}\rightarrow X_j$ for all $j\in[N]\backslash\{k\}$.
In this deterministic case, the recursive nature of the definition entails that every
subset of parties must admit at least one party whose outcome is independent of the other parties' settings in this subset for all settings of parties in the complementary subset.\footnote{The complementary subset is empty if we are considering the full set of parties.}

The causal correlations in a correlational scenario define its \textit{causal polytope}, \textit{i.e.}, the set of all correlations that admit decompositions of the form in Eq.~\eqref{eq:causalcorr}. Facets of the causal polytope define the set of (facet) causal inequalities whose satisfaction is necessary and sufficient for membership in the causal polytope. More generally, a causal inequality (facet or otherwise) provides a necessary condition that all causal correlations must satisfy. Hence, violation of a causal inequality rules out membership in the causal polytope, thus witnessing indefinite causal order in a device-independent manner.

\subsection{Noncausal correlations}
Correlations that do not admit a decomposition of the type in Eq.~\eqref{eq:causalcorr} are said to be \textit{noncausal}, \textit{i.e.}, they lie outside the causal polytope. Below we give two examples of causal inequalities---one bipartite and one tripartite---and review their violations \cite{BAF15, BW16}.

\subsubsection{Bipartite quantum violation:} We consider the Guess Your Neighbour's Input (GYNI) inequality as our illustrative example. Each party $S_k$ (where $k=1,2$) receives a setting $x_k\in\{0,1\}$ and must produce an outcome $a_k\in\{0,1\}$ such that
\begin{align}
	x_1=a_2, x_2=a_1,
\end{align}
that is, each party must perfectly guess the other party's input setting. Assuming uniformly random choices of settings, the winning probability in the GYNI game is given by 
\begin{align}\label{gyni}
	p_{\rm gyni}:=\frac{1}{4}\sum_{a_1,a_2,x_1,x_2}\delta_{x_1,a_2}\delta_{x_2,a_1}p(x_1,x_2|a_1,a_2)
\end{align}
An optimal causal strategy in this game is the following: party $S_1$ communicates its setting to party $S_2$ who then copies it and reports $x_2=a_1$; party $S_1$ is then forced to make a uniformly random guess for $a_2$ in assigning $x_1\in\{0,1\}$. This gives us the causal inequality
\begin{align}
	p_{\rm gyni}\leq\frac{1}{2}.
\end{align}
This inequality can be violated by a bipartite process matrix wherein each party receives and sends out a qubit system, implementing the appropriate local interventions; this protocol is described in Ref.~\cite{BAF15} and we will not repeat it here; it achieves a value of $p_{\rm gyni}=\frac{5}{16}\left(1+\frac{1}{\sqrt{2}}\right)\approx 0.53347$. 

\subsubsection{Tripartite classical violation:}
Consider the following tripartite game \cite{BW16}: on receiving the settings $a_1,a_2,a_3$ uniformly randomly, the parties must produce outcomes $x_1,x_2,x_3$ that satisfy the constraints
\begin{align}
	&x_1=a_3, x_2=a_1, x_3=a_2, \textrm{ when } {\rm maj}(a_1,a_2,a_3)=0,\nonumber\\
	&x_1=\bar{a}_2, x_2=\bar{a}_3, x_3=\bar{a}_1, \textrm{ when } {\rm maj}(a_1,a_2,a_3)=1,
\end{align}
where ${\rm maj}(a_1,a_2,a_3)$ assigns value $0$ if the majority of the three settings are $0$, and $1$ otherwise. No causal strategy can win this game perfectly \cite{BW16} and we have the following causal inequality bounding the winning probability:
\begin{align}\label{eq:afbwcausineq}
	p_{\rm afbw}:=&\frac{1}{2}\sum_{\vec{x},\vec{a}}p(x_1,x_2,x_3|a_1,a_2,a_3)
	\Big(\delta_{x_1,a_3}\delta_{x_2,a_1}\delta_{x_3,a_2}\delta_{{\rm maj}(a_1,a_2,a_3),0}\nonumber\\
	+&\delta_{x_1,\bar{a}_2}\delta_{x_2,\bar{a}_3}\delta_{x_3,\bar{a}_1}\delta_{{\rm maj}(a_1,a_2,a_3),1}\Big)\\
	\leq&\frac{3}{4}.
\end{align}
The AF/BW (or Lugano) process function violates this inequality maximally, achieving $p_{\rm afbw}=1$. This process function is given by $p(\vec{i}|\vec{o})=\delta_{i_1,\bar{o}_2o_3}\delta_{i_2,\bar{o}_3o_1}\delta_{i_3,\bar{o}_1o_2}$ and the interventions under which it wins the tripartite game above perfectly are given by $p(x_k,o_k|a_k,i_k)=\delta_{x_k,i_k}\delta_{o_k,a_k}$ (\textit{cf}.~Eq.~\eqref{eq:loccons1})

\section{Trading causal order for locality}

\subsection{Quantum nonlocality without entanglement}

The phenomenon of quantum nonlocality without entanglement (QNLWE) was first discussed by Bennett \textit{et al.}~\cite{BDF99}. QNLWE is the following property of an ensemble of multipartite product states: although the elements of this ensemble can all be prepared via local operations and classical communication (LOCC), they cannot be perfectly discriminated via LOCC. That is, a QNLWE ensemble exhibits a form of ``nonlocality" (not to be confused with Bell nonlocality \cite{BCP14}) wherein locally preparable states cannot be perfectly discriminated locally. We consider a three-qubit example of a QNLWE ensemble that Bennett \textit{et al.}~referred to as the SHIFT ensemble, given by 
\begin{align}
{\rm SHIFT}:=\begin{split}
		\{&\ket{000},\ket{111},\ket{{+}01},\ket{{-}01},\\&\ket{1{+}0},\ket{1{-}0},\ket{01{+}},\ket{01{-}}\}
		\,.
	\end{split}
	\label{eq:shift}
\end{align} 

A causal intuition for why this ensemble cannot be perfectly discriminated via LOCC is as follows: To perfectly discriminate elements of the SHIFT ensemble, it is necessary and sufficient to implement a projective measurement in the basis given by elements of this ensemble.\footnote{Since the SHIFT ensemble forms an orthonormal basis.} Now, in any LOCC protocol there must be at least one party that initiates the protocol.\footnote{This is because the CC is implicitly presumed to be causally ordered in LOCC.} To implement the SHIFT measurement via LOCC then requires that the party initiating the protocol must measure in a fixed basis, independent of other parties' subsequent local operations. However, a simple inspection of the SHIFT basis elements reveal that each party switches its measurement basis depending on the remaining parties. Hence, none of the parties can initiate an LOCC protocol that implements the SHIFT measurement and the SHIFT ensemble therefore exhibits QNLWE. 

This causal intuition comes in handy when we consider process functions and how they help implement operations impossible with LOCC.

\subsection{Local Operations and Process Functions (LOPF)}
Consider now a scenario where, although each party can implement arbitrary local quantum operations (``LO"), the classical communication between parties is modelled by a process function (``PF"), \textit{i.e.}, the parties are allowed to signal noncausally while still being subject to logical consistency. We will refer to the set of operations that the parties can implement in this scenario as LOPF (Local Operations and Process Functions).

Although the SHIFT measurement cannot be implemented with LOCC, Ref.~\cite{KB22} showed that it can be implemented with LOPF, using the AF/BW process function \cite{BW16} given by 
\begin{align}
	i_1=\bar{o}_2o_3, i_2=\bar{o}_3o_1, i_3=\bar{o}_1o_2,
\end{align}
where $i_k\in I_k:=\{0,1\}, o_k\in O_k:=\{0,1\}$ for all $k\in\{1,2,3\}$. 
\begin{figure}
	\centering
	\includegraphics[scale=0.4]{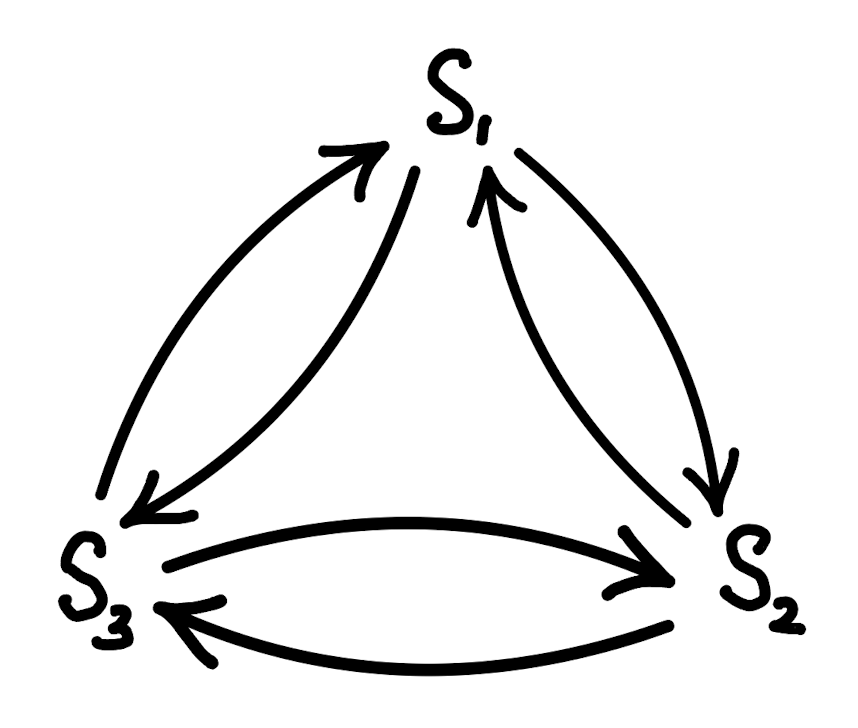}
	\caption{Causal structure of the AF/BW process \cite{BW16}: Each party is in the causal future of the other two parties.}
	\label{fig:afbw}
\end{figure}
Operationally, this amounts to the following perfect discrimination task:\\

\textit{Each party is given a qubit with the promise that the three-qubit state they hold is from the SHIFT ensemble. The parties must then infer, with probability $1$, which of the eight states they share.}\\

Our LOPF protocol works as follows:
\begin{enumerate}
	\item On receiving input $i_k$, party $S_k$ applies the gate $H^{i_k}$ to its qubit, where $H^0=\id, H^1=H$, $H$ being the Hadamard gate.
	\item Party $S_k$ then implements a $Z$ measurement on the qubit and records its outcome as $o_k\in\{0,1\}$, which is sent to the environment.\footnote{As usual, the projection $\ketbra{0}{0}$ is labelled $o_k=0$ and $\ketbra{1}{1}$ is labelled $o_k=1$.}	
	\item Party $S_k$ then applies $H^{i_k}$ to its qubit again.
\end{enumerate}
This protocol implements the following transformation on any three-qubit state $\ket{\psi}$ that the parties share:
\begin{align}
	\sum_{o_1,o_2,o_3}
	\left|\bra{o_1o_2o_3}H^{(i_1,i_2,i_3)}\ket\psi\right|^2
	H^{(i_1,i_2,i_3)}
	\ket{o_1o_2o_3}
	\bra{o_1o_2o_3}
	H^{(i_1,i_2,i_3)},
	\label{eq:final}
\end{align}	
where $H^{(i_1,i_2,i_3)}:=H^{i_1}\otimes H^{i_2}\otimes H^{i_3}$. Assuming $\ket{\psi}\in{\rm SHIFT}$, it is easy to verify that the choice of local measurements (labelled by $\vec{i}$) and their outcomes (labelled by $\vec{o}$) perfectly identify $\ket{\psi}$ via the following correspondence:
\begin{center}
	\begin{tabular}{ |c|c| } 
		\hline
		$(\vec{i},\vec{o})$& $\ket{\psi}$ \\
		\hline
		$(000,000)$& $\ket{000}$ \\
		\hline 
		$(000,111)$& $\ket{111}$ \\
		\hline
		$(100,001)$& $\ket{+01}$ \\
		\hline
		$(100,101)$& $\ket{-01}$ \\
		\hline
		$(010,100)$& $\ket{1+0}$ \\
		\hline
		$(010,110)$& $\ket{1-0}$ \\
		\hline
		$(001,010)$& $\ket{01+}$ \\
		\hline
		$(001,011)$& $\ket{01-}$ \\
		\hline
	\end{tabular}
\end{center}

We now generalize this observation to Boolean process functions (\textit{i.e.}, those with $I_k=O_k=\{0,1\}$ for all $k=1,2,\dots,N$) that maximally violate causal order in the following sense: no party is in the global past of the other parties, \textit{i.e.}, each party can receive a signal from at least one other party.
Going beyond the specific example of the SHIFT basis, Ref.~\cite{KB22} proves the following general theorem for the multiqubit case:

\begin{Th}\label{thm:booleanpf}
If~$\omega:=(\omega_1,\omega_2,\dots,\omega_n):\vec{O}\rightarrow\vec{I}$ is a Boolean~$n$-party process function without global past,
then
\begin{align}
	\mathcal S_{\omega}:=
	\left\{
	H^{(\omega(\vec{o}))}\ket{\vec{o}} \mid \vec{o}\in\{0,1\}^n
	\right\}
\end{align}
is a basis of orthonormal states that exhibits QNLWE.	
\end{Th}
We refer the interested reader to Ref.~\cite{KB22} for the proof of Theorem \ref{thm:booleanpf}.
\subsection{The gap between LOCC and SEP}
What the case of the SHIFT basis and the more general Theorem \ref{thm:booleanpf} demonstrate is the strict inclusion ${\rm LOCC}\subsetneq {\rm LOPF}$, \textit{i.e.}, process functions allow parties to implement operations that are impossible with LOCC. It is also well-known that ${\rm LOCC}\subsetneq {\rm SEP}$. Hence, LOPF allows us to bridge some of the gap between LOCC and SEP operations. Does it, however, cover all of SEP?  No.

The fact that there is a strict gap between LOPF and SEP, \textit{i.e.}, ${\rm LOCC}\subsetneq {\rm LOPF}\subsetneq{\rm SEP}$, follows from the example of the bipartite \textit{domino states}, also presented in Ref.~\cite{BDF99}. These are two-qutrit states that form an orthonormal basis given by
\begin{align}
	\{\ket{0+},\ket{0-},\ket{+2},\ket{-2},\ket{2+'}\ket{2-'},\ket{+'0},\ket{-'0}, \ket{11}\},
\end{align}
where $\{\ket{0},\ket{1},\ket{2}\}$ is the qutrit computational basis,  $\ket{+}:=\frac{1}{\sqrt{2}}\left(\ket{0}+\ket{1}\right)$, $\ket{-}:=\frac{1}{\sqrt{2}}\left(\ket{0}-\ket{1}\right)$,  $\ket{+'}:=\frac{1}{\sqrt{2}}\left(\ket{1}+\ket{2}\right)$, and $\ket{-'}:=\frac{1}{\sqrt{2}}\left(\ket{1}-\ket{2}\right)$. One way to see that the domino basis is an example of QNLWE is again via a causal intuition: in any LOCC protocol, one of the parties must have a fixed measurement basis; however, in the domino basis, each party makes a change of measurement basis depending on the other party; this means the domino basis cannot be implemented via an LOCC protocol. Since we know from Ref.~\cite{OCB12} that all bipartite process functions always yield causal correlations, they cannot be used to go outside LOCC. Hence, we have that LOPF cannot bridge the gap between LOCC and SEP.\footnote{See also Refs.~\cite{AOKM17,Akibue22} and their discussion in Ref.~\cite{KB22}.}

\section{Antinomicity, or nonclassicality in correlations without causal order}

Standard notions of nonclassicality for quantum correlations (\textit{e.g.}, contextuality, Bell or network nonlocality) presume an underlying causal structure---with a definite causal order---as a starting point. In a contextuality scenario, the underlying causal structure is one of a prepare-and-measure experiment,
where the preparation procedure is in the causal past of the measurement procedure. In a Bell scenario, the underlying causal structure is one where all the parties share a common cause in their global past and, furthermore, they cannot causally influence each other.
In network scenarios generalizing Bell, the underlying causal structure can be any directed acyclic graph (DAG).

However, in the absence of a definite causal order, how might one assess nonclassicality? We consider multipartite scenarios where no global causal order between the parties is assumed, following the operational paradigm we outlined in Section \ref{sec:oplpar}. The process-matrix framework \cite{OCB12} falls within this operational paradigm and it allows for the violation of causal inequalities. However, causal inequality violations occur even in the classical deterministic limit of the process-matrix framework \cite{BW16} (\textit{cf.}~Eq.~\eqref{eq:afbwcausineq}), contrary to expectations that they witness nonclassicality akin to Bell inequality violations \cite{OCB12}. So if it isn't the violation of a causal inequality, what other property of multipartite correlations might witness their nonclassicality?

We motivate \textit{antinomicity} as a property of correlations that certifies their nonclassicality in such scenarios \cite{KO23, KO24}. Intuitively, an antinomic correlation necessitates time-travel antinomies in any classical explanation, thus ruling out such explanations on pain of logical contradictions in one's physical theory. Unlike the classical case, we show that quantum processes without time-travel antinomies can exhibit antinomic correlations, but crucially, their ability to do so is limited. When parties do not communicate, antinomicity reduces to Bell nonlocality. 

In the following sections, we provide an overview of the results reported in Refs.~\cite{KO23, KO24}.

\subsection{A notion of (non)classicality in the presence of indefinite causal order}
For nonsignaling correlations $p(\vec{x}|\vec{a})$, local causality \cite{Bell76, Wiseman14} requires that $p(\vec{x}|\vec{a})=\sum_{\vec{i}}\prod_{k=1}^Np(x_k|a_k,i_k)p(\vec{i})$, where $\vec{i}=(i_1,i_2,\dots,i_N)$ denotes a source of classical shared randomness that is distributed among the parties and $p(x_k|a_k,i_k)$ denotes the local strategy of party $S_k$ with the key feature that it is independent of the settings and outcomes of other (spacelike separated) parties. Here $p(\vec{i})$ lives in a probability simplex with the vertices of the simplex denoting deterministic assignments to $\vec{i}$.
In terms of classical processes, local causality is mathematically equivalent to requiring that the parties cannot signal to each other via the environment, \textit{i.e.},  $p(\vec{i}|\vec{o})=p(\vec{i})$ for all $\vec{o}$, so that $\vec{o}$ in Eq.~\eqref{eq:loccons1} can be marginalized and we recover correlations within the Bell polytope. Hence, $p(\vec{i})$ is a nonsignaling classical process. It can be understood as a probabilistic mixture of deterministic nonsignaling classical processes, \textit{i.e.}, $p(\vec{i})=\sum_lp(l)\delta_{\vec{i},\vec{i}_l}$, where $l$ labels deterministic assignments $\vec{i}_l$ to $\vec{i}$. 

The above approach is consistent with the idea that any indeterminism in classical physical theories (like special or general relativity) can always be understood as one's lack of knowledge about an underlying physics that is fundamentally deterministic. In keeping with the same idea when we move to our correlational scenario (rather than Bell scenarios), we propose that the most general correlations achievable in a classical physical theory without definite causal order are those that can be understood as arising from probabilistic mixtures of process functions. We refer to this notion of classicality as \textit{deterministic consistency}, or simply, \textit{nomicity}. 

Deterministic consistency (or nomicity) can be viewed as a conjunction of two assumptions on the realizability of a correlation via some classical quasi-process under local interventions:\footnote{As we will show further on, \textit{every} correlation admits a realization with a classical quasi-process under local interventions if no further assumptions are imposed on the realization.} firstly, that the classical quasi-process satisfies \textit{logical consistency}, \textit{i.e.}, it is a classical process, and, secondly, that it satisfies \textit{determinism}, \textit{i.e.}, it lies within the deterministic-extrema polytope \cite{BW16}. We refer to the failure of deterministic consistency or nomicity (analogous to the failure of local causality) as \textit{antinomicity} (analogous to nonlocality), \textit{i.e.}, any correlation that fails to be nomic is \textit{antinomic}. This terminology is motivated by the fact that antinomicity entails the presence of time-travel antinomies \cite{Baumeler17, BT21} in any underlying classical explanation of the correlation.

\subsection{Strict hierarchy of correlation sets}
We can now define a hierarchy of sets of correlations as follows: i) Deterministically Consistent (nomic) correlations $\mathcal{DC}$ (achievable by convex mixtures of process functions), ii) Probabilistically Consistent correlations $\mathcal{PC}$ (achievable by classical processes), iii) Quantum Process correlations $\mathcal{QP}$ (achievable by process matrices), and iv) Quasi-consistent correlations $\mathcal{qC}$ (achievable by classical quasi-processes). Our main result establishes the following strict inclusions: 
\begin{equation}
	\mathcal{DC}\subsetneq\mathcal{PC}\subsetneq\mathcal{QP}\subsetneq\mathcal{qC}.	
\end{equation}
It is easy to see that $\mathcal{qC}$ is the set of all multipartite correlations. Given any multipartite correlation $p(\vec{x}|\vec{a})$, one can simply encode this correlation in a classical quasi-process defined via 
\begin{align}
	p(\vec{i}|\vec{o}):=\sum_{\vec{x},\vec{a}}p(\vec{x}|\vec{a})\delta_{\vec{i},\vec{x}}\delta_{\vec{o},\vec{a}}.
\end{align}
Given such a classical quasi-process, the correlation $p(\vec{x}|\vec{a})$ can be recovered via the following local interventions: each party simply passes the value $i_k$ it receives from this classical quasi-process to the outcome $x_k$ and the value of the setting $a_k$ it receives (from a referee) to the value of the output $o_k$ it sends to the environment; mathematically,
\begin{align}
	\forall k\in [N]: p(x_k,o_k|a_k,i_k)=\delta_{x_k,i_k}\delta_{o_k,a_k},
\end{align}
so that we have 
\begin{align}
	\sum_{\vec{i},\vec{o}}\prod_{k=1}^Np(x_k,o_k|a_k,i_k)p(\vec{i}|\vec{o})=p(\vec{x}|\vec{a}).
\end{align}
The following theorem then is key to these strict inclusions.
\begin{Th}\label{thm:det}
	Every deterministic correlation that can be realized by a process matrix can also be realized by a process function.
\end{Th}
Theorem \ref{thm:det} follows from Theorem 4 in Ref.~\cite{KO23}. 
It can be viewed as a generalization of the following observation that holds in Bell scenarios: every deterministic nonsignaling correlation (vertices of the Bell polytope) that can be realized by a quantum state can also be realized by a local hidden variable model. The logic of the strict inclusions is then as follows. 

1) $\mathcal{QP}\subsetneq\mathcal{qC}$: The bipartite Guess Your Neighbor's Input (GYNI) game requires each party to guess (as \textit{outcome}) the other party's input (\textit{setting}) \cite{BAF15}. A correlation that wins the GYNI game perfectly entails that for $a_1,a_2,x_1,x_2\in \{0,1\}$, the parties $S_1$ and $S_2$ should guess each other's inputs (settings) deterministically, \textit{i.e.}, $x_1=a_2$ and $x_2=a_1$. In the bipartite case, perfect GYNI correlation is unachievable by any process function since there are no causal inequality violations in the bipartite diagonal limit of the process-matrix framework \cite{OCB12}. Hence, by Theorem \ref{thm:det}, perfect GYNI correlation is impossible with process matrices, \textit{i.e.}, $\mathcal{QP}\subsetneq \mathcal{qC}$.

2) $\mathcal{PC}\subsetneq\mathcal{QP}$: This follows from the fact that in the bipartite case $\mathcal{DC}=\mathcal{PC}$ and that bipartite causal inequalities are violated by process matrices \cite{OCB12, BW16}.

3) $\mathcal{DC}\subsetneq\mathcal{PC}$: This strict inclusion follows from our construction of the tripartite Guess Your Neighbor's Input, or NOT (GYNIN) inequality and its violation, as we demonstrate below.

\textit{GYNIN game:} Three parties $S_1,S_2,S_3$ receive settings $a_1,a_2,a_3$ (respectively) and report outcomes $x_1,x_2,x_3$ (respectively) with the winning condition that either $(x_1,x_2,x_3)=(a_3,a_1,a_2)$ or $(x_1,x_2,x_3)=(\bar{a}_3,\bar{a}_1,\bar{a}_2)$. The winning probability when the settings are drawn uniformly at random is given by $p_{\rm gynin}:=\frac{1}{8}\sum_{\vec{x},\vec{a}}p(\vec{x}|\vec{a})\left(\delta_{x_1,a_3}\delta_{x_2,a_1}\delta_{x_3,a_2}+\delta_{x_1,\bar{a}_3}\delta_{x_2,\bar{a}_1}\delta_{x_3,\bar{a}_2}\right)$.

The following GYNIN inequality then serves as our witness of antinomicity:
\begin{align}\label{eq:gynin}
	p_{\rm gynin}\leq\frac{5}{8}.
\end{align}
This inequality is saturated by the deterministic AF/BW process \cite{BW16, KB22} but not by any causal strategy since the causal bound on the winning probability is $\frac{1}{2}$. Finally, this game can be won perfectly, \textit{i.e.}, with $p_{\rm gynin}=1$, for a probabilistically consistent correlation realized by the Baumeler-Feix-Wolf (BFW) process \cite{BFW14}. This establishes the strict inclusion $\mathcal{DC}\subsetneq\mathcal{PC}$. The causal bound follows quite similarly as in the case of other causal inequalities, \textit{e.g.}, GYNI inequality \cite{BAF15}.\footnote{For the proof, see Ref.~\cite{KO23}} The BFW process \cite{BFW14} can be expressed as a conditional probability distribution given by $p(\vec{i}|\vec{o}):=\frac{1}{2}\delta_{i_1,o_3}\delta_{i_2,o_1}\delta_{i_3,o_2}+\frac{1}{2}\delta_{i_1,\bar{o}_3}\delta_{i_2,\bar{o}_1}\delta_{i_3,\bar{o}_2}$, where $i_k,o_k\in\{0,1\}$ for all $k\in\{1,2,3\}$. The interventions on this process that, via Eq.~\eqref{eq:loccons1}, win the GYNIN game perfectly are given by $p(x_k,o_k|a_k,i_k)=\delta_{x_k,i_k}\delta_{o_k,a_k}$ for all $k\in\{1,2,3\}$.

\begin{figure}
	\centering
	\includegraphics[scale=0.5]{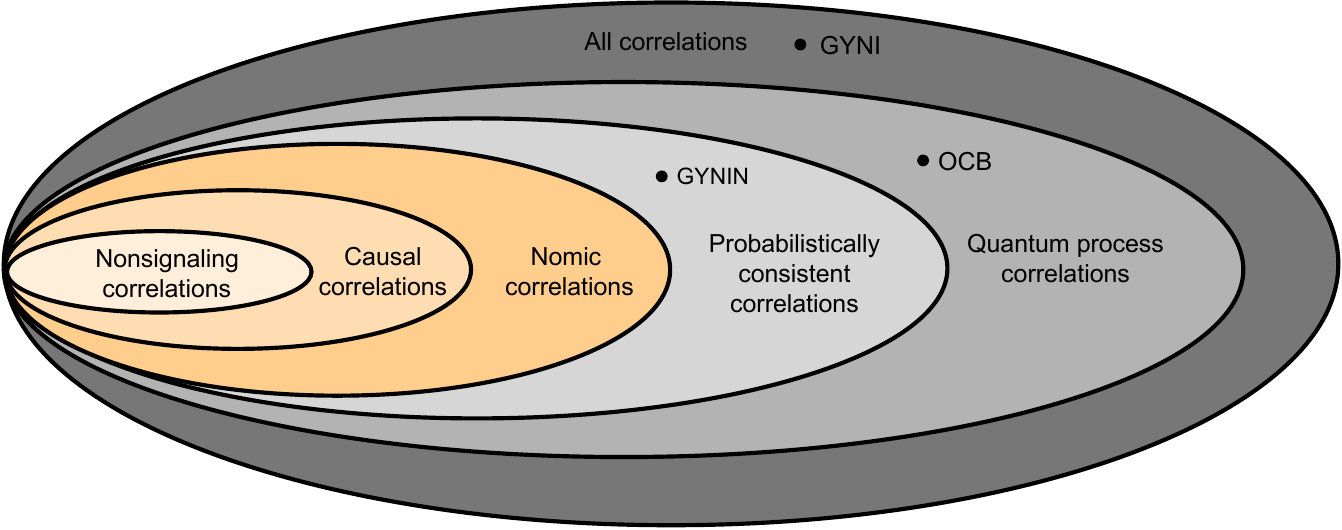}
	\caption{Inclusion relations between different sets of correlations. The point GYNIN refers to the correlation for which $p_{\rm gynin}=1$, achieved by the BFW process \cite{BFW14}. The point OCB refers to the correlation that wins the Oreshkov-Costa-Brukner game with probability $\frac{2+\sqrt{2}}{4}$ \cite{OCB12}. The point GYNI \cite{BAF15} refers to the correlation that wins the GYNI game with probability $1$. Antinomicity witnesses such as Eq.~\eqref{eq:gynin} separate nomic correlations from the rest.}\label{fig:setinclusions}
\end{figure}

\section{Conclusion}
By focussing on the properties of multipartite correlations, we have obtained results that demonstrate at least two senses in which there is an information-theoretic advantage from strategies that use indefinite causal order over strategies that assume a definite causal order (\textit{i.e.}, the existence of a DAG, even if unknown, that explains observed correlations). 
While on the one hand we have opened the door wide open to generalizing the dictionary between QNLWE and process functions, on the other hand, our introduction of antinomicity, and our proof that it cannot be arbitrarily strong in the process-matrix frameowrk, has opened the door to asking whether one might be able to obtain information-theoretic security guarantees in multipartite protocols where an adversary is not limited by the laws of causality (at least to the extent allowed by the process-matrix framework, \textit{cf.}~Figure \ref{fig:setinclusions}).
\newpage

\chapter{Joint measurability structures}\label{chap:CorrelationsJMS}

\toabstract{
We study the patterns of incompatibility relations between quantum measurements. Given a set of measurements, their incompatibility relations can be faithfully captured by a hypergraph---termed a \textit{joint measurability structure}---whose vertices represent the measurements and hyperedges represent exactly those subsets of vertices that are compatible. We survey three sets of results: 1) using a particular joint measurability structure to discriminate quantum correlations from almost quantum correlations, 2) exploring the set of joint measurability structures that admit qubit realizations, and 3) showing that every non-trivial joint measurability structure is capable of Bell inequality violations.}

\clearpage

\section{Introduction}
The incompatibility of measurements---namely, the impossibility of implementing arbitrary sets of them simultaneously---is a central feature of quantum theory that sets it apart from classical theories of physics. In the case of projective measurements, their compatibility is exactly captured by their commutativity, \textit{i.e.}, a set of projective measurements is compatible if and only if every pair in the set commutes. In the case of non-projective measurements (or POVMs), commutativity implies compatibility but the converse is not true, \textit{i.e.}, it is possible to have noncommuting POVMs that are nonetheless compatible. Here compatibility simply means that these POVMs can be recovered by classically coarse graining a single POVM.

In this chapter, we review results from Refs.~\cite{GKS18,AK20,YAK24} that all rely on a hypergraph representation of the incompatibility of a set of measurements, namely, its \textit{joint measurability structure}. Before we go on to review these results, we provide some definitions that will be essential for the exposition that follows.

We begin with the definition of joint measurability: 

\begin{Def}[Compatibility, or joint measurability, of POVMs]\label{jmdef}
	A set of POVMs, $\mathcal{M}=\{M_x\}_{x=1}^N$, each with outcome set $\mathcal{O}_x$, is said to be compatible or jointly measurable if each POVM in the set can be obtained by marginalizing the outcomes of a single POVM, $G$, with outcome set $\mathcal{O}:=\mathcal{O}_1\times\mathcal{O}_2\times\dots\times\mathcal{O}_N$, \textit{i.e.}, $M_{a|x}=\sum_{\vec{a}\in\mathcal{O}}^{a_x=a}G(\vec{a})$, for all $a_x=a\in \mathcal{O}_x$, $x\in\{1,2,\dots,N\}$. Here $\vec{a}=(a_x)_{x=1}^N\in \mathcal{O}$.	
\end{Def}

For any given set of POVMs, the pattern of incompatibility relations between the POVMs can be represented by its \textit{joint measurability structure}, which we define below (in a manner that is agnostic about the measurements being quantum):

\begin{Def}[Joint measurability structure]\label{jmsdef} A joint measurability structure on a set of measurements $\mathcal{M}$ is a hypergraph $(V_\mathcal{M},E_\mathcal{M})$, with the set of vertices $V_\mathcal{M}$, each vertex representing a different measurement in $\mathcal{M}$, and a set of hyperedges $E_\mathcal{M}=\{e|e\subseteq V^\mathcal{M}\}$ denoting all and only compatible (or jointly measurable) subsets of $\mathcal{M}$. Since every subset of a compatible set of POVMs is also compatible, in a valid joint measurability structure we must have $e'\subset e\in E_\mathcal{M}\Rightarrow e'\in E_\mathcal{M}$.
\end{Def}
A joint measurability structure is said to be \textit{non-trivial} if and only if it admits subset of incompatible vertices. It is said to be \textit{quantum-realizable} if and only if there exist quantum measurements that can be assigned to its vertices such that these measurements satisfy all the (in)compatibility relations specified by the joint measurability structure. In Ref.~\cite{KHF14} it was shown that all joint measurability structures admit quantum realizations via an explicit construction. Crucial to this construction is a particular class of joint measurability structures called $N$-Specker scenarios.
\begin{Def}[$N$-Specker scenario]\label{def:nspecker} An $N$-Specker scenario is a joint measurability structure on a set of $N\geq2$ incompatible measurements where every $(N-1)$-element subset of the set is compatible. 
\end{Def}
Here, a $2$-Specker scenario corresponds to a pair of incompatible measurements, a $3$-Specker scenario is the original Specker's scenario \cite{Specker60, Specker60_1, Specker60_2, LSW11, KG14}, while more general $N$-Specker scenarios (for $N\geq 4$) have also been studied in the literature \cite{ULMH16,AK20}.
Another family of joint measurability structures that are of interest to us in this chapter are $N$-cycle scenarios.

\begin{Def}[$N$-cycle scenario]\label{def:ncycle} An $N$-cycle scenario is a joint measurability structure on a set of $N\geq3$ measurements arranged in a cycle such that adjacent pairs in the cycle are compatible and all other subsets of measurements are incompatible.
\end{Def}
The key contributions on joint measurability that we will discuss in this chapter are the following:

\begin{enumerate}
	\item The first contribution \cite{GKS18} proves that the set of almost quantum correlations \cite{NGH15} can be discriminated from the set of quantum correlations in a principled way, \textit{i.e.}, while quantum correlations arise from quantum theory (which satisfies Specker's principle), any theory that aims to recover almost quantum correlations must necessarily fail to satisfy Specker's principle.\footnote{We will state it more formally later, but Specker's principle is simply the idea that pairwise compatibility of a set of measurements implies their global compatibility.}
		
	\item The second contribution \cite{AK20} addresses the following question: Are all joint measurability structures realizable with qubit POVMs? 
	
	By introducing a technique we term \textit{marginal surgery}, we provide concrete qubit realizations of several joint measurability structures, including infinite families like $N$-cycle and $N$-Specker scenarios. However, we do not settle the question in its full generality, providing potential counter-example(s) to an affirmative answer.
		
	\item The third contribution \cite{YAK24} considers the connection between incompatibility and Bell nonlocality: 
	
	Given an arbitrary non-trivial joint measurability structure on Alice's wing of a bipartite Bell experiment, does there always exist a quantum strategy---that is, POVMs for Alice that respect the joint measurability structure, an entangled state shared between Alice and Bob, and POVMs on Bob's part of the state---that permits a Bell inequality violation? 
	
	We answer this question in the affirmative by providing an explicit construction of such Bell inequality violations.
\end{enumerate}

\section{Almost quantum correlations are inconsistent with Specker's principle}
We now review the central result of Ref.~\cite{GKS18}, which answers the following question: \\

\textit{Is there a physical principle that can discriminate quantum theory from an almost quantum theory, \textit{i.e.}, a principle that is satisfied by one theory but fails for the other?}\\

Before we proceed, let us define what we mean by \textit{almost quantum} correlations in the language of probabilistic models on contextuality scenarios, as we did in Chapter \ref{chap:hypergraphframeworks}, Sec.~\ref{sec:probmodels}. Recall that a contextuality scenario is a hypergraph $H$ whose vertices $v\in V(H)$ represent measurement outcomes and whose hyperedges $e\in E(H)$ represent measurement settings, \textit{i.e.}, the vertices $v\in e$ represent a set of mutually exclusive and jointly exhaustive outcomes for the measurement setting represented by the hyperedge $e\in E(H)$. A probabilistic model on $H$ is defined as $p(v):V(H)\rightarrow [0,1]$ such that $\sum_{v\in e} p(v)=1$ for all $e\in E(H)$. The full set of probabilistic models on $H$ is $\mathcal{G}(H)$, the set of quantum models on $H$ is $\mathcal{Q}(H)$, and the set of classical models on $H$ is $\mathcal{C}(H)$ (\textit{cf.}~Sec.~\ref{sec:probmodels} for precise definitions). 

An \textit{almost quantum model} on $H$ \cite{GKS18} is a probabilistic model $p\in\mathcal{G}(H)$ that can be realized via some separable Hilbert space $\mathcal{H}$ such that (i) for each $v\in V(H)$, $\exists$ a projector $\Pi_v$, so that $\sum_{v\in e}\Pi_v\leq I_\mathcal{H}$ for all $e\in E(H)$, and (ii) $\exists$ a quantum state $\rho$ on $\mathcal{H}$ such that $p(v)=\Tr(\Pi_v\rho)$ for all $v\in V(H)$. We will denote by $\mathcal{Q}_1(H)$ the set of \textit{almost quantum models} on $H$, so that we now have $\mathcal{C}(H)\subsetneq \mathcal{Q}(H)\subsetneq\mathcal{Q}_1(H)\subsetneq\mathcal{G}(H)$.
The term ``almost quantum" originates in the fact that when this correlation set was first defined for Bell scenarios \cite{NGH15}, it was motivated as a set of correlations that satisfies all the physical principles that had until then been proposed for limiting non-signalling correlations to exactly the quantum set of correlations. In this sense, the almost quantum correlations provided a counter-point to the idea that such physical principles (\textit{e.g.}, consistent exclusivity \cite{AFL15}) could single out quantum theory.

An \textit{almost quantum theory} is any physical theory that can realize the almost quantum set $\mathcal{Q}_1(H)$ of correlations for any contextuality scenario $H$. Our goal is to obtain a principled distinction between quantum theory (which realizes the quantum set of correlations) and an almost quantum theory (which realizes the almost quantum set). In Ref.~\cite{SGAN18}, prior to our work \cite{GKS18}, it was shown that any generalized probabilistic theory that reproduces almost quantum correlations in Bell scenarios must fail the \textit{no-restriction hypothesis} (which is satisfied by quantum theory).\footnote{Roughly speaking, the no-restriction hypothesis requires that, given the set of measurements in a theory, all states that yield valid probabilities for these measurements are valid states in the theory, \textit{i.e.}, there are no restrictions on the set of allowed states beyond the fact that they yield valid probabilities for measurements in the theory.} In our work \cite{GKS18}, on the other hand, we show that even outside of Bell scenarios (e.g., the single-system setting of contextuality scenarios), one can show that an almost quantum theory fails to satisfy a principle---namely, Specker's principle---that is satisfied by quantum theory. 

As formalized in Ref.~\cite{GKS18}, Specker's principle is the principle of \textit{pairwise sufficiency for measurements}, \textit{i.e.},

\begin{quotation}
If in a set of measurements every pair is compatible, then all the measurements are compatible.	
\end{quotation}

This principle holds for sharp, \textit{i.e.}, projective, measurements in quantum theory: if in a set of projective measurements every pair is compatible,\footnote{Compatibility of a pair of projective measurements is equivalent to their commutativity.} then all the measurements are compatible.\footnote{The global joint measurement is simply the product of the commuting measurements.} Specker's principle also holds in any classical theory since there are all measurements are compatible in such a theory. 

To assess the validity of this principle in an almost quantum theory, we consider a set of $4$ measurements, $\{X_1,X_2,X_3,X_4\}$, that are pairwise compatible, \textit{i.e.}, the following subsets are compatible: $\{X_1,X_2\}$,$\{X_2,X_3\}$, $\{X_3,X_4\}$, $\{X_4,X_1\}$, $\{X_1,X_3\}$, $\{X_2,X_4\}$. Clearly, in quantum and classical theories, these $4$ measurements would also be globally compatible (on account of Specker's principle). This means that the outcome probabilities of these measurements would be such that the set of classical and quantum models coincide for any contextuality scenario arising from them: this is simply because the contextuality scenario would have exactly one hyperedge, namely, one that contains all the measurement outcomes.

What about an almost quantum theory? To answer this question, we need to consider a concrete contextuality scenario obtained from four pairwise compatible measurements. It suffices to assume that each measurement $X_i$ has binary outcomes $a_i\in O:=\{0,1\}$, where $i\in\{1,2,3,4\}$. Following is a facet inequality (termed a \textit{pentagonal inequality} \cite{RQS17, GKS18}) of the classical polytope in this contextuality scenario:

\begin{align}\label{eq:pent}
	I_{\mathrm{pent}} =& - \langle X_1X_2\rangle - \langle X_1X_3\rangle + \langle X_1X_4\rangle + \langle X_1 \rangle - \langle X_2X_3\rangle \\ \nonumber
	&+\langle X_2X_4\rangle+ \langle X_2\rangle + \langle X_3X_4\rangle + \langle X_3 \rangle - \langle X_4\rangle \leq 2\,,
\end{align}
where $\langle X_iX_j\rangle = \sum_{a_i,a_j \in O} (-1)^{a_i+a_j} p(a_ia_j|ij)$ and $\langle X_j\rangle = \sum_{a_j \in O} (-1)^{a_j} p(a_j|j)$. The upper bound of $2$ is respected by quantum models since quantum theory satisfies Specker's principle. However, almost quantum models can beat this upper bound, achieving $I_{\rm pent}=2.5$: hence, any almost quantum theory realizing such probabilistic models is inconsistent with Specker's principle.\footnote{A maximization over arbitrary probabilistic models on the resulting contextuality scenario yields $I_{\rm pent}=6$.} 

We have restricted our discussion here to describing the key result reported in Ref.~\cite{GKS18}. The interested reader may find more results and further technical details of the argument in Ref.~\cite{GKS18}.

\section{Qubit-realizable joint measurability structures}
The goal of this section is to review key results of Ref.~\cite{AK20}, which aimed to address the following broad question:\\

\textit{Are all joint measurability structures realizable via qubit measurements?}\\

The roots of this question lie in the general quantum realization for arbitrary joint measurability structures obtained in Ref.~\cite{KHF14}. The construction of Ref.~\cite{KHF14} requires a Hilbert space dimension that grows with the size of the number of vertices in the joint measurability structure. We therefore considered in Ref.~\cite{AK20} the possibility that \textit{all} joint measurability structures can be realized quantumly on a Hilbert space of constant dimension, the smallest case being a qubit Hilbert space. Key to the results of Ref.~\cite{AK20} is an approach we term \textit{marginal surgery} for obtaining desired incompatibility relations between subsets of measurements given a set of compatible measurements. We briefly describe this approach below and then highlight the main results from Ref.~\cite{AK20}.

\begin{figure}
	\centering
	\includegraphics[scale=0.2]{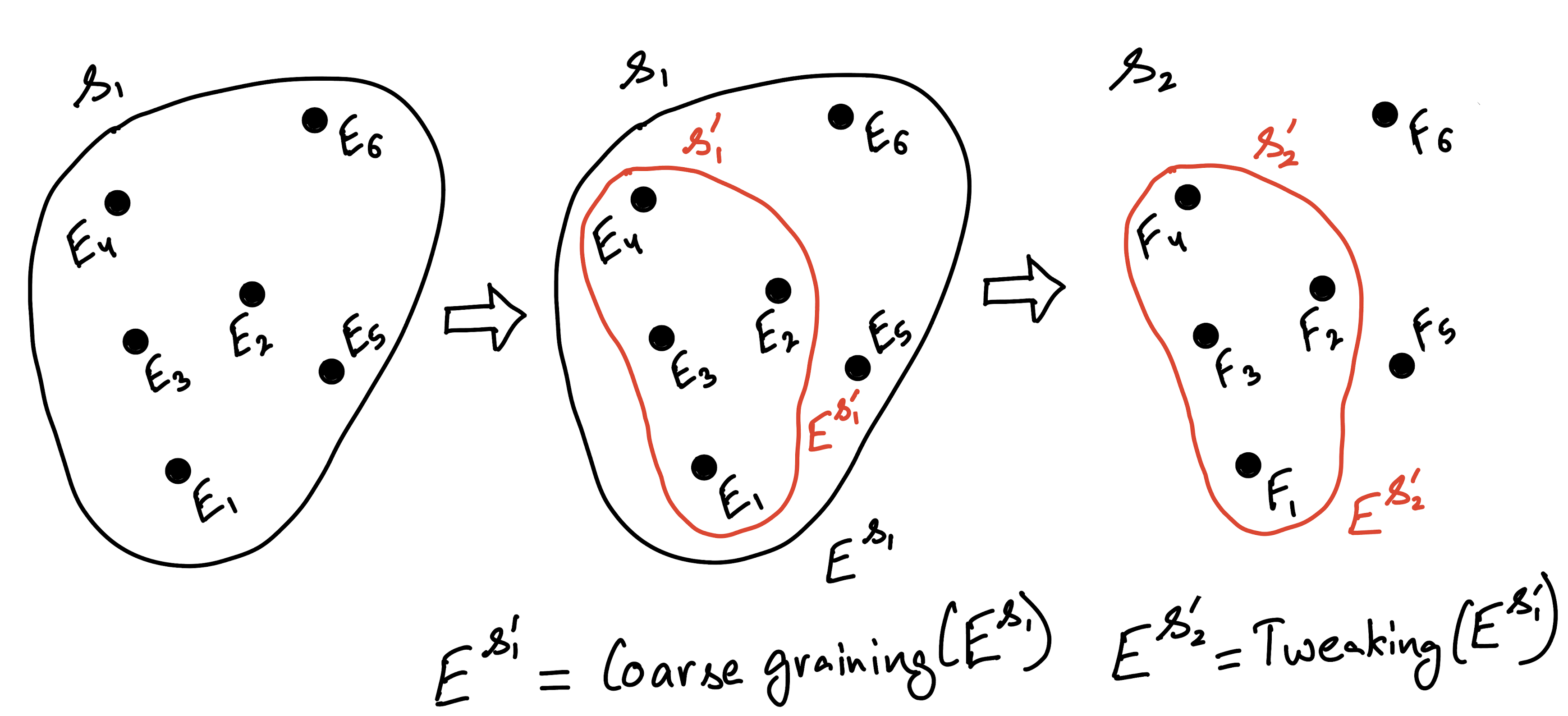}
	\caption{Illustration of the procedure of marginal surgery.}
	\label{fig:marginalsurgery}
\end{figure}

\textbf{General schema of marginal surgery:} We illustrate marginal surgery in Figure \ref{fig:marginalsurgery}: Consider a set of compatible POVMs $s_1:=\{E_1,E_2,\dots,E_6\}$ and a proper subset $s'_1\subsetneq s_1$, where $s'_1:=\{E_1,E_2,E_3,E_4\}$. Any joint POVM for $s_1$, given by $E^{s_1}$, defines a joint POVM for $s'_1$ (denoted $E^{s'_1}$), obtained by marginalizing (or coarse graining) the outcomes of all POVMs outside the set $s'_1$. Marginal surgery involves tweaking the marginal POVM $E^{s'_1}$ to obtain a new POVM $E^{s'_2}$ that is a joint POVM for a compatible set of POVMs $s'_2:=\{F_1,F_2,F_3,F_4\}$; however, this new compatible set $s'_2$ (of the same cardinality as $s'_1$) is now incompatible with two other POVMs $\{F_5, F_6\}$ (which are also modifications of $\{E_5,E_6\}$). 

\textit{Example with noisy Paulis:} Here's a more concrete example following the general scheme above: Consider the three noisy qubit Pauli measurements given by 
\begin{align}
	E^k_{\pm}:=\frac{1}{2}\left(\id\pm \eta \sigma_k\right),
\end{align}
where $k\in\{1,2,3\}$, $\eta\in [0,1]$, $\sigma_1=\begin{pmatrix} 0 & 1 \\ 1 & 0 \end{pmatrix}$, $\sigma_2=\begin{pmatrix} 0 & -i \\ -i & 0 \end{pmatrix}$, and $\sigma_3=\begin{pmatrix} 1 & 0 \\ 0 & -1 \end{pmatrix}$. We denote the POVMs by $M_k:=\{E^k_+,E^k_-\}$. The case $\eta=1$ is the case of projective Pauli measurements (every pair is incompatible) and the case $\eta=0$ is the case of trivial (coin flip) POVMs (the measurements are trivially compatible). In between these extreme cases, we have more interesting joint measurability conditions. The necessary and sufficient condition for the three POVMs to be compatible is $\eta \leq \frac{1}{\sqrt{3}}$ while the necessary and sufficient condition for any pair out of the three POVMs to be compatible is $\eta\leq \frac{1}{\sqrt{2}}$ \cite{HRS08,LSW11}.

Let us assume we are given a compatible set $s_1:=\{M_1,M_2,M_3\}$ where $\eta \leq \frac{1}{\sqrt{3}}$. Let us consider the compatible subset $s'_1:=\{M_1,M_2\}$. We modify this subset to another compatible set $s'_2:=\{M^*_1,M^*_2\}$ by increasing the value of $\eta$ to satisfy $\frac{1}{\sqrt{3}}<\eta\leq\frac{1}{\sqrt{2}}$; this ensures that the subset $s'_2$ becomes incompatible with respect to the third POVM $M_3$ (even though each member of $s'_2$ is individually compatible with $M_3$). In this way, we use this ``marginal surgery" (here it involves tweaking $\eta$ in the marginals) to obtain an incompatible set of measurements $\{M^*_1,M^*_2,M_3\}$. We could also further modify $M_3$ to, for example, define $M^*_3$ with $\eta=1$ (i.e., a Pauli measurement); this would ensure that both elements of the set $s'_2$ are individually incompatible with $M^*_3$, giving a different incompatible set 
$\{M^*_1,M^*_2,M^*_3\}$ (and a different joint measurability structure).

\textbf{Results:} Besides the introduction of marginal surgery, the main results obtained in Ref.~\cite{AK20} are the following:
\begin{enumerate}
	
	\item We use marginal surgery to construct quantum realizations of two families of joint measurability structures: $N$-cycle and $N$-Specker scenarios for any integer $N\geq 3$.
	
	\item We show that all joint measurability structures up to $N=4$ vertices are quantum-realizable. 
	
	\item We also obtain a sufficient condition for the joint measurability of any finite set of qubit POVMs with binary outcomes.\footnote{Obtaining joint measurability conditions applicable to multiple measurements (but also in higher dimensions) is known to be a hard problem in general \cite{HRS08,YLL10,LSW11,PG11}.}
\end{enumerate}
We also conjecture that \textit{all} joint measurability structures for any finite $N$ are realizable with qubit POVMs. To our knowledge, this conjecture is still open.

\section{Incompatibility vs.~Bell nonlocality}

The goal of this section is to review the central result of Ref.~\cite{YAK24}, which addresses the following question: \textit{Do all non-trivial joint measurability structures admit Bell-violating quantum realizations?}
§We first motivate this question and then describe how Ref.~\cite{YAK24} answers it in the affirmative. 

Bell scenarios consist of at least two spacelike separated parties that share an entangled state and implement local measurements on their share of the state. While entanglement and incompatibility are necessary for Bell inequality violations, they are not sufficient. We know that there exist entangled states that do not violate Bell inequalities \cite{Werner89, Barrett02}. It has also been shown that there exist incompatible measurements that are useless for Bell inequality violations \cite{BV18,HQB18}.

What does it mean to ask whether incompatibility is sufficient for Bell nonlocality? It means the following: given any set of incompatible POVMs that Alice can implement on her quantum system, does there always exist a bipartite entangled state that she can share with Bob and a set of POVMs that Bob can implement on his part of the state such that their joint statistics violates some Bell inequality? This question was recently settled in the negative via explicit counter-examples \cite{BV18,HQB18}. One of the counter-examples consists of three measurements that are pairwise compatible but triplewise incompatible (forming Specker's scenario) \cite{BV18} and the other involves an uncountably infinite number of measurements \cite{HQB18}. Thus, measurement incompatibility does not imply Bell nonlocality, similar to how there exist entangled states that do not violate Bell inequalities.

Although Specker's scenario admits a set of POVMs that do not violate any Bell inequality \cite{BV18}, it also admits sets of POVMs that do violate a Bell inequality \cite{BV18,QVB14}. The necessity of incompatibility for Bell nonlocality implies the necessity of a non-trivial joint measurability structure on each wing of a Bell experiment, \textit{i.e.}, the joint measurability structure of the measurements on each wing should contain a subset of incompatible vertices. Is a non-trivial joint measurability structure, however, sufficient for a Bell inequality violation? That is, despite the inequivalence of incompatibility and Bell nonlocality in a quantitative sense \cite{BV18,HQB18}, we ask whether a \textit{qualitative} equivalence between incompatibility and Bell nonlocality nevertheless holds: given any non-trivial joint measurability structure, can Alice identify a set of POVMs satisfying it such that there exists a bipartite entangled state that she can share with Bob and a set of POVMs that Bob can implement on his part of the state so that their joint statistics violates a Bell inequality?

We answer this question in the affirmative in the following steps:
\begin{enumerate}
	\item Given a $v$-vertex joint measurability structure $\mathcal{J}$, we first decompose it into $N$-Specker scenarios (for $2\leq N \leq v$) following the method proposed in Ref.~\cite{KHF14}.\footnote{The fact that this can be done for any joint measurability structure is a consequence of the partial order under set inclusion over subsets of incompatible vertices. We refer the interested reader to Section I.A of the Supplemental Material of Ref.~\cite{YAK24}.}
	\item For any particular $N$-Specker scenario in the above decomposition, we use the family of Bell-violating realizations constructed in Ref.~\cite{BV18} for all $N$-Specker scenarios, where $N\geq 3$. The case $N=2$ always admits a Bell-violating realization, \textit{e.g.}, it can violate the Bell-CHSH inequality.
	\item We combine the Bell-violating realizations of the component $N$-Specker scenarios to construct a Bell-violating realization of the given joint measurability structure $\mathcal{J}$. The Bell inequality that this realization violates is the $I_{vv22}$ Bell inequality \cite{CG2004I3322,Cereceda01}.\footnote{For $v=2$, this coincides with the CHSH inequality \cite{CHSH,CH74,Cereceda01}.}
\end{enumerate}

Our proof in Ref.~\cite{YAK24} is constructive, providing an explicit realization for any joint measurability structure using quantum measurements that are useful for a Bell inequality violation, building on the results of Refs.~\cite{KHF14,BV18}. It is also modular in that future constructions of Bell inequality violations using $N$-Specker scenarios can be easily incorporated in our construction to build Bell-violating realizations for arbitrary (non-trivial) joint measurability structures. Hence, we show a qualitative equivalence between incompatibility and Bell nonlocality, \textit{i.e.}, a non-trivial joint measurability structure is not only necessary but also sufficient for Bell nonlocality. Put differently, there are no non-trivial joint measurability structures that are ``useless" for a Bell inequality violation.

A fundamental question our work raises is the following: given a quantum realization of a non-trivial joint measurability structure, which features of this realization are responsible for a Bell inequality violation? A characterization of this type would help us obtain a finer handle on the relationship between incompatibility and Bell nonlocality by allowing us to target, when required, those measurements that are useful for Bell inequality violations. While the general characterization problem may be difficult to solve for arbitrary joint measurability structures, a lot of insight can be gained by studying Bell-violating vs.~Bell non-violating realizations of the simplest joint measurability structure beyond a pair of incompatible measurements, \textit{i.e.}, Specker's scenario with three binary outcome measurements. In Section IV of the Supplemental Material of Ref.~\cite{YAK24}, we make numerical progress towards characterizing qubit measurements realizing Specker's scenario that are also useful for Bell inequality violations.

\section{Conclusion}
In this chapter we have reviewed the main results from three contributions \cite{GKS18,AK20,YAK24} on joint measurability structures and their interplay with (almost) quantum correlations. Key to this conception of joint measurability is the representation of incompatibility relations using hypergraphs and using properties of these hypergraphs to infer the possibilities and limitations of their quantum realizations. In future work, we expect to investigate operational tasks that provide a more fine-grained understanding of incompatibility as a resource for quantum information and computation, \textit{e.g.}, tasks where the relevant resource is the \textit{pattern} of incompatibility relations exhibited by a set of measurements rather than the mere fact of their incompatiblity.
\newpage

\chapter{Conclusion}\label{chap:conclusion}

We have reviewed work on the role of discrete structures in characterizing quantumness via contextuality, causality, and the incompatibility of quantum measurements:

\begin{itemize}
	\item For contextuality, the discrete structures of interest are (undirected) graphs and hypergraphs; the invariants associated with these graphs and hypergraphs put bounds on the correlations achievable classically and the violation of these bounds certifies quantumness in the sense of contextuality.
	
	\item For causality, the discrete structures of interest are directed acyclic graphs (DAGs) as well as directed cyclic graphs. The cyclicity in directed graphs allows us to make sense of indefinite causal order in the process-matrix framework.
	
	\item For incompatibility, the discrete structures of interest are hypergraphs that represent abstract simplicial complexes \cite{KHF14}. Their decomposability into $N$-Specker scenarios plays a crucial role in the results we reviewed.
\end{itemize}

Other areas that I have worked on during my postdoctoral research include the framework of generalized probabilistic theories, how contextuality can be characterized within it, and the subtle relationship between contextuality and incompatibility \cite{SSW21,SSWSK21}. It also includes work on the resource theories of Bell nonclassicality and of entanglement under local operations and shared randomness (LOSR) as the set of free operations \cite{WSS20,SFK20}, a paper on a notion of spacetime entanglement entropy in quantum field theories \cite{CHK20}, and work on anomalous weak values and their relationship with generalized contextuality \cite{KLP19}.

The doctoral and postdoctoral phases of my research have so far largely focussed on using the tools of quantum information theory for a better understanding of quantum foundations. In the current phase of my research, and going forward, I aim to put quantum foundations into quantum practice, \textit{i.e.}, use the concepts developed in quantum foundations over the last couple of decades to better assess quantumness and its role in quantum information and quantum computation. This is crucial if we hope to obtain a principled account of the power of quantum computation and to guarantee that the quantum speedups it promises are not ephemeral \cite{Tang19}. 

\backmatter

\singlespace\pagestyle{plain}\pagenumbering{alph}

\bibliographystyle{unsrt}
\bibliography{masterbibfilev2}
\end{document}